\documentclass[fleqn,usenatbib]{mnras}

\usepackage{newtxtext,newtxmath}
\usepackage[T1]{fontenc}
\usepackage{xspace}
\DeclareRobustCommand{\VAN}[3]{#2}
\let\VANthebibliography\thebibliography
\def\thebibliography{\DeclareRobustCommand{\VAN}[3]{##3}\VANthebibliography}

\usepackage{graphicx}	% Including figure files
\usepackage{amsmath}	% Advanced maths commands
\usepackage{cprotect}
\usepackage{verbatim}
\usepackage{booktabs,siunitx,caption}
\usepackage{placeins}
\usepackage{float}
\newcommand{\microns}{\textmu m\xspace}

\newcommand{\Rj}{${R_{\mathrm {J}} }$\xspace}
\newcommand{\Mj}{${M_{\mathrm {J}} }$\xspace}
\title[Exoplanet spectroscopy with JWST NIRSpec]{Exoplanet Transit Spectroscopy with JWST NIRSpec: Diagnostics and   Homogeneous Case Study of WASP-39 b}

\author[S. Sarkar et al.]{
Subhajit Sarkar,$^{1}$\thanks{E-mail: subhajit.sarkar@astro.cf.ac.uk (SS)}
Nikku Madhusudhan,$^{2}$
Savvas Constantinou,$^{2}$
Måns Holmberg,$^{2}$
\\
$^{1}$School of Physics and Astronomy, Cardiff University, Queen’s Buildings, The Parade, Cardiff, CF24 3AA, UK\\
$^{2}$Institute of Astronomy, University of Cambridge, Madingley Road, Cambridge, CB3 0HA, UK
}

\date{Accepted XXX. Received YYY; in original form ZZZ}

\pubyear{2015}

\begin{document}
\label{firstpage}
\pagerange{\pageref{firstpage}--\pageref{lastpage}}
\maketitle

\begin{abstract}

The JWST has ushered in a new era of exoplanet transit spectroscopy. Among the JWST instruments, the Near-Infrared Spectrograph (NIRSpec) has the most extensive set of configurations for exoplanet time series observations.  The NIRSpec Prism and G395H grating represent two extremes in NIRSpec instrument modes, with the Prism spanning a wider spectral range (0.6-5.3 \textmu m) at lower resolution (R$\sim$100) compared to G395H (2.87-5.14 \textmu m; R$\sim$2700). In this work, we develop a new data reduction framework, JexoPipe, to conduct a homogeneous assessment of the two NIRSpec modes for exoplanet spectroscopy. We use observations of the hot Saturn WASP-39 b obtained as part of the JWST Transiting Exoplanets ERS program to assess the spectral quality and stability between the two instrument modes at different epochs. We explore the noise sources, effect of saturation, and offsets in transmission spectra between the different instrument modes and also between the two G395H NRS detectors. We find an inter-detector offset in G395H of $\sim$ 40-50 ppm, consistent with recent studies. We find evidence for correlated noise in the Prism white light curve. We find the G395H spectrum to be of higher precision compared to the Prism at the same resolution.  We also compare the JexoPipe  spectra with those reported from other pipelines. Our work underscores the need for robust assessment of instrument performance and identification of optimal practices for JWST data reduction and analyses. 

\end{abstract}


%%%%%%%%%%%%%%%%%%%%%%%%%%%%%%%%%%%%%%%%%%%%%%%%%%

%%%%%%%%%%%%%%%%% BODY OF PAPER %%%%%%%%%%%%%%%%%%

\section{Introduction}

Exoplanetary atmospheres are key to understanding the physical conditions, chemical composition and origins of exoplanets. Transmission spectroscopy \citep{Seager2000, Brown2001} has emerged over the past two decades as the most widely applied technique used to obtain exoplanet spectra.  Since the first detection of sodium in the upper atmosphere of the hot Jupiter HD 209458 b \citep{Charbonneau2002}, transmission spectroscopy has progressed with dozens of exoplanet atmospheres probed, and remains an extremely promising method of planetary remote sensing \citep{Madhusudhan2019}.  In particular over the last decade the Hubble Wide Field Camera 3 (WFC3) near-infrared (NIR) instrument G141 grism (1.1-1.7 \textmu m) has delivered high precision detections of H$_2$O in numerous hot Jupiters  \citep[e.g.,][]{Deming2013, McCullough2014, Sing2016} as well as in the sub-Neptune K2-18 b \citep{Benneke2019, Tsiaras2019}. 

Transmission spectroscopy depends on  wavelength-dependent absorption and scattering of stellar light by atmospheric atomic and molecular species as the planet transits in front of its host star. During the transit, a proportion of the star's light is effectively blocked in the line-of-sight by the atmosphere which can be represented as an opaque annulus which adds an apparent extra height to the bulk radius of the planet. Due to the wavelength dependence of the various opacities, the transit depth varies with wavelength and thus the apparent $(R_p/R_s)^2$ (where $R_p$ is the  planet radius and $R_s$ is the star radius). This technique probes the high altitude atmosphere at the planet day-night terminator.  In addition, atmospheric modelling and spectral retrieval methods have progressed to allow increasingly more sophisticated interpretation of these spectra \citep{MadhusudhanSeager2009, Madhusudhan2018}. Transmission spectroscopy is particularly powerful in the visible and near-infrared wavelength ranges, where the host star flux is maximal (minimising fractional photon noise), and where there are numerous atomic and molecular spectral signatures of molecules expected in planetary atmospheres.

The James Webb Space Telescope (JWST) promises to revolutionize our understanding of exoplanet atmospheres providing the highest quality transmission spectra ever obtained 
in terms of both precision (through its 6.5 m primary mirror in comparison to the 2.4 m Hubble primary mirror) and unparalleled wavelength coverage through its suite of four instruments: Near Infra-Red Slitless Spectrograph (NIRISS) \citep{Doyon2012}, Near Infra-red Camera (NIRCam) \citep{Beichman2012}, Near Infra-red Spectrograph (NIRSpec) \citep{Ferruit2014} and Mid-Infrared Instrument (MIRI) \citep{Rieke2015}.  Combined these instruments provide a potential wavelength coverage ranging from 0.6 to 24 \microns. In the past year, JWST has delivered several transmission spectra of exoplanets \citep[e.g.][]{Lustig-Yaeger2023, Lim2023, May2023, Madhusudhan2023, Bell2023, Kempton2023, Grant2023, Moran2023, Dyrek2024, Kirk2024} in addition to those completed under the Early Release Science (ERS) \citep{Bean2018} and Early Release Observations (ERO) \citep{Pontoppidan2022} programs \citep{ERS2023, Ahrer2023, Feinstein2023, Alderson2023, Rustamkulov2023, Radica2023, Taylor2023, Fournier-Tondreau2024}. These observations have led to confident detections of several prominent molecules, such as CO$_2$, H$_2$O and SO$_2$ in the hot Saturn WASP-39~b \citep[e.g.][]{Alderson2023,Feinstein2023,Rustamkulov2023,Tsai2023,Constantinou2023}, CH$_4$ in the warm Jupiter WASP-80 b \citep{Bell2023}, and CO$_2$ and CH$_4$ in the candidate Hycean world K2-18~b \citep{Madhusudhan2023}.

While only MIRI provides substantial wavelength coverage in the mid-infrared (MIR) beyond 5 \microns, there is a choice of three instruments in the near-infared (NIR) range: $\sim$ 1-5 \textmu m.  In addition, each of these NIR instruments have different configurations, i.e. combinations of dispersion element, filter, detector subarray  and readout pattern, that further expands the choices presented to an observer.  While the choice of configuration may at least partly depend on wavelength coverage, there are overlapping wavelength ranges between the different NIR instrument configurations: see Fig. 1  in \cite{Sarkar2021}. One of the key questions in this early stage of JWST operations is the relative performance of the different instruments and their configurations particularly in regions of overlapping wavelength and also the optimal data reduction methods to process each configuration. 

Of the available NIR instruments, NIRSpec presents the largest number of possible configurations with 7 different dispersive elements available.  Two of these elements: the Prism and the G395H grating, present two extremes.  The Prism mode gives the broadest wavelength coverage in a single pass (0.6-5.3 \textmu m) with high pixel count rates, but has the lowest spectral resolving power (R) of $\sim 100$, with the spectrum concentrated within a 512 pixel-wide detector subarray on the NRS1 detector (the SUB512 subarray which is 32 $\times$ 512 pixels in size) which causes it to saturate more easily than other modes. In contrast G395H has a high R of $\sim2700$ but covers about half the wavelength range of Prism (2.87–5.14 \textmu m), with the spectrum dispersed widely over the two NRS (2048 $\times$ 2048 pixel) detectors (using the SUB2048 subarray which is 32 $\times$ 2048 pixels in size over each of the two detectors).

In this paper we aim to compare these two key NIRSpec modes and uncover how similar the final results are between the two.  We do this by  performing a homogenized analysis of two observations taken of the hot Saturn mass planet, WASP-39 b, obtained as part of the JWST Transiting Exoplanet Community ERS program 1366 (PI: N. Batalha), one with NIRSpec Prism and the other with G395H.
To facilitate this work we developed an end-to-end data processing framework, JexoPipe, which takes the raw uncalibrated files and produces final transmission spectrum.  We minimise differences in processing between the two configurations, allowing us to better control and assess the causes of any differences between the final spectra in their overlapped regions.
 We further examine JexoPipe against previously developed pipelines by comparing the final spectra obtained.  
The ERS NIRSpec Prism observation of WASP-39b was originally presented in \cite{ERS2023} and \cite{Rustamkulov2023} and the G395H observation in \cite{Alderson2023}. 

%JexoPipe is one of a set of pipelines under current development for processing data in our own GO Cycle 1 and Cycle 2 JWST observations (programs 2722 and 3557) that includes JExoRES \citep{Holmberg2023}.  It will also eventually provide compatibility with the JWST time-domain simulator, JexoSim \citep{Sarkar2021}.  We will use JexoPipe and JexoRES to provide cross-checks on final results.  
 
In the following we first describe WASP-39 b and summarise previous results in Section 2 and then review NIRSpec and its performance in Section 3.  We describe JexoPipe in Section 4.  In Section 5, we present the final spectra obtained and compare these to those from other pipelines.  In Section 6 we discuss any differences in the spectra obtained from the different NIRSpec modes.  Section 7 describes application of an atmospheric forward model to the data.  We summarise our findings and conclusions in Section 8.

\section{WASP-39 {b}}
WASP-39 b is a highly inflated Saturn-mass  planet discovered in 2011 \citep{Faedi2011} transiting a G8 star ([Fe/H] = -0.12) located at a distance of 213.982 pc\footnote{https://exoplanetarchive.ipac.caltech.edu/overview/WASP-39}. It has an equilibrium temperature of 1166 K with an orbital period of 4.0552941 days \citep{Mancini2018}.  With a radius of 1.279 \Rj and mass of 0.281 \Mj, its bulk density is roughly four times less than of Saturn at 0.167 g/cm$^3$ \citep{Mancini2018}. 

Prior to JWST, NIR transmission spectra had been obtained with Hubble WFC3 \cite{Wakeford2018}, STIS and Spitzer IRAC \cite{Fischer2016, Sing2016}. Ground-based facilities have also contributed with a UV-optical transmission spectrum from VLT-FORS2 \cite{Nikolov2016} and UV-optical multi-band photometry \citep{Ricci2015}.  These spectra indicated a cloud-free atmosphere with the detection of sodium and potassium \citep{Fischer2016, Sing2016, Nikolov2016}. Water vapour was also detected with an estimated metallicity of $151^{+48}_{-46}$ $\times$ solar \citep{Wakeford2018}.  

A suite of JWST ERS NIR observations have been performed on WASP-39 b.  Using NIRISS SOSS (0.6-2.8 \textmu m) \citep{Feinstein2023} reported a super-solar metallicity with results varying from 10-30 $\times$ solar dependent on the atmospheric models applied, a sub-solar C/O ratio and super-solar K/O ratio.  Using the G395H grating \cite{Alderson2023} reported a somewhat lower metallicity of 3-10 $\times$  solar, with a sub-solar to solar C/O ratio.  They also report detection of CO$_2$, CO, H$_2$O and SO$_2$.  \cite{Rustamkulov2023} and \cite{ERS2023} reported results using the Prism configuration (0.6-5 microns), with detection of CO$_2$, CO, H$_2$O SO$_2$ and Na.  The best-fitting model corresponded super-solar metallicity and super-solar C/O ratio with moderate cloud opacity.  The SO$_2$ feature has also been discussed in \cite{Tsai2023} who reported that the feature could be explained by the photochemical breakdown of H$_2$S. 
Using NIRCam F322W2 (2-4 \microns), \cite{Ahrer2023} reported detection of water vapour and that best-fit chemical equilibrium models favoured a metallicity of 1-100 $\times$ solar with a substellar C/O ratio.

\section{NIRSpec}

NIRSpec is a complex instrument with multiple configurations which cover (in total) the wavelength range from 0.6-5.3 \textmu m.   Due to the wide wavelength coverage in the NIR (covering the spectral features of key atmospheric molecules at high SNR as well as the Rayleigh scattering slope) and the many choices of configuration (giving flexibity for different targets and brightnesses), NIRSpec will be one of the principal instruments used in JWST transmission spectroscopy. 
It incorporates an integral field unit (IFU) and a micro-shutter array (MSA) for multi-object spectroscopy and five slits (apertures) for individual spectroscopy.   The observing template for exoplanet time-series is the bright object time-series (BOTS) mode.  This utilises the large aperture S1600A (1600 x 1600 mas$^2$) minimizing slit losses, in combination with up to nine different disperser/filter combinations.  There are three high resolution (R $\sim$ 2700), each with an associated filter, and three medium resolution (R$\sim$1000) gratings, where G140H and G140M have two possible filters, and the others have one filter.  This gives eight possible grating/filter configurations.  In addition, the Prism mode is combined with the CLEAR filter to give the ninth configuration.  Two Teledyne H2RG 2048 $\times$ 2048 HgCdTe detector arrays are used called NRS1 and NRS2.   Spectra from the medium resolution gratings and Prism project only onto NRS1.  

A choice of detector subarrays exist with different frame times.  The detectors are read up-the-ramp producing non-destructive reads (NDRs) separated by the frame time. Up-the-ramp sampling allows for cosmic ray detection and possible recovery of affected slopes \citep{Giardino2019}. Most time-series observations are expected to use the NRSRAPID read pattern, where there is no on-board frame averaging, giving one NDR or frame per `group', and where one group time per integration ramp is lost to reset.  The observation proceeds as a sequence of integrations constituting an exposure. Thermal settling at the beginning of the exposure can occur  \citep{Birkmann2022}.   If the SUB2048 subarray is used (with the high resolution gratings), both detectors are implemented and a small gap will appear in any spectrum due to the physical separation between the two detectors.  While the detectors are identical in manufacture, the commissioning study by \cite{Espinoza2022} using G395H and HAT-P-14 b as the target found differences in the slope of the systematic trend (being stronger in NRS1) and how closely the measured scattered in light curves matched the calculated noise (being a closer match in NRS1).  All NIRSpec spectra have a slight curvature and tilt on the detector.

One of the main sources of systematics in NIRSpec is `1/f noise' which correlates counts during readout mainly the fast readout direction (and to a lesser extent in the slow direction) and manifests as  vertical banding seen in raw images.  \cite{Espinoza2022} recommended producing light curves at the sampling resolution of the instrument (rather than binning across pixel columns) to minimize the degradation in SNR from correlated 1/f noise across columns, and then binning the results at a post-processing stage. However, \cite{Holmberg2023} tested this strategy for NIRISS and found that the result is the same, regardless of the order of the binning, if one takes into account the covariance.

While pointing jitter and drift combined with intra-pixel variations was considered another possible correlated noise source, commissioning studies found that the line-of-sight pointing stability is very good $\sim$ 1 mas radial \citep{Lallo2022}.  This minimizes the need to apply de-jittering algorithms to the data.

While persistence is expected in H2RG detectors producing a similar form of temporal effect as seen in the Hubble WFC3 detectors, the effect is expected to be smaller in the newer JWST detectors \citep{Birkmann2022}.  Commissioning studies indicate the effect of persistence may not be a major concern for science programs \citep{Boker2023}.

Bad pixels can exist of various types, e.g. `hot' pixels with high dark signal, pixels with poor response to light, or dead pixels. From commissioning \cite{Boker2023} found an operability rate of pixels of 99.59 \% on NRS1 and 99.80 \% on NRS2.   These give 16948 and 8275 non-operable pixels on NRS1 and NRS2 respectively.   Cosmic rays (CRs) are another source of abnormal pixel counts during an observation.  \cite{Boker2023} report an 
 average CR hit rate of about 5.5 cm$^{-2}s^{-1}$
 with a typical hit area of about 10.5 pixels.  In addition the so-called "snowballs" \citep{Birkmann2022b} have been identified: cosmic ray events with a heavily saturated core of 2-5 pixels in radius surrounded by a halo.
The large number of bad pixels combined with CR hits require pipelines to have a robust management method for dealing with bad pixel counts.

Time-dependent systematics and time-correlated noise are important to characterise for time-series exoplanet observations. \cite{Espinoza2022} performed an Allan deviation analysis of the band-integrated light curve fit residuals using NIRSpec G395H. These were consistent with uncorrelated noise. 
The \cite{Alderson2023} analysis of the G395H data for WASP-39 b found minimal systematics but there was a mirror segment tilt event that caused a change point spread function (PSF) and a jump in flux. For the Prism \cite{Rustamkulov2023} found a high-gain antenna (HGA) movement event that affected a few integrations during the WASP-39 b Prism observation.

\section{JexoPipe}
JexoPipe is a recently developed data reduction framework that incorporates selected JWST Science Calibration Pipeline steps which it combines with its own customised steps, procedures and pathways.  JexoPipe remains under development and will evolve to be applied for different instrument configurations.  In this paper we describe its use for the Prism and G395H configurations.

We applied JexoPipe to the Prism and G395H observations  of WASP-39 b from the ERS program 1366 (PI: N. Batalha).  Details of these observations are given in \cite{ERS2023}, \cite{Alderson2023} and \cite{Rustamkulov2023}.  To summarize, the Prism observation of WASP-39 b began at 15:30:59 on 10th July 2022 and ended at 23:37:13 UTC.  The NRSRAPID readout mode was used with group time of 0.22616 seconds and 5 groups per integration ramp,  with a total of 21500 integrations.  The G395H observation began at 22:04:06 on 30th July 2022 and ended at 06:20:26 on 31st July 2022 UTC.  NRSRAPID was used with a group time of 0.902 seconds and 70 groups per integration, with a total of 465 integrations. The duty cycle efficiencies were 82\% and 98.6\% respectively for Prism and G395H.   

\begin{figure*}
	\includegraphics[trim={0cm 2cm 7cm 0cm}, clip,width=\textwidth]{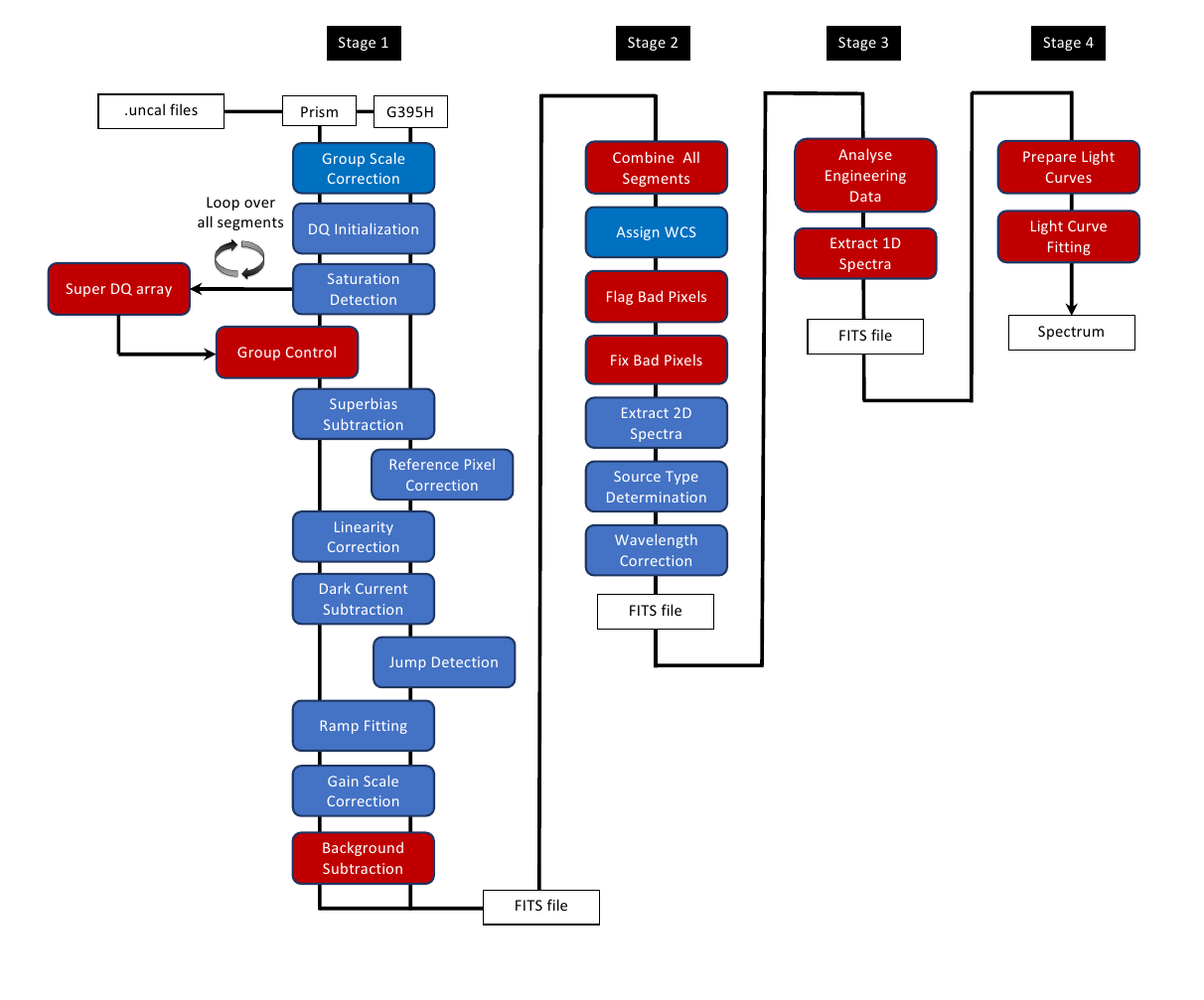}
    \caption{JexoPipe: the data reduction framework used in this work. Blue boxes are steps utilised from the JWST Science Calibration Pipeline. Red boxes are customized steps.  In Stage 1 the Prism and G395H pathways differ in the handling of saturated pixels, the application of reference pixels and jump detection.  Note that for G395H NRS2 the dark step was applied but automatically `skipped' because the dark reference frame does not have enough groups.
     We expect this framework to evolve as our understanding of JWST performance improves and to be tailored for different instrument configurations.}
    \label{fig:JexoPipe}
\end{figure*}

The pipeline is summarised in Figure \ref{fig:JexoPipe}, with four stages of processing.  To allow a homogenised approach to facilitate comparison of spectra from the two configurations we keep pathways for the two configurations as similar as possible.   The main differences arise in Stage 1 due to management of saturated pixels in the Prism data.  The G395H pathway also applies a \textit{Reference Pixel Correction} step and \textit{Jump Detection} step in Stage 1 which are not applied in the Prism pathway.  Reference files will also differ between the two configurations (and between the two G395H NRS detectors), such as the superbias file.  The final transmission spectra were produced using steps from version 1.11.3 of the JWST Science Calibration Pipeline.
 
%the uncalibrated NDRs (.uncal files) and produces the count rate per integration (.rateints files).  In this stage we use steps from the official JWST pipeline with some modifications described below.  Stage 2 takes these .rateints files and produces 1-D stellar spectra for each integration again with a combination of official JWST pipeline steps and customised steps.  Stage 3 consists of producing light curves and light curve fitting to extract the planet spectrum..

\subsection{Stage 1}
\label{stage_1}

Stage 1 begins with .uncal FITS files obtained from the Mikulski Archive for Space Telescopes (MAST) archive\footnote{https://mast.stsci.edu/portal/Mashup/Clients/Mast/Portal.html}.  These contain the uncalibrated non-destructive reads  (i.e. group level images) per integration, and are provided as contiguous segments: four for the Prism and three for the G395H grating data. The G395H data is additionally divided between the two detectors: NRS1 and NRS2. The end products of Stage 1 are .rateints FITS files containing the count rate (in DN/s) per pixel per integration. 

For both Prism and G395H we process each segment through the steps shown in Figure \ref{fig:JexoPipe}.  The functions of the official JWST pipeline (blue and grey) steps are described in the JWST pipeline package documentation\footnote{https://jwst-pipeline.readthedocs.io/en/latest/index.html}.  We use the default settings for these steps except where described below.

For the Prism we first run the pipeline for all segments up to completion of the \textit{Saturation Detection} step.  At that point the group data quality (DQ) arrays from all the segments are combined into a `super DQ array'.  This super DQ array is subsequently used in the \textit{Group Control} step.  

The pipeline then proceeds to the \textit{Superbias Subtraction} step.  We used the default superbias files in all cases.  Given previous concern with the G395H NRS1 superbias file as cited in \cite{Alderson2023} for this mode, we tried using a custom superbias created from the median of all first group images. However we found no significant difference in the final spectrum offset or noise for G395H compared to using the default file (Table \ref{table: spectrum comparisons}). For Prism, there was a very small increase in the average spectrum transit depth when using the custom file (Table \ref{table: spectrum comparisons}). 

The G395H pipeline applies the \textit{Reference Pixel Correction} step as side reference pixels are available, however the top and bottom reference pixels are not available due to the subarray cutting these off. 
The SUB512 subarray used in the Prism observation has no reference pixels available (either at the sides or at the top and bottom) to apply the \textit{Reference Pixel Correction} step. The \textit{Linearity Correction} and \textit{Dark Current Subtraction} steps are then applied.  We found no significant difference in the spectral baseline if the dark current step was omitted compared to that if it was included  (Table \ref{table: spectrum comparisons}).

In common with the pipelines in \cite{Rustamkulov2023}, the \textit{Jump Detection} step is omitted for Prism (but not for G395H), as it results in a large number of false positives.  Instead cosmic rays in Prism are managed by detecting outliers in Stage 2 (in the \textit{Flag Bad Pixels} step which is also applied to G395H).  For G395H we trialled the default rejection threshold of 4$\sigma$ and also a threshold of 15$\sigma$, as used in \cite{Alderson2023}.    There were   statistically insignificant differences in the final spectrum (Table \ref{table: spectrum comparisons}). For the final baseline case  we use the results from the 15$\sigma$ threshold.

The \textit{Background Subtraction} step
is performed  to mitigate 1/f noise and remove diffuse and sky backgrounds.   This is applied at the group level in both the Prism and G395H modes.    
Column-by-column background subtraction at the group level has been proposed to mitigate the vertical banding that arises from 1/f noise particularly for the Prism where no reference pixels are available \citep{Birkmann2022}.   Background subtraction is performed as follows.  For the Prism, a median image is first obtained from the median of all the final groups in each integration (per data segment).  A spatial profile of the median image is obtained by taking the sum in the y-axis and is used to identify the pixel row with the maximum count.  In each group image, we mask out the spectrum, leaving the peripheral 5 rows on the top and bottom of the image,  We also mask out bad pixels identified in the group DQ array and the pixel DQ array.  Then we mask out outliers to minimize the impact of cosmic rays:  half of the 16th-84th percentile range of all unmasked pixel values is taken to be $\sigma_{bg}$ and pixels $\pm$ 10$\sigma_{bg}$ beyond the median value of all unmasked pixels are then masked. The column mean of unmasked pixels  is then subtracted from all pixels in that column.  We use this method since the spectral trace for the Prism is nearly linear and parallel to the x-axis of the detector.

For G395H, a median image is obtained in the same way as for the Prism.  While the Prism spectral trace is roughly linear and parallel to the x-axis of the detector, the trace for G395H is slightly curved.  As a result the application of the mask for G395H images required first that the trace be mapped on the detector.  On the median image we divide the subarray into 10 pixel-wide column-wise slices and obtain a median profile in the y-axis for each slice.  The pixel row with the maximum count in this profile is identified for each slice, and this is used to assign an approximate initial maximum point for each individual pixel column.  We then fit a 4th order polynomial to these points. This polynomial refines the pixel row in each column closest to the maximum of the spectrum and also allows the trace to be extended to the edges of the image where the signal is low.  In each group-level image, a mask is applied $\pm$ 10 pixels around this central pixel per column.  Bad pixels and outliers are masked out in the same way as for the Prism, and the mean of the unmasked pixels in each column is then subtracted from all pixels in that column.

As noted in \cite{Radica2023} while photons from diffuse backgrounds are affected by detector non-linearity, 1/f noise is not, so ideally 1/f noise correction should occur before the \textit{Linearity Correction} step and diffuse background subtraction after.  For NIRISS SOSS, \cite{Radica2023} present a way to do this with multiple steps, however since the main purpose of this study was to compare two data sets with the same pipeline we decided implementing a single stage of background subtraction which happens after the \textit{Linearity Correction} step, to be sufficient, thus conflating 1/f noise and diffuse background subtraction.  \cite{Radica2023} noted that if 1/f correction was performed after the non-linearity correction it did not result in any biases in the spectrum but increased noise.  Thus this single-step approach may result in increased noise.

%We have omitted steps to inter-pixel capaticance (IPC) correction and persistence correction steps as these appear not to have been applied to produce the official .rateints files for both the Prism and G395H, indicating these steps need further refinement before being applied.

\subsubsection{Saturated pixels}
\begin{figure*}
	\includegraphics[trim={5cm 0.5cm 9.5cm 0cm}, clip,width=1.0\textwidth]{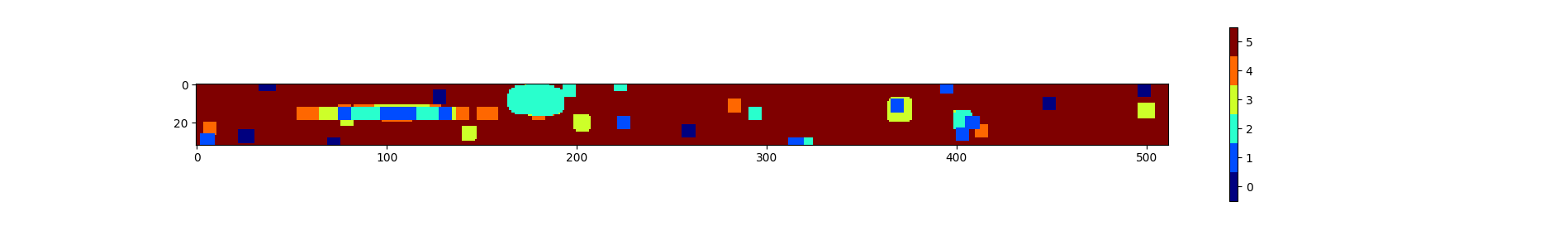}
 \includegraphics[trim={3.8cm 0.5cm 1cm 0cm}, clip,width=1.0\textwidth]
 {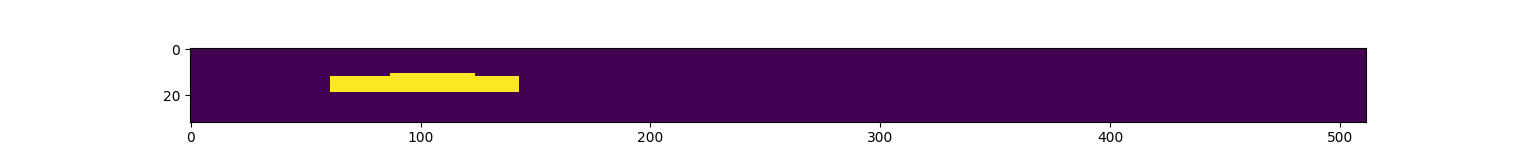}
    \caption{Determing a `persistently saturated region' on the Prism subarray.  Top: per pixel, the minimum number of unsaturated groups in any integration after completing the \textit{Saturation Detection} step as implemented in this framework.  Bottom: yellow region shows pixels where $\geq$ 10\% of the timeline have at least one saturated group.  We term this the `persistently saturated region'.}

    \label{fig:PSR}
\end{figure*}

Prism data presents a challenge in that a large number of pixels near the peak of the stellar spectrum are saturated.  We identify a `persistently saturated region' of pixels, which are pixels where at least 10\% of the integrations in the timeline have at least one saturated group.  This defines a region around the peak of the spectrum between 0.69 and 1.91 \textmu m (Figure \ref{fig:PSR}). %0.688 and 1.909 
The management of this region is challenging and final results from this saturated region need to be interpreted with added caution. 
%Another reason for caution is that for wavelengths below 1.4 \microns there is a finite probability of the quantum yield releasing two electrons per photon rather than one \citep{Jakobsen2022, Rauscher2014}.

We take a different approach to managing saturation compared to the pipelines in \cite{Rustamkulov2023}.  All four pipelines used in that paper expand the saturation flags along entire columns if a pixel in that column was saturated.  Instead, here we make use of the \textit{Saturation Detection} step argument \verb'n_pix_grow_sat' which we describe further below.

The \textit{Saturation Detection} step compares the count on each pixel (before linearity correction) to a pre-determined saturation level in a reference file.  If during an integration a pixel exceeds its reference level in a particular group, then it is flagged as saturated in the DQ array (DQ flag = 2) in that group and all remaining groups after that, \textit{but not in the preceding groups}.  Then at the \textit{Ramp Fitting} step, the saturated groups are ignored when fitting the ramp.
%If all groups are saturated, the pixel is given zero count in the ramp fit step and is flagged in the Stage 2 DQ array as "DO NOT USE" as well as "SATURATED".
If all but the first group is saturated, there is the option of using this first group alone to obtain a `ramp' value.  To do this the \textit{Ramp Fitting} step argument \verb'suppress_one_group' must be changed to False from the default of True.  Since the `slope' value using just one group may be quite different compared to that from two or more groups, this can potentially lead to noise in pixel timelines if a pixel flips between a single group and two groups in different integrations in its timeline.  To reduce the noise impact of such an effect we implement a \textit{Group Control} step for the Prism pipeline which is described further below.

%However there will be a pedestal value to the ramp of each pixel such that first group alone is does not give an accurate absolute ramp value (and the pedestal itself will vary from integration to integration somewhat due to kTC noise).  However, since in a transit light curve, we are ultimately utilising the proportional change in count rates over time, rather than the absolute count rates, it might be acceptable to use single groups as long as the entire timeline for the pixel is fixed consistently to a single group in every integration. To facilitate this we implement the \textit{Group Control} step for the Prism pipeline which is described further below.

If a pixel becomes saturated, exceeding its full well capacity, charge can leak from the pixel diffusing into surrounding pixels. This horizontal charge overflow from saturated pixels is known as `blooming' \citep{Cohen2020}.
These neighbouring pixels may not be truly saturated (as defined by exceeding their saturation thresholds) however due to this leaked charge their electron counts now become unreliable in relation to their incident photon counts.  By default the \textit{Saturation Detection} step flags all neighbouring pixels within 1 pixel of a truly saturated pixel as `saturated' in the same groups as the truly saturated pixel, i.e. 3 x 3 box of pixels centred on the affected pixel is flagged as saturated. The size of this box is controlled by the step argument \verb'n_pix_grow_sat', which has the default value of 1.  

\begin{figure*}
\includegraphics[trim={0cm 0cm 0cm 0cm}, clip, width=0.95\textwidth]{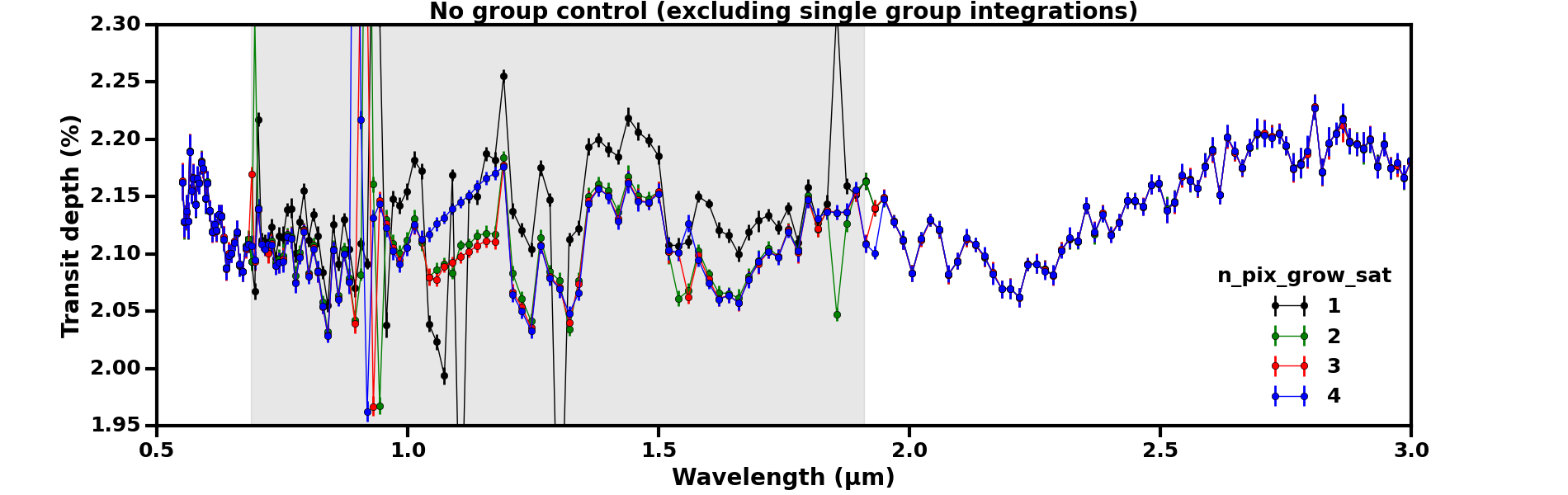}
	\includegraphics[trim={0cm 0cm 0cm 0cm}, clip, width=0.95\textwidth]{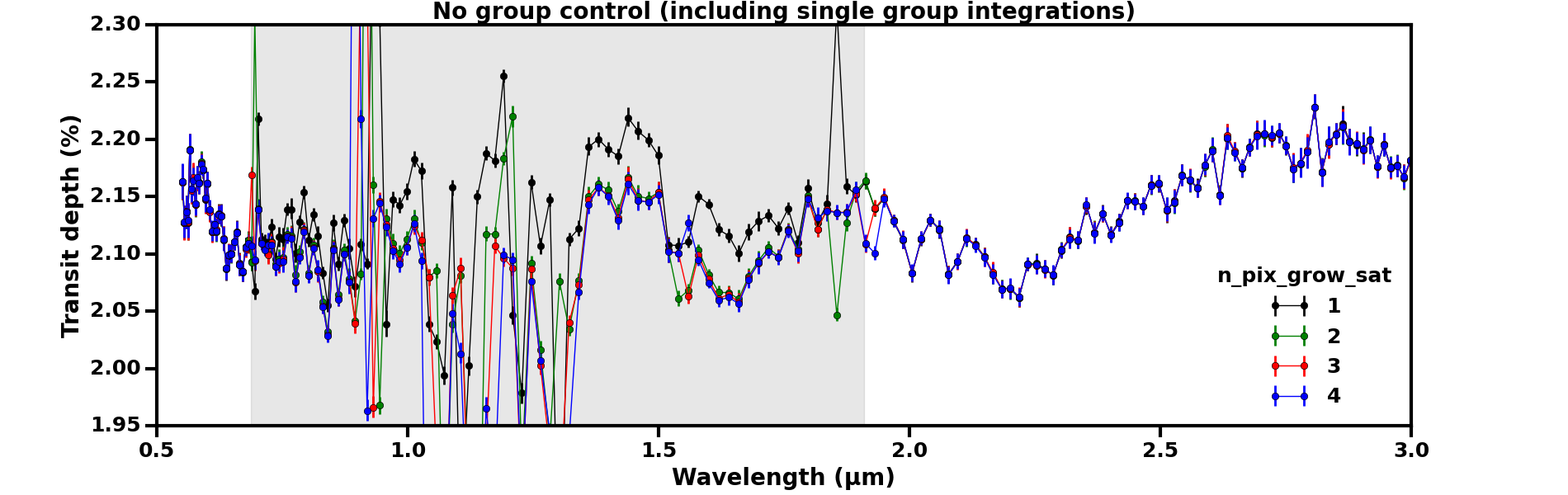}
 	\includegraphics[trim={0cm 0cm 0cm 0cm}, clip, width=0.95\textwidth]{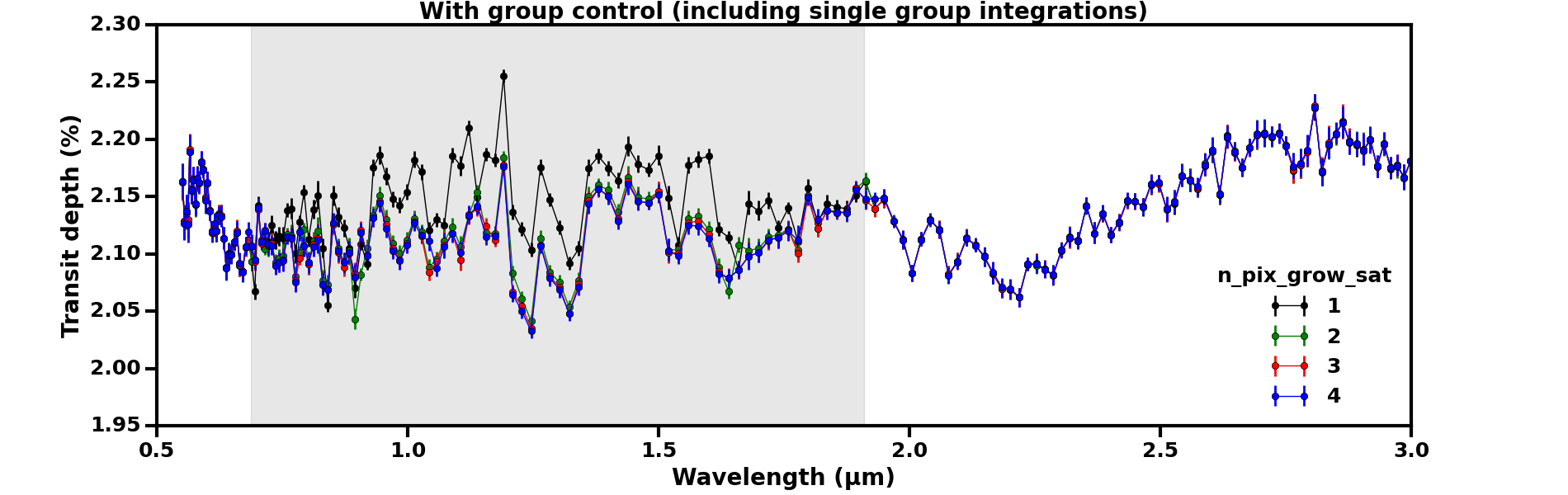}
    \cprotect\caption{The effect of changing the \textit{Saturation Detection} step argument, \verb"n_pix_grow_sat", on the transmission spectrum for Prism.  The shaded area shows the persistently saturated region which extends from 0.69 to 1.91 \textmu m.  Top: case a) no group control and \verb'suppress_one_group' set to True.  Middle: case b) no group control and \verb'suppress_one_group' set to False. Bottom: case a) with group control and \verb'suppress_one_group' set to False.  We settle on using an \verb"n_pix_grow_sat" value of 3 for our final reduction.}
    \label{fig:npix}
\end{figure*}

We ran our pipeline for the Prism data varying the value of \verb'n_pix_grow_sat' between 1 and 4, and the final spectra obtained are shown in Figure \ref{fig:npix} up to 3 \textmu m for three different conditions: a) no group control and \verb'suppress_one_group' set to True (excludes single group integrations), b) no group control and \verb'suppress_one_group' set to False (includes single group integrations), c) with group control and \verb'suppress_one_group' set to False  (includes single group integrations).  We explain group control below.  In both cases a) and b) there are wide point-to-point variations in the spectrum within the persistently saturated region for all values of \verb'n_pix_grow_sat', while outside this region, all values of \verb'n_pix_grow_sat' result in similar final spectra.  In case c), implementing group control greatly reduces the point-to-point variations, however we also notice that  increasing \verb'n_pix_grow_sat' to values > 1 reduces the spectrum baseline which leads to the emergence of distinct peaks at  1.4 and 1.9 \microns consistent with water bands (as well as further reducing point-to-point variation).  The fact that increasing \verb'n_pix_grow_sat' to values $>1$ appears to bring out some spectral features would suggest that the effect of the saturated pixels influences the counts on neighbouring pixels greater than one pixel away.   We realise however that this a somewhat subjective assessment of the improvement in the spectrum, and thus we exercise caution in the interpretation of the saturated region. 

We decided to proceed in the final analysis using a value of \verb'n_pix_grow_sat' = 3 for all saturated pixels in the Prism data.  However we use the default value of 1 for the G395H data.  This is because count rates in the G395H data are much lower than in the Prism, and rare persistently saturated pixels are those isolated pixels with unusually low saturation thresholds rather than those with excessively high photon counts in large groups.  Thus we assume that the leakage into surrounding pixels is less of an issue.  However to be sure we also ran the pipeline with \verb'n_pix_grow_sat' = 3 for G395H and found no significant difference in the spectrum transit depths or noise (Table \ref{table: spectrum comparisons}).

\begin{figure*}
	\includegraphics[trim={0cm 0.5cm 0cm 0.5cm}, clip,width=1\textwidth]{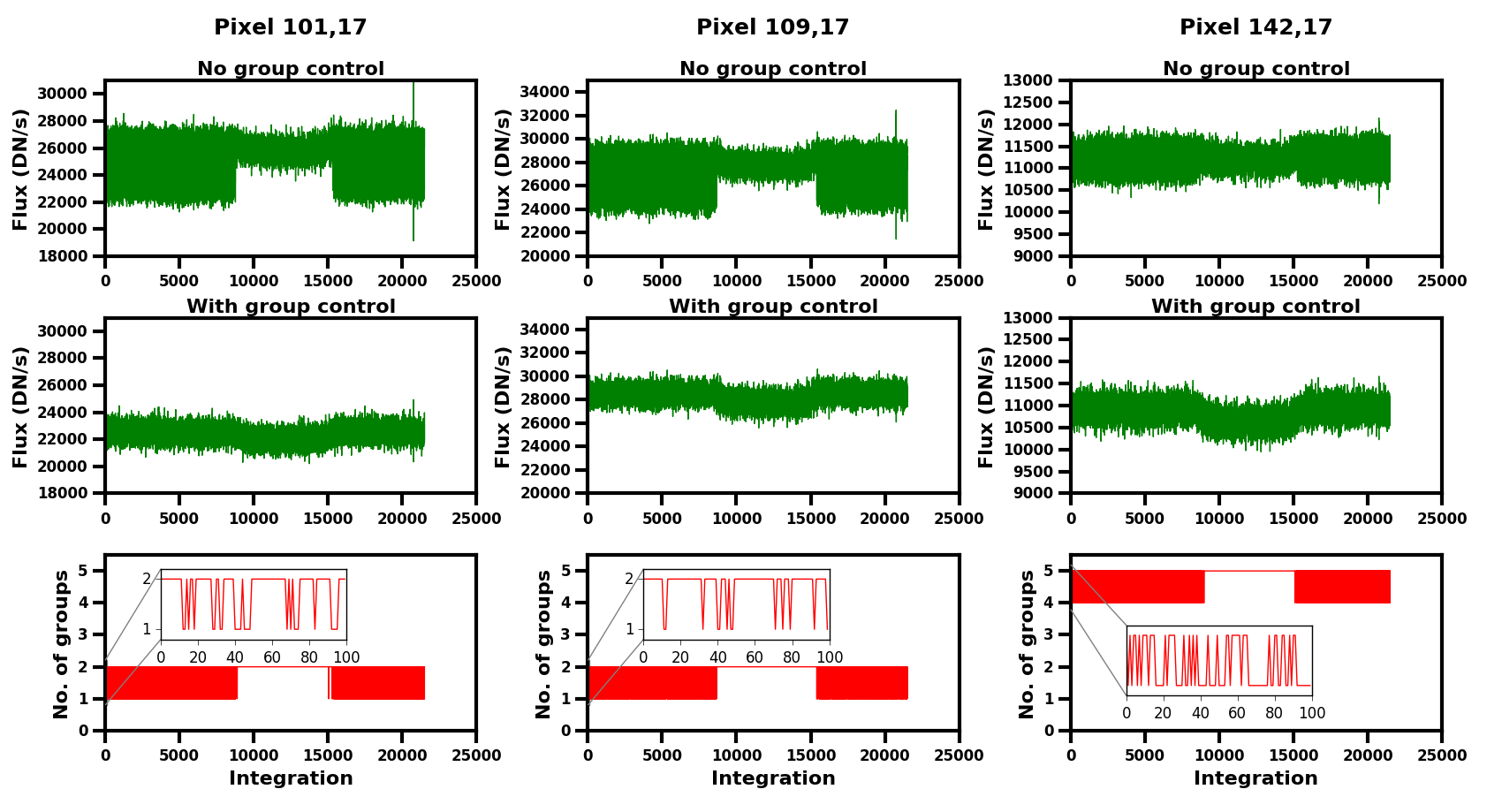}
    \cprotect\caption{`Flipping' noise as the rationale for the \textit{Group Control} step in Prism processing.  Three pixels are shown, all of which fall in the persistently saturated region. \verb|suppress_one_group| was set to False. The top plots show the timelines of pixel counts without any group control.  Middle plots show the timelines with the \textit{Group Control} step implemented.  The timelines are shown after the \textit{Fix Bad Pixels} step in Stage 2. The bottom panel shows the number of unsaturated groups per integration (and thus the default ramp length).  For pixel (101,17), the number of groups flips between 1 and 2 during the out-of-transit (OOT) phase. As explained in the text, and as can be seen in the top row of figures, this leads to noise in the timeline. The number of single group integrations is $\geq$ 10\% (20\%) of the timeline for this pixel, so per the rules adopted we force all the integrations to use only 1 group. This gives the result shown in the middle panel, where we see the noise has been mitigated.  Pixel (109,17) also flips between 1 and 2 groups with the same kind of associated noise in the OOT phase.  However it has $<$ 10\% (8.4\%) of its timeline with single group integrations.  For such cases we flag the single group integrations as `bad' by giving them NaN values in Stage 1, and these are later filled in during the Stage 2 \textit{Fix Bad Pixels} correction step as described in the text.  Pixel (142,17) occurs at the edge of the persistently saturated region, and flips between 4 and 5 unsaturated groups in its OOT phase.  This again results in noise which is controlled by fixing all integrations to 4 groups (middle panel).  The noise resulting from flipping between different groups per integration may be due in part to the small number of total groups per integration operating near the saturation threshold, giving noticeably different slopes per integration. }
    \label{fig:group control}
\end{figure*}

\subsubsection{Group control step and `flipping' noise} 
We find that some pixels in the Prism data exhibit a type of noise in their timelines that seems to result from flipping of the number of unsaturated groups per integration, and which can be controlled if the number of groups fitted for in the \textit{Ramp Fitting} step is kept constant for the entire timeline (Figure \ref{fig:group control}).  This also appears to give the wide point to point variations seen in Figure \ref{fig:npix} in cases a) and b).  For flips between 1 and 2 groups, this `flipping noise'  is also controlled by the default ramp fit step argument \verb|suppress_one_group|, which is by default set to True, where 1 group integrations are given NaN values, and which explain the slightly less noisy spectra seen for case a) in Figure \ref{fig:group control} compared to case b). 

For this reason, and also to manage pixels where there are a significant number of single unsaturated groups, we apply a \textit{Group Control} step for the Prism data.   For pixels in the persistently saturated region, the super DQ array is used to find the minimum number of unsaturated groups in any integration (Figure \ref{fig:PSR}, top).
If the minimum number of unsaturated groups is 2 or more, all integrations of that pixel are forced to use this number of groups.  For pixels where the minimum number of unsaturated groups is 1, if the pixel exhibits $\geq$ 10\% of its timeline as single unsaturated groups, then the full timeline is fixed to a single group per integration.  If a pixel has single groups for $<$ 10\% of its timeline these groups are given NaN values (resulting in NaN values for the integration after the ramp fit step) (and flagged as `bad' in the group DQ array) and the number of groups fixed at 2 for the remainder of the timeline. 

The NaNs act as effective flags for correction in the Stage 2 \textit{Fix Bad Pixels} step.  Outside of the persistently saturated region, we do not apply group control as saturation is less frequent. If a pixel outside the persistently saturated region has a rare integration where the number of groups is different from the majority of its integrations, this may manifest as an outlier in the timeline, and thus managed in the same way as for other outliers in the Stage 2 \textit{Flag Bad Pixels} step. There are a few pixels outside the persistently saturated region that have single unsaturated groups but these are for only 1-2 integrations (possibly from cosmic ray hits) and are flagged by applying NaN values at this stage to these groups. 

When we apply group control, the point to point variations seen in the persistently saturated region  are largely suppressed: Figure \ref{fig:npix} case c).   
The \textit{Group Control} step was not applied to the G395H data as saturation was much less frequent, and the longer ramp probably favours more stability against changes in the ramp gradient (and thus flipping noise) if the number of groups changes from integration to integration.

\subsection{Stage 2}

Stage 2 begins with the .rateints FITS files for each segment and first combines them into one file for the entire observation (Figure \ref{fig:JexoPipe}).  This facilitates the production of a rolling median image used in the \textit{Fix Bad Pixels} step (see below).  On completion of Stage 2 a .calints FITS file is produced containing calibrated, wavelength-assigned 2-D slope images in flux units of DN/s.

\subsubsection{Flag bad pixels}

After combining all segments, the \textit{Assign WCS} step is applied,  but does not change the science data.  We then apply a custom \textit{Flag Bad Pixels} step.  We flag all pixels that have abnormal DQ flags.  We however do not include pixels flagged just for saturation or jumps as these effects will have been managed in the \textit{Ramp Fitting} step, %and we assume the values in these pixels are reliable %  
except in cases where additional flags, e.g. "DO NOT USE" have been generated.  NaN values are applied to these flagged pixels. 
  Next we check for outliers not picked up in DQ flagging in each integration image on a row-by-row basis.  For each row in each integration image we obtain a rolling median and rolling standard deviation ($\sigma_1$) $\pm$ 5 pixels around a given pixel in the x-direction.  We also find a line "sigma" ($\sigma_2$) based on the half of the 16th-84th percentile range in the entire row.  We flag as outliers (and give NaN values to) any pixel $\pm$ 3$\sigma_1$ or  $\pm$ 3$\sigma_2$ from its corresponding rolling median value.   This is iterated 3 times. For our baseline case, 
  before this step the proportion of NaN values (for all pixels over all integrations) is  0.6\% in Prism,
  0.8\% in G395H NRS1 and 0.9\% in G395H NRS2.  After the \textit{Flag Bad Pixels} this increases to 3.5\% in Prism,
  3.6\% in G395H NRS1 and 1.6\% in G395H NRS2.

\subsubsection{Fix bad pixels}

We adopt the approach of filling in NaNs which have flagged bad pixels and outliers, rather than leaving these values open.  This 
contrasts with the approach where such values are not filled and bad pixels are de-weighted.  We fill in any NaN values in two ways.  Firstly if a pixel timeline has $<$ 10\% NaN values these are filled in by linear interpolation of good values.  If a pixel timeline has $\geq$ 10\% NaN values, then the full timeline for that pixel is made NaN.  Secondly, any remaining NaNs are filled in spatially by linear interpolation of good values  on a row-by-row basis in each integration image.   
Finally, we deal with any remaining pixel-level light curve outliers thus far not identified.  For each pixel on each integration image, a rolling median value is produced from neighbouring images spanning $\pm$ 100 integrations for the Prism data set and $\pm$ 10 integrations for the G395H data.  In addition for each pixel, we obtain a rolling standard deviation   ($\sigma_{roll}$) over the same range, and also the median of the rolling   standard deviation values ($\sigma_{median}$).  
We identify pixel values which are either $\pm$ 5$\sigma_{roll}$ or $\pm$ 5$\sigma_{median}$ beyond the corresponding rolling median value. These outlier pixel values are replaced by their value in the corresponding rolling median image.

\subsubsection{Remaining steps}
 
We then proceed to utilise steps from the JWST Science Calibration Pipeline that apply the 2-D wavelength solution(Figure \ref{fig:JexoPipe}) ending with the   \textit{Wavelength Correction} step.  The flux units after this step remain as DN/s.  The \textit{Photometric Calibration} step in the JWST Science Calibration Pipeline 
is not strictly required since we are interested in relative flux changes.

In common with other pipelines we have not applied a flat field step as the reference files were not complete \citep{Alderson2023, Lustig-Yaeger2023}. The effect of not applying a flat field may result in increased noise due to uncorrected pixel quantum efficiency and gain variations that manifests during movement (jitter) of the spectral image on the detector \citep{Sarkar2021}.
The JWST pointing system has line-of-sight pointing stability of $\sim$1 mas \citep{Lallo2022} which is 1/100th the pixel scale of NIRSpec (0.1"). The jitter noise impact of not applying the flat field should thus be negligible.

%, although effects due to intra-pixel sensitivity variations could still cause noise.  The impact of jitter noise has been estimated to be < 25 ppm for all modes \citep{Birkmann2022}. 

%We did run the pipeline applying just the D-FLAT step and found that....  Due to the low jitter of the JWST pointing system the impact of not applying the D-FLAT may be small, although effects due to intra-pixel sensitivity variations could still cause noise.  The Photom step converts counts into units of astronomical flux.  

\begin{figure}
	\includegraphics[trim={0cm 0cm 0cm 0cm}, clip,width=\columnwidth]{ 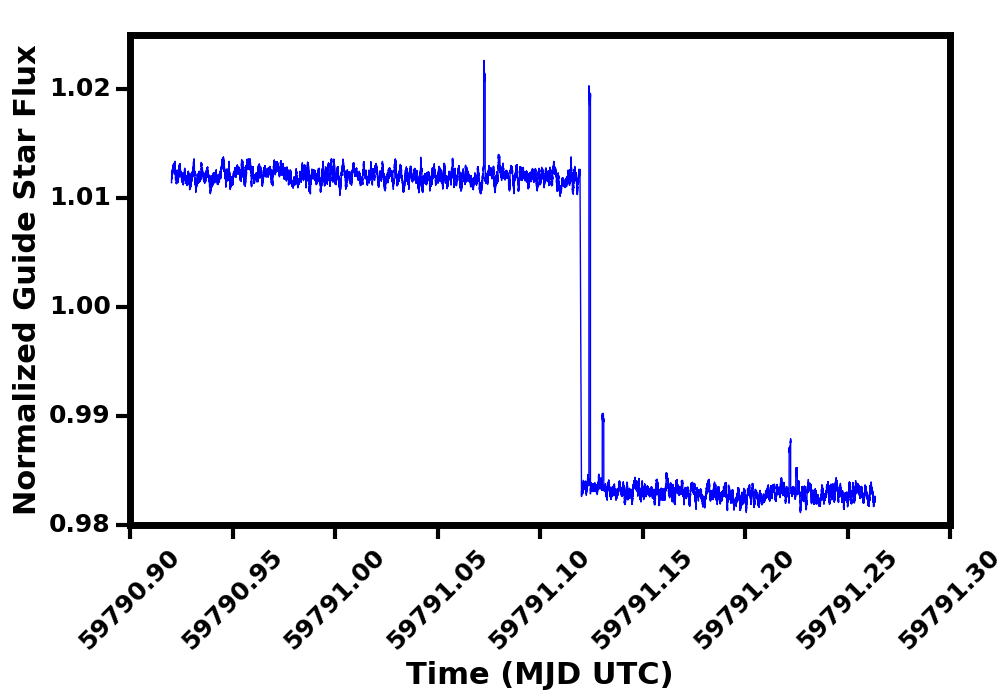}
    \caption{Guide star flux during the G395H observation. Flux has been normalized to the mean value, and smoothed by convolving with a 1000 step box-shaped kernel for clarity.   }
    \label{fig:gs flux}
\end{figure}

\subsection{Stage 3}

Stage 3 begins with a manual examination of engineering data to identify relevant events in the timeline that may require addressing at the light curve level.  The .calints file from Stage 2 is then opened and extraction of the 1-D stellar spectra performed.  Stage generates a FITS file containing 1-D stellar spectra per integration.

\subsubsection{Analysis of engineering data}

\begin{figure}
	\includegraphics[trim={0cm 0cm 0cm 0cm}, clip,width=\columnwidth]{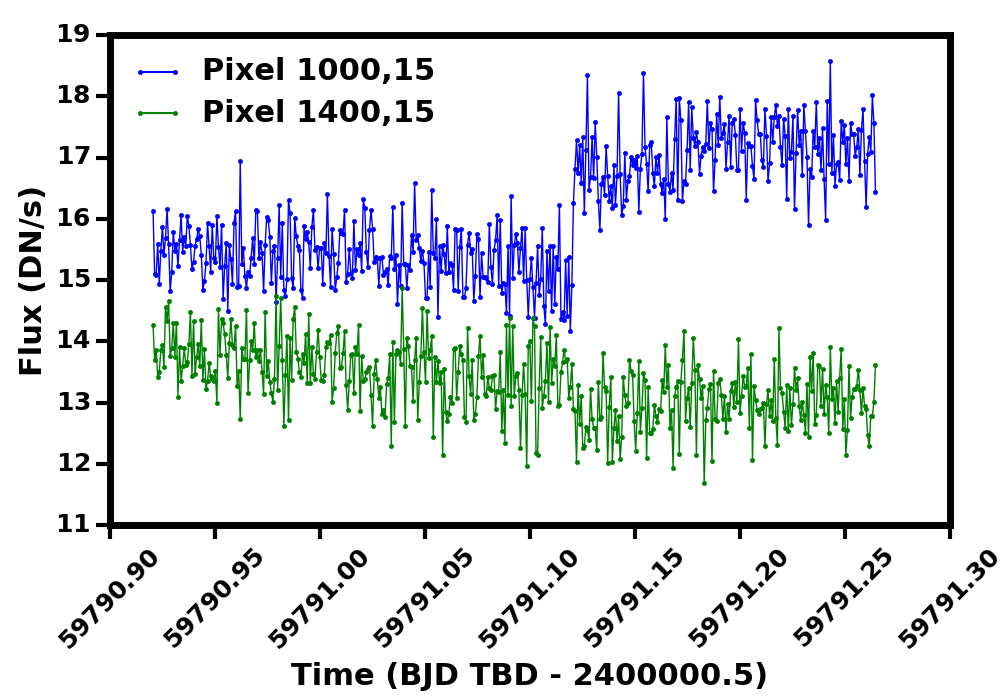}
    \caption{G395H mirror segment tilt event at pixel level. Two example pixel timelines from NRS2 are shown, with a step clearly visible and corresponding to that in the guide star flux timeline.  In pixel (1000, 15), there is an increase in flux and in pixel (1400, 15) there is a decrease in flux after the event.  The amount and direction of flux change is thus pixel-dependent.  }
    \label{fig:step}
\end{figure}

We analyse the engineering data associated with each observation\footnote{  https://jwst-docs.stsci.edu/methods-and-roadmaps/jwst-time-series-observations/jwst-time-series-observations-noise-sources}, in particular the guide star x and y centroids, guide star flux, and HGA movement flags.  As previously noted in \cite{Rustamkulov2023} in the Prism data there is HGA movement, and we find this occurs between 59770.9722 and 59770.9723 MJD UTC. Therefore in Stage 4 we exclude any light curve points that fall into this time period. As explained below, we bin the Prism timeline to every 25 integrations, so this amounts to exclusion of two of the final binned light curve points.  In the G395H data, there are no HGA movements, however we can see the impact of the previously noted \citep{Alderson2023} mirror segment tilt event as a step change in the guide star flux (Figure \ref{fig:gs flux}).  This is reflected in the light curves of individual pixels, where the change in count can be positive or negative and varies between pixels (Figure \ref{fig:step}).  We did not detect an associated change in the guide star centroid positions.  The tilt event is corrected in the Stage 4 \textit{Prepare Light Curves} step.

%\begin{figure}
%	\includegraphics[trim={0cm 0cm 0cm 0cm}, clip,width=\columnwidth]{figs/wlc_nrs1_corr.png}
%		\includegraphics[trim={0cm 0cm 0cm 0cm}, clip,width=\columnwidth]{figs/wlc_nrs2_corr.png}
%   \caption{G395H mirror segment tilt event at white light curve level. A step change occurs in the white light curves in both detectors (top: NRS1, bottom: NRS2). Also shown are the curves after implementing the correction applied in the Stage 4 \textit{Prepare Light Curves} step.}
    \label{fig:wlc step}
%\end{figure}

\subsubsection{1-D spectral extraction}
Next, a custom \textit{Extract 1-D Spectra} step is applied.  
Extraction is obtained through a `box' extraction method and also an optimal extraction method.

\begin{figure}
	\includegraphics[trim={0cm 0cm 0cm 0cm}, clip,width=\columnwidth]{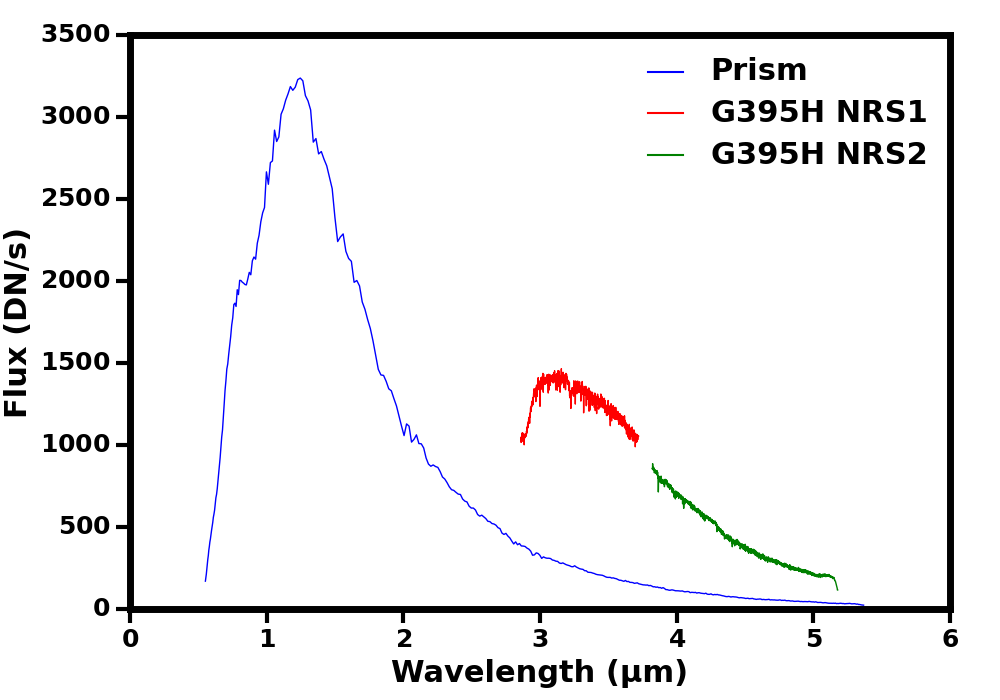}
    \caption{Example 1-D stellar spectra from the first integration of each time series. These are the end result of Stage 3 of the pipeline.  The flux counts on the Prism have been reduced by a factor of 100 to allow all spectra to be shown on same plot.}
    \label{fig:stellar spectra}
\end{figure}

For box extraction, using the assigned 2-D wavelength solution per pixel, we obtain the mean wavelength per pixel column giving us a nominal 1-D wavelength grid.  We then assume wavelength bins centred on these values with boundaries being the mean of adjacent values.  We then sum up the counts on all pixels that fall in a given wavelength bin. This method allows flexibility to obtain the 1-D spectra from potentially tilted or curved spectral traces. We found the wavelength variations over a pixel column to be as follows: in Prism the standard deviation of the wavelength ranged from 0.002 to 0.013 \textmu m per column, and for G395H this was $\sim$ 0.0003 \textmu m per column in both NRS1 and NRS2. The resulting 1-D spectra look virtually identical to those obtained through simple summation of the counts per pixel column.  These 1-D spectra, the  wavelength grid, and timing information are saved together in a FITS file which forms the output of Stage 3.  Example 1-D spectra from the first integration in each time series are shown in Figure \ref{fig:stellar spectra}. We use the box extraction results for our baseline case.

For optimal extraction, we apply the principles in \citep{Horne1986} to obtain the spectra and variance for each integration image.  We use a rolling median image of 12 integrations for G395 and 100 integrations for Prism to provide the column-wise profile for each integration image.  No background subtraction is included, and 5$\sigma$ outliers were rejected iteratively.  The final optimal extraction transmission spectra did not show any significant differences in offset and comparable or slightly noise compared to the baseline cases (Table \ref{table: spectrum comparisons}).

\subsubsection{Error propagation}
Regarding error propagation up to the end of stage 3, the ERR array in stages 1 and 2 propagates the calculated photon noise and read noise errors through the different steps.  We do not adjust the error in the custom \textit{Background Subtraction} step since the proportion of signal removed is small will not impact the photon noise substantially.  In the \textit{Fix Bad Pixels} step, the variances of bad pixels are corrected in the same way as the pixel values themselves.  In Stage 3, box extraction of the 1-D spectra proceeds with quadrature addition of the corresponding ERR array values, whereas in optimal extraction, the errors are obtaining directly from the optimal extraction algorithm.

%In contrast, JexoRES applies an optimal extraction algorithm to extract 1-D spectra.

%After this process bad pixels identified through the DQ array and outliers have been flagged.  We then proceed to replace the values in these flagged pixels.  To do this we first replaced the values with NaN values.  We then produce a median image for each integration image, where the median image is the rolling median image of images +/- 2 images in the time sequence.  These median images are similarly sigma-clipped, and NaN values applied to any outliers. In each median image, the profile of counts per row is obtained.  Any missing values (where there are NaNs) are filled through linear interpolation.  The image is once more sigma clipped now in comparison to the median image +/- 3 sigma1, and NaNs applied to any outliers.  Finally the image NaN values are replaced by those from the corresponding median image.  An example of the final cleaned imaged is 

\subsection{Stage 4}

In Stage 4, the 1-D spectra from Stage 3 are extracted from the Stage 3 FITS file to produce a data cube of 1-D spectra vs time for each dataset.  In the time axis, this constitutes the set of native (i.e. pixel-level) resolution spectral light curves.  These spectral light curves are then fitted with light curve models to extract the transit depth per wavelength, resulting in the final output consisting of the planet transmission spectrum.

 In Stage 4, When calculating the errors on white light curve data points, these are obtained by taking the quadrature sum of the errors on the spectral data points which are summed to produce the white light curve point.  Further error handling is discussed below.

\subsubsection{Prepare light curves}

Prior to light curve fitting we prepare the light curves in various ways, e.g. removal of any unwanted sections or data points, and/or binning in time or wavelength space as needed. 

We perform one more stage of outlier removal by identifying white light curve outliers.  A rolling median and rolling standard deviation ($\sigma$) of 200 data points in Prism and 20 data points in G395H are obtained, and outliers identified as being those $\pm$ 4$\sigma$ from the median.  No outliers were identified for G395H this way and five data points identified in Prism.  The 1-D spectrum corresponding to the outlying integrations were rejected and replaced by the average of the two integrations either side of it.  

For the Prism, the white light curve shows a clear non-linear trend in the out-of-transit (OOT) data.  By excluding the first 1500 points, we found we could reasonably fit this trend with a second order polynomial, so this clipping was applied to all spectral light curves as well as the white light curve. Given the high cadence rate of the Prism data, we bin down each spectral light curve in time every 25 points, obtaining the mean mid-BJD time stamp of the 25 binned points. The time step between the binned points is 34.438 seconds. Two points are then excluded around the HGA movement event as previously mentioned.

For G395H, we do not bin down the timelines, the time step between points being 64.056 seconds.  Although systematic trend appears more linear than in the Prism light curves, we fitted the trend with a second-order polynomial to keep the pipeline as consistent with the Prism pipeline as possible.

%No significant differences if fitting for a linear trend or quadratic for G395H in the final spectrum.  Using a custom superbias consisting of the median image of all first groups did not change the final result significantly
%Correcting the mirror segment tilt event at the pixel level vs the spectral light curve level also did not change the final spectrum.  We note that there have been updates to the default superbias files since the observations were first performed.  Here we the default superbias files were.... (date)
%We also applied more recent superbias files 0427 (date) and 0429 (date) (which are not selected by default)..
%We changed cosmic ray detection threshold....
%Applying or not applying the dark current step did not significantly change the final grating spectrum.
A short `hook' appears at the start of the G395H white light curves, which we remove by excluding the first 10 integrations. 
To correct for the mirror-segment tilt event in G395H we first identify visually the event on the white light curves, and exclude three integrations around the tilt event.  The event of the event is then corrected during light curve fitting as explained below. 

% >2 um PRISM wlc fit (no GP)
% delta $0.021328_{-0.000042}^{+0.000041}$
% rp/rs $0.14604_{-0.00015}^{+0.00014}$
% a/Rs $11.430_{-0.020}^{+0.020}$
% inc $87.790_{-0.024}^{+0.025}$
% t0 $59770.835663_{-0.000013}^{+0.000012}$
% c1 $0.059_{-0.019}^{+0.019}$
% c2 $0.223_{-0.035}^{+0.034}$
% poly0 $170584500_{-2900}^{+2800}$
% poly1 $-0.00820_{-0.00031}^{+0.00032}$
% poly2 $0.00890_{-0.00097}^{+0.00095}$

% >2 um PRISM wlc fit (with GP)
% delta $0.021343_{-0.000062}^{+0.000059}$
% rp/rs $0.14609_{-0.00021}^{+0.00020}$
% a/Rs $11.417_{-0.029}^{+0.028}$
% inc $87.777_{-0.031}^{+0.031}$
% t0 $59770.835664_{-0.000025}^{+0.000024}$
% c1 $0.061_{-0.031}^{+0.033}$
% c2 $0.218_{-0.054}^{+0.051}$
% poly0 $170583200_{-6000}^{+6100}$
% poly1 $-0.00802_{-0.00067}^{+0.00067}$
% poly2 $0.0083_{-0.0020}^{+0.0020}$

% The transit + GP model is preferred over the transit model at 6.5 sigma (>2 um PRISM wlc data). Computed using the Bayes factor (Multinest). 

%%%%%%%%%%%%%%%%%%%%%%% wlc tables
\begin{table*}
\begin{center}
\caption{Summary of retrieved Prism white light curve parameters.}
\label{table: wlc table prism}
\begin{tabular}
%{cccccc} 
 {p{1.2cm}p{3cm}p{3cm}	}
\hline
\hline
\multicolumn{1}{c}{Parameter} &
\multicolumn{1}{c}{Prism} &
\multicolumn{1}{c}{Prism} 
\\
\multicolumn{1}{c}{} &
\multicolumn{1}{c}{all wavelengths} &
\multicolumn{1}{c}{>2 \textmu m only} 
\\
\hline
\multicolumn{1}{c}{$R_p/R_s$} &\multicolumn{1}{c}{ 0.14522 $\pm$ 0.00015 } &\multicolumn{1}{c}{ 0.14604 $\pm$ 0.00014} \\
\multicolumn{1}{c}{$a'/R_s$} &\multicolumn{1}{c}{ 11.46 $\pm$ 0.02 } &\multicolumn{1}{c}{ 11.43 $\pm$ 0.02 } \\
\multicolumn{1}{c}{$i$ ($^\circ$)} &\multicolumn{1}{c}{ 87.83 $\pm$ 0.02 } &\multicolumn{1}{c}{ 87.792 $\pm$ 0.025 } \\
\multicolumn{1}{c}{$t_0$ (BJD TDB - 2400000.5)} &\multicolumn{1}{c}{ 59770.83563 $\pm$ 0.00001} &\multicolumn{1}{c}{ 59770.83566 $\pm$ 0.00001} \\
\multicolumn{1}{c}{$c_1$} &\multicolumn{1}{c}{ 0.17 $\pm$ 0.01} &\multicolumn{1}{c}{ 0.06 $\pm$ 0.02 } \\
\multicolumn{1}{c}{$c_2$} &\multicolumn{1}{c}{ 0.28 $\pm$ 0.03 } &\multicolumn{1}{c}{ 0.22 $\pm$ 0.03 } \\
\multicolumn{1}{c}{$a$ (DN/s)} &\multicolumn{1}{c}{ 6.84575 $\pm$ 0.00009 $ \times 10^8$ } &\multicolumn{1}{c}{ 1.70584 $\pm$ 0.00003 $ \times 10^8$ } \\
\multicolumn{1}{c}{$b$ (s$^{-1}$)} &\multicolumn{1}{c}{-0.0057 $\pm$ 0.0003 } &\multicolumn{1}{c}{-0.0081 $\pm$ 0.0003 } \\
\multicolumn{1}{c}{$c$ (s$^{-2}$)} &\multicolumn{1}{c}{ 0.0067 $\pm$  0.0008} &\multicolumn{1}{c}{ 0.009 $\pm$ 0.001 } \\
\hline
\hline
\end{tabular}
\end{center}
\end{table*}

\begin{table*}
\begin{center}
\caption{Summary of retrieved G395H white light curve parameters.}
\label{table: wlc table grating}
\begin{tabular}
%{cccccc} 
 {p{1.1cm}p{3cm}p{3cm}p{3cm}p{3cm}p{3cm}	}
\hline
\hline
\multicolumn{1}{c}{Parameter} &
\multicolumn{1}{c}{G395H NRS1} &
\multicolumn{1}{c}{G395H NRS2}&
\multicolumn{1}{c}{G395H NRS1} &
\multicolumn{1}{c}{G395H NRS2}
\\
\multicolumn{1}{c}{} &
\multicolumn{1}{c}{} &
\multicolumn{1}{c}{}&
\multicolumn{1}{c}{independent} &
\multicolumn{1}{c}{independent}
\\
\hline
\multicolumn{1}{c}{$R_p/R_s$} &\multicolumn{1}{c}{ 0.14585 $\pm$ 0.00007} &\multicolumn{1}{c}{ 0.1465 $\pm$ 0.0001} &\multicolumn{1}{c}{ 0.1463 $\pm$ 0.0002} &\multicolumn{1}{c}{ 0.1466 $\pm$ 0.0002  } \\
\multicolumn{1}{c}{$a'/R_s$} &\multicolumn{1}{c}{fixed to Prism value } &\multicolumn{1}{c}{ fixed to Prism value} &\multicolumn{1}{c}{ 11.41 $\pm$ 0.03 } &\multicolumn{1}{c}{ 11.39 $\pm$ 0.04 } \\
\multicolumn{1}{c}{$i$ ($^\circ$} &\multicolumn{1}{c}{ fixed to Prism value} &\multicolumn{1}{c}{fixed to Prism value } &\multicolumn{1}{c}{ 87.74 $\pm$ 0.03} &\multicolumn{1}{c}{ 87.75 $\pm$ 0.04 } \\
\multicolumn{1}{c}{$t_0$ (BJD TDB - 2400000.5)} &\multicolumn{1}{c}{ 59791.11203 $\pm$ 0.00002
 } &\multicolumn{1}{c}{ 59791.11214 $\pm$ 0.00003  } &\multicolumn{1}{c}{ 59791.11203 $\pm$ 0.00002 } &\multicolumn{1}{c}{ 59791.11214 $\pm$ 0.00003 } \\
\multicolumn{1}{c}{$c_1$} &\multicolumn{1}{c}{ 0.04 $\pm$ 0.02} &\multicolumn{1}{c}{ 0.06 $\pm$ 0.02} &\multicolumn{1}{c}{ 0.09 $\pm$ 0.03 } &\multicolumn{1}{c}{$ 0.07^{+ 0.04}_{- 0.03} $} \\
\multicolumn{1}{c}{$c_2$} &\multicolumn{1}{c}{ 0.21 $\pm$ 0.02} &\multicolumn{1}{c}{ 0.13 $\pm$ 0.03} &\multicolumn{1}{c}{ 0.10 $\pm$ 0.05} &\multicolumn{1}{c}{ 0.12 $\pm$ 0.06 } \\
\multicolumn{1}{c}{$a$ (DN/s)} &\multicolumn{1}{c}{ 1.60934 $\pm$ 0.00002 $ \times 10^6$ } &\multicolumn{1}{c}{ 8.7624 $\pm$ 0.0002 $ \times 10^5$ } &\multicolumn{1}{c}{ 1.60934 $\pm$ 0.00002 $ \times 10^6$ } &\multicolumn{1}{c}{ 8.7624 $\pm$ 0.0002 $ \times 10^5$ } \\
\multicolumn{1}{c}{$b$ (s$^{-1}$)} &\multicolumn{1}{c}{ $0.00001^{ + 0.00023}_{- 0.00001} $ } &\multicolumn{1}{c}{ $0.0002^{+ 0.0003}_{- 0.0002} $} &\multicolumn{1}{c}{ $0.00001^{ + 0.00020}_{- 0.00001}$ } &\multicolumn{1}{c}{ 0.0003 $\pm$ 0.0003 } \\

\multicolumn{1}{c}{$c$ (s$^{-2}$)} &\multicolumn{1}{c}{ $-0.0011^{+ 0.0007}_{ - 0.0008} $ } &\multicolumn{1}{c}{ $ -0.00002^{+ 0.00001}_{ - 0.00020} $} &\multicolumn{1}{c}{$-0.0009^{ + 0.0007}_{ - 0.0008} $} &\multicolumn{1}{c}{$- 0.00002^{+ 0.00002}_{- 0.00024} $} \\

\multicolumn{1}{c}{shift} &\multicolumn{1}{c}{ $0.00097^{ + 0.00004}_{ - 0.00003} $} &\multicolumn{1}{c}{ $0.00058^{+ 0.00006}_{- 0.00004} $} &\multicolumn{1}{c}{ 0.00099 $\pm$ 0.00004 } &\multicolumn{1}{c}{ 0.00059 $\pm$ 0.00005 } \\
\hline
\hline
\end{tabular}
\end{center}
\end{table*}

%%%%%%%%%%%%%%%%%%% prism wlc figures

\begin{figure*}
	\includegraphics[trim={0cm 0cm 0cm 0cm}, clip,width=0.9\textwidth]{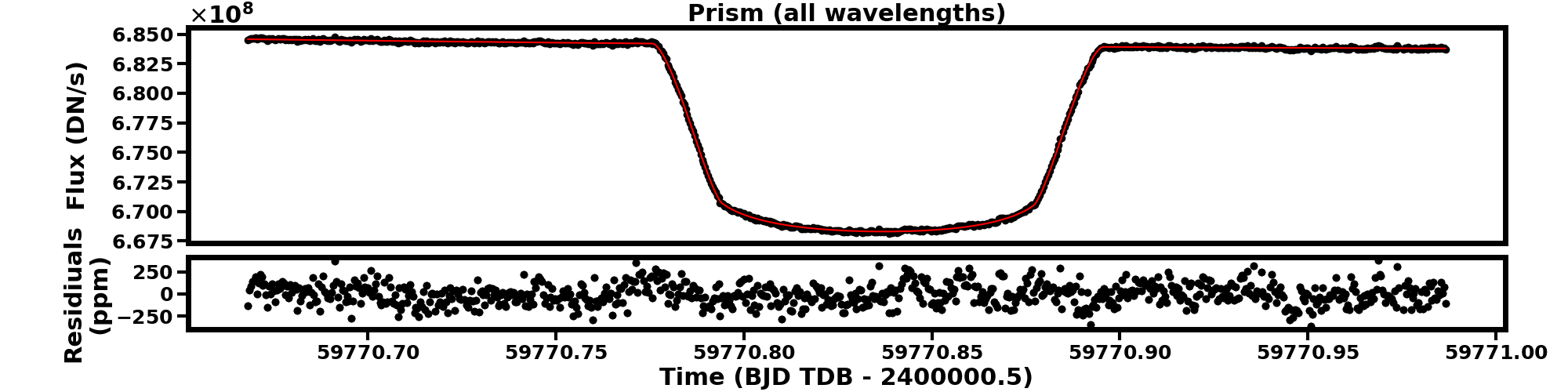}
 	\includegraphics[trim={0cm 0cm 0cm 0cm}, clip,width=0.9\textwidth]{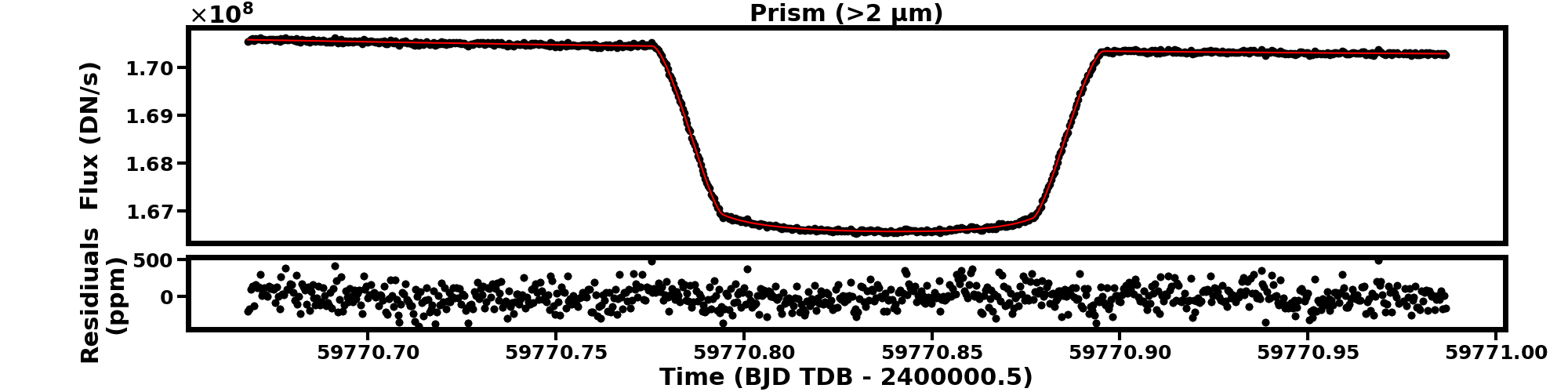}
    \caption{Prism White light curves and best fit solutions using median values from the MCMC posterior distribution, with residuals.}
    \label{fig:fits prism wlc}
\end{figure*}
%%%%%%%%%%%%%%%%%%%%%%%%%%%%%%%%%

%%%%%%%%%%%%%%%%%% wlc g395H figures

\begin{figure*}
 	\includegraphics[trim={0cm 0cm 0cm 0cm}, clip,width=0.9\textwidth]{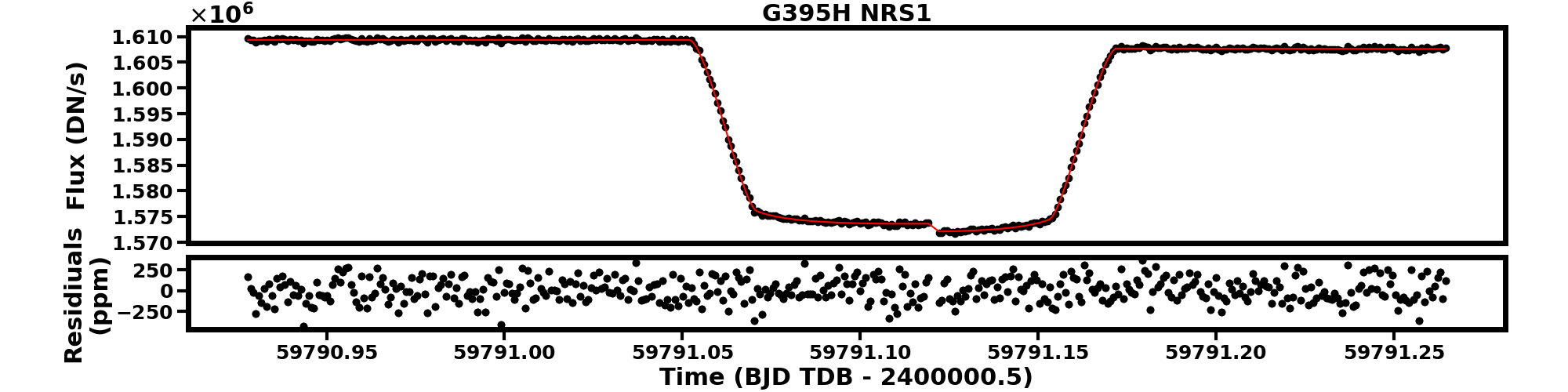}
 	\includegraphics[trim={0cm 0cm 0cm 0cm}, clip,width=0.9\textwidth]{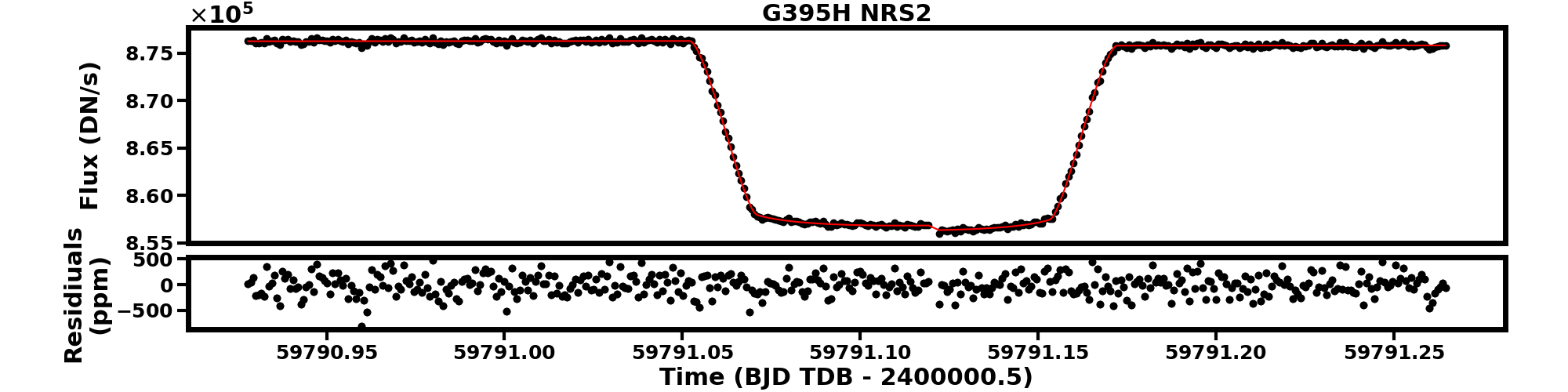}
 	\includegraphics[trim={0cm 0cm 0cm 0cm}, clip,width=0.9\textwidth]{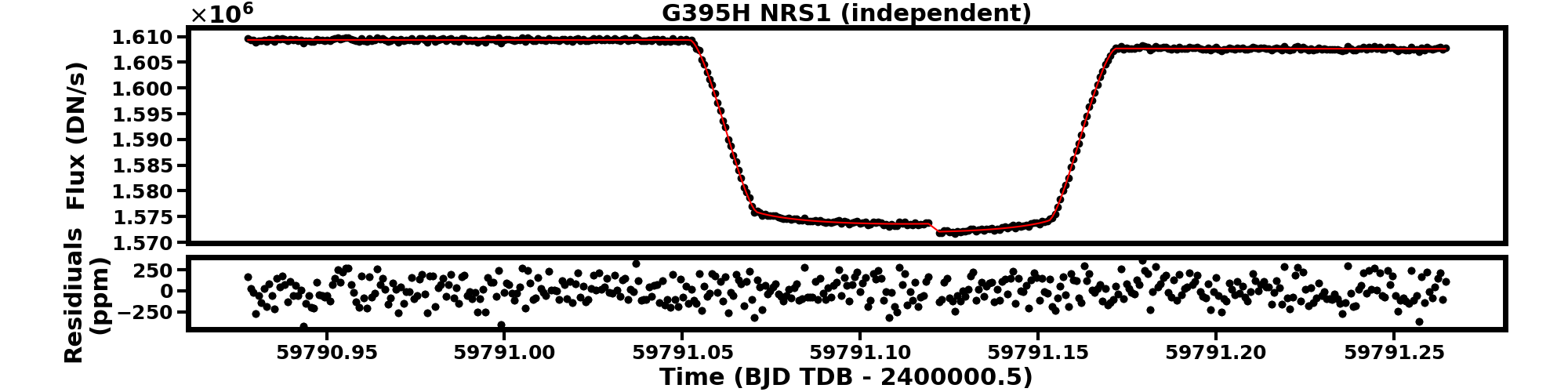}
 	\includegraphics[trim={0cm 0cm 0cm 0cm}, clip,width=0.9\textwidth]{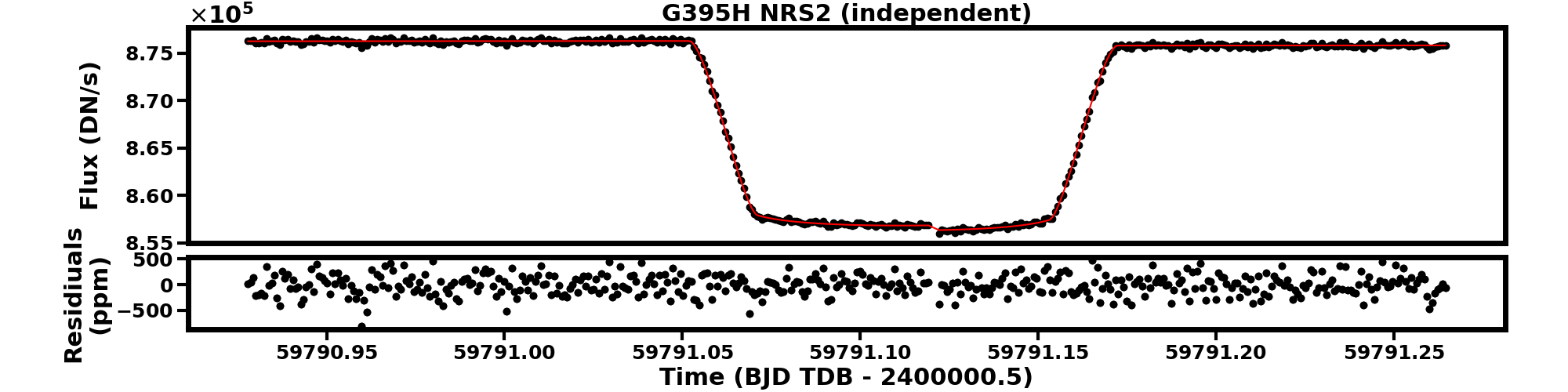}
    \caption{G395H white light curves and best fit solutions using median values from the MCMC posterior distribution, with residuals for G395H.}
    \label{fig:fits grating wlc}
\end{figure*}
%%%%%%%%%%%%%%%%%% 

\subsubsection{White light curve fitting}

The white light curve for the Prism was obtained by co-adding the Stage 4 spectral light curves.  We trialled this twice: 1) including only wavelengths above 2 \textmu m (and thus excluding the persistently saturated region), and 2) including all wavelengths.  For light curve fitting we use a Mandel-Agol transit model applied using PyLightcurve\footnote{https://github.com/ucl-exoplanets/pylightcurve} within a Monte Carlo Markov Chain (MCMC) algorithm applied using emcee \citep{Foreman-Mackey2013}. 
In addition to the planet-star radius ratio ($R_p/R_s$), we fit for mid-transit time ($t_0$), the ratio of semi-major axis to star radius ($a'/R_s$), the inclination angle ($i$), and two quadratic limb darkening coefficients ($c_1$ and $c_2$).  We also fit for the systematic trend and out-of-transit baseline using a second-order polynomial (and thus three coefficients, $a$, $b$ and $c$, where the trend is represented by $a (1 + bt + ct^2)$ (where $t$ is time since the first integration included).  The error on the data is obtained by fitting the OOT data only with the systematic model (using a least squares method), dividing this out and obtaining the standard deviation of the resulting points. The error bars on the individual light curve data points are then scaled by a factor such that the average error bar matches the standard deviation of points. 

This initial fit to the systematic trend also gives the signs of the coefficients $c$ and $b$.  These signs are held in variables and applied later in the final model fit, while the positive magnitudes of $c$ and $b$ are converted to natural logarithmic values for the MCMC algorithm.  A log-likelihood function is used with uniform priors for all free parameters.  We fit for the natural logarithms of all free parameters except $t_0$, $c_1$ and $c_2$.   
The period ($P$) is fixed to 4.0552941 days \citep{Mancini2018}. Eccentricity is set to 0 and argument of periastron to 90$^\circ$.  We use 64 walkers with a burn-in of 1000 steps, followed by 4000 steps for the production run. Mean acceptance fractions of 44 and 43 \%s respectively for the all wavelength and the $>2$ \textmu m case are obtained consistent with good convergence.  Figure \ref{fig:corner prism wlc} shows the posterior distributions for these cases and Figure \ref{fig:fits prism wlc} shows the white light curves with best-fitting model and residuals. Table \ref{table: wlc table prism} summarises the parameter estimation results from the Prism white light fits.  We find that there are no significant differences between in the system parameters obtained ($a'/Rs$ and $i$) using all wavelengths vs just $>$ 2 \textmu m, however given the uncertainties in managing the saturated region, we choose to use the $>$ 2 \textmu m values for the rest of the study.  The residuals for the all wavelength case have a standard deviation of 120 ppm which is $\sim$ 1.8 $\times$ the estimated noise based on propagating the ERR array values, and for the $>2$ \textmu m case, the residuals are 140 ppm, 1.6 $\times$ the estimated noise.

For G395H, the white light curves for NRS1 and NRS2 were obtained by co-adding all spectral light curves.  We performed two sets of fits 1) where the system parameters  $a'/Rs$ and $i$ are fixed to those from the Prism $>2$ \textmu m result, and 2) `independent' white light curve fits  where $a'/Rs$ and $i$ were not fixed to the Prism values but obtained directly from the fits.   In both cases we fitted for  $t_0$, $c_1$, $c_2$ and the natural logarithms of $R_p/R_s$ and three polynomial coefficients, $a$, $b$ and $c$. To correct for the tilt event, a `shift' parameter is added to the light curve model that adds an offset to the post-event section of the light curve model prior to multiplication by the systematic model.

Other aspects of the fit were as for the Prism above. We obtain mean acceptance rates of 31 and 32\% for NRS1 and NRS2 fixed-to-Prism cases and 29\% for the independent cases.  The results are summarised in  Table \ref{table: wlc table grating}. Figure \ref{fig:corner grating wlc} shows the posterior distributions and Figure \ref{fig:fits grating wlc} shows the light curve fits.  For the fixed-to-Prism cases, the residuals for NRS1 and NRS2 have standard deviations of 142 and 195 ppm respectively (1.4 and 1.5 $\times$ the estimated noise). For the independent cases these are almost the same, 141 ppm and 195 ppm respectively (1.4 and 1.5 $\times$ the estimated noise).
 
We find that the system parameters $a'/R_s$ and $i$ 
have somewhat lower median values in the `independent' G395H fit than in the Prism fit, however these are both highly correlated in the corner plots.
To allow an unbiased comparison of Prism and G395H final spectra we thus need to control for $a'/R_s$ and $i$, which we do by adopting the Prism $>2$ \textmu m values for  $a'/R_s$ and $i$ for all final spectral light curve fits.  The median $R_p/R_s$ results for the G395H results are somewhat greater than for the Prism all wavelength result, but closer to the Prism $>2$ \textmu m result.  This could be explainable by the high amplitude spectral features in the G395H range that pushing up the average apparent radius over NRS1 and NRS2 wavelength ranges.   

\subsubsection{Allan deviation analysis}

\begin{figure}
   	\includegraphics[trim={0cm 0cm 0cm 0cm}, clip,width=1\columnwidth]{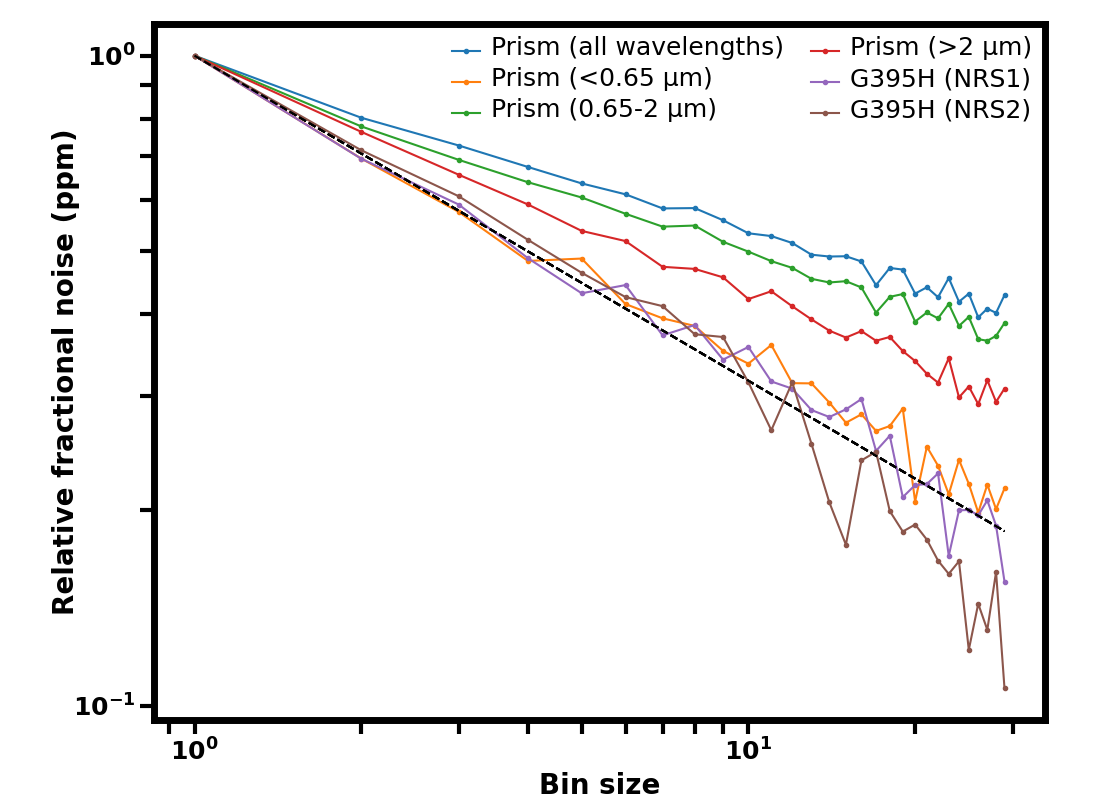}
    \caption{Allan deviation analysis of white light curve residuals.  The fractional noise is normalised to the first data point for comparison purposes.  
    The dotted black line 
    has a gradient of -0.5 in log-log space and indicates how uncorrelated noise would appear to integrate down.}
    \label{fig:Allan}
\end{figure}

We also performed an Allan deviation analysis of white light curves to examine for correlated noise.  We look at the following cases: 1) Prism all wavelengths, 2) Prism $>$ 2 \textmu m, 3) Prism 0.65-2 \textmu m, 4) Prism $<$0.65 \textmu m, 5) G395H independent NRS1, 6) G395H independent NRS 2.  The Allan deviation plots are shown in Figure \ref{fig:Allan}.  To produce these plots, the residuals are binned into progressively larger bin sizes.  At each bin size the fractional noise is calculated from the standard deviation of the binned counts divided by the binned signal. The latter is obtained from the coefficient $a$ multiplied by the number of binned points per bin.   For uncorrelated noise the gradient of the log of the fractional noise vs the log of the bin size should be -0.5. 
Shallower gradients may indicate correlated noise in the time line.  
In Figure \ref{fig:Allan}, the plots have been normalised to the first point, to allow easier comparison of the relative deviation from the uncorrelated expectation.  The dotted line with a gradient of -0.5 is shown as an indicator of how uncorrelated noise would bin down.

We find that the G395H noise bins down fairly closely to what we would expect for uncorrelated noise, consistent with \cite{Espinoza2022}.  For the Prism, when we include all wavelengths in the white light curve, there is a markedly shallower gradient than expected for uncorrelated noise.  For the fractionated white light curves, we see that the deviation is worst for 0.65-2 \textmu m, which encompasses the persistently saturated region.  The deviation is less for $>$ 2 \textmu m and lesser still for $<$ 0.65 \textmu m. These results suggest that the saturated region suffers the greatest correlated noise which may be related to the saturation or our data reduction methodology for that region.  However some correlated noise extends beyond this region and this is therefore not attributable to saturation.  It may be related to 1/f noise, our chosen systematic model or an additional unidentified cause
but further investigation is needed to elucidate this.  In Figure \ref{fig:fits prism wlc}, we see a slight upward `bump' in the residuals around the start of ingress (more noticeable in the all wavelengths case).  We repeated the Prism white light curve fits with this region (containing 43 light curve points) removed, and found the Allan deviations for the different cases to be unaffected, indicating that the correlated noise is not due to a poorer model fit in this region.

We use the Monte Carlo prayer bead method \citep{Gillon2007, Manjavacas2018} to estimate the error inflation if we account for the correlated noise.  For each case we perform 50 trials.  In each trial we produce 1000 model realisations of the white light curve and then fit for all system (except period, eccentricity and argument of periastron which are set to the previously stated values), light curve and systematic parameters using LMFIT \citep{Newville2016}.  Each realisation is produced as follows. The residuals from the MCMC best-fit model and the light curve are obtained. The residuals are then shifted in sequence by a random value picked from a uniform distribution ranging from zero up to the total number of light curve points.  The shifted residuals are added to the best-fit model to construct a `new' light curve.  This light curve is then fitted using  LMFIT to extract the planet-star radius ratio.  1000 values of the planet-star radius ratio are thus obtained and we take the standard deviation of the distribution as an estimate of the 1$\sigma$ uncertainty that incorporates correlated noise.  We then take the ratio of this to the error given by LMFIT on the original light curve, to obtain an error inflation factor.   We then obtain the mean and standard deviation of the inflation factor from the 50 trials.
%The inflation in errors on the transit depth as a result of this are as follows:  Prism (all wavelengths) $\times$ 1.72, Prism ($>2$ \textmu m) $\times$ 1.36, and Prism ($0.65-2$ \textmu m) $\times$ 1.57. 
The inflation in errors on $R_p/R_s$ as a result of this are as follows:  Prism (all wavelengths) $\times$ 1.57 $\pm$ 0.03, Prism ($>2$ \textmu m) $\times$ 1.36 $\pm$ 0.03, Prism ($0.65-2$ \textmu m) $\times$ 1.45 $\pm$ 0.03,
Prism ($<$ 0.65 \textmu m) $\times$ 1.29 $\pm$ 0.03.
We confirm the relative lack of correlated noise in the G395H data by finding inflation factors of $\times$ 1.09 $\pm$ 0.02 in NRS1 and $\times$ 1.119 $\pm$ 0.025 in NRS2.

\subsubsection{Use of model 4-factor limb-darkening coefficients}
\begin{figure*}
 	\includegraphics[trim={0cm 0cm 0cm 0cm}, clip,width=0.9\textwidth]{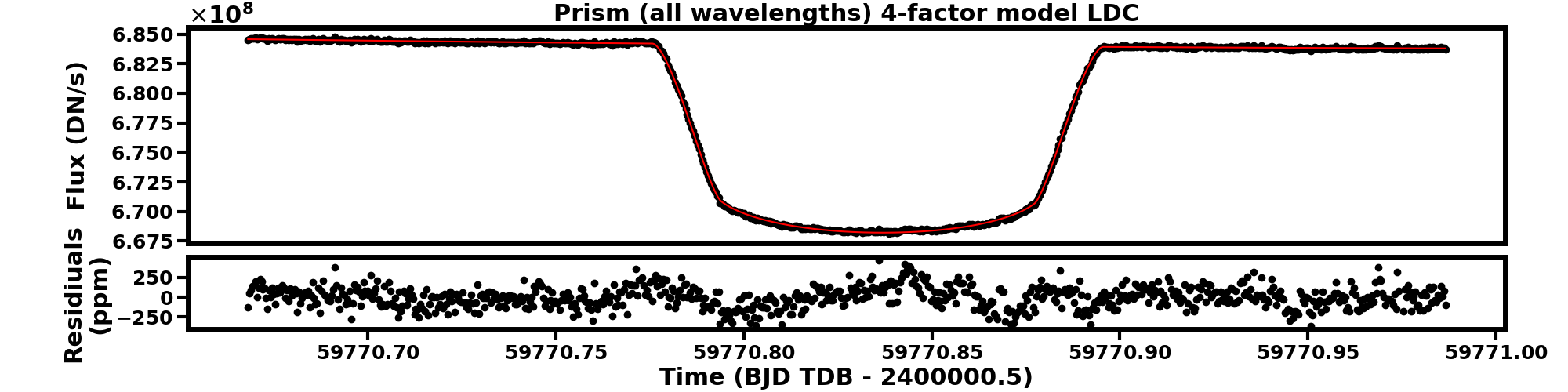}
 	\includegraphics[trim={0cm 0cm 0cm 0cm}, clip,width=0.9\textwidth]{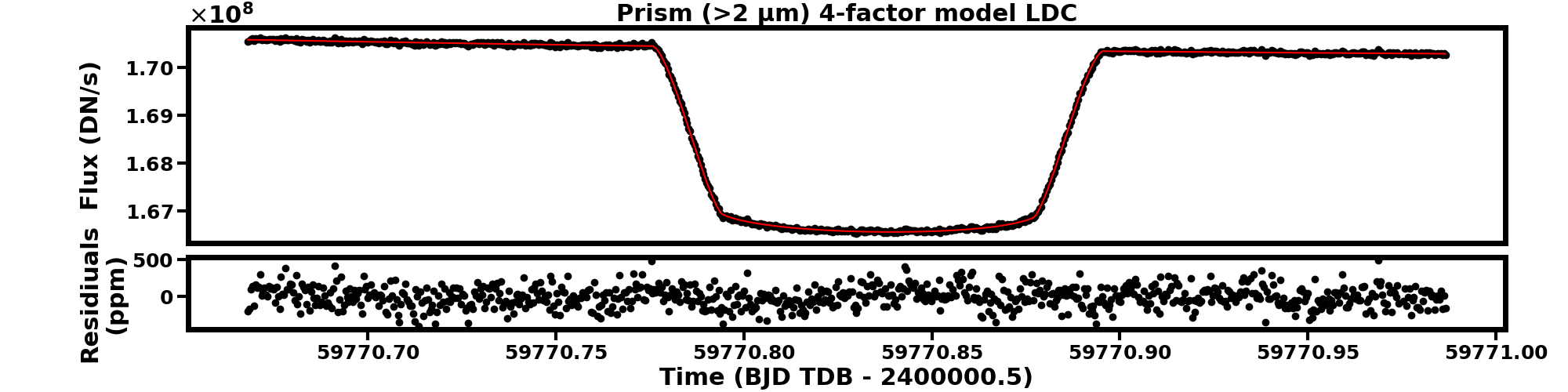}
 	\includegraphics[trim={0cm 0cm 0cm 0cm}, clip,width=0.9\textwidth]{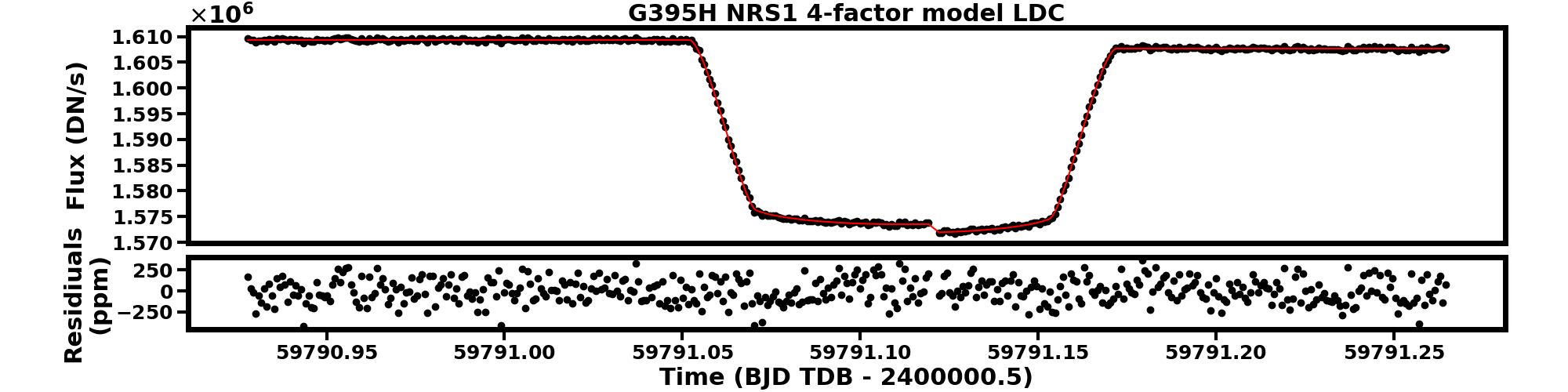}
 	\includegraphics[trim={0cm 0cm 0cm 0cm}, clip,width=0.9\textwidth]{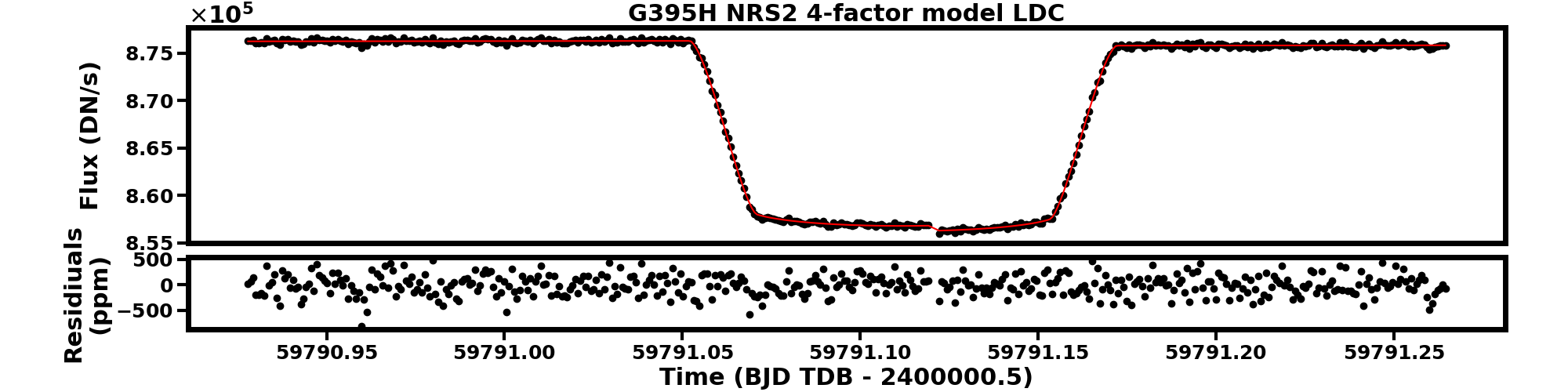}
    \caption{White light curves and best fit solutions when using model 4-factor LDCs.}
    \label{fig:fits claret}
\end{figure*}

To see if any improvement in accuracy or precision occurs using model limb-darkening coefficients (LDCs) we repeated the following cases using model 4-factor LDCs obtained from ExoCTK \footnote{https://exoctk.stsci.edu/limb\textunderscore darkening}: Prism (all wavelengths), Prism ($>$ 2 \textmu m), G395H NRS1 and NRS2 (independent of Prism values).  The resulting corner plots and light curve fits are shown in Figures \ref{fig:corner claret} and \ref{fig:fits claret}, with the parameters summarised in Table \ref{table: wlc table claret}  For the Prism, there is close agreement with the baseline case in the $>$ 2 \textmu m case.  In the all wavelengths case, the transit depth, $a'/R_s$ and $i$ are distinct at 1$\sigma$ but not at 2$\sigma$.  
In G395H NRS1, the transit depths again agree at 2$\sigma$, while in NRS2 they agree at 1$\sigma$. The residuals on the Prism show more systematic variation than in the baseline case  indicating a poorer fit to the data: the standard deviation of the residuals are 133 ppm and 141 ppm for the all wavelength and $>$2 \textmu m cases respectively. For G395H, the fits appears similar to the baseline cases with residual standard deviations of 145 and 196 ppm for NRS1 and NRS2 respectively.  As a goodness-of-fit test to compare the baseline case with the 4-factor LDC case, we employ the reduced chi-squared, keeping the noise constant between the two comparisons.  We use the absolute noise from the baseline cases for this purpose.
 (i.e. the standard deviation of the residuals). Table \ref{table: wlc goodness of fit} summarises the results.  The reduced chi-square is larger in the 4-factor model cases in all four cases, with the greatest increase being in the Prism all wavelength case.  This would indicate that the empirically-derived quadratic LDCs give a better fit to the data than the model 4-factor LDCs.

%%%%%%%%%%%%%%%%%% claret table
\begin{table*}
\begin{center}
\caption{Summary of retrieved white light curve parameters using model 4-factor LDCs}
\label{table: wlc table claret}
\begin{tabular}
%{cccccc} 
 {p{1.1cm}p{3cm}p{3cm}p{3cm}p{3cm}p{3cm}	}
\hline
\hline
\multicolumn{1}{c}{Parameter} &
\multicolumn{1}{c}{Prism } &
\multicolumn{1}{c}{Prism}&
\multicolumn{1}{c}{G395H NRS1} &
\multicolumn{1}{c}{G395H NRS2}
\\
\multicolumn{1}{c}{} &
\multicolumn{1}{c}{all wavelengths} &
\multicolumn{1}{c}{>2 \textmu m only} &
\multicolumn{1}{c}{ } &
\multicolumn{1}{c}{ }
\\
\hline
\multicolumn{1}{c}{$R_p/R_s$} &\multicolumn{1}{c}{ 0.14487 $\pm$ 0.00006 } &\multicolumn{1}{c}{ 0.14609 $\pm$ 0.00006 } &\multicolumn{1}{c}{ 0.14584 $^{+0.00007}_{-0.00006}$ } &\multicolumn{1}{c}{ 0.14653 $\pm$ 0.00009 } \\
\multicolumn{1}{c}{$a'/R_s$} &\multicolumn{1}{c}{ 11.51 $\pm$ 0.02 } &\multicolumn{1}{c}{ 11.43 $\pm$ 0.02  } &\multicolumn{1}{c}{ 11.448 $\pm$ 0.025  } &\multicolumn{1}{c}{ 11.41 $\pm$ 0.03 } \\
\multicolumn{1}{c}{$i$ ($^\circ$} &\multicolumn{1}{c}{ 87.90 + 0.02 } &\multicolumn{1}{c}{ 87.79 $\pm$ 0.02 } &\multicolumn{1}{c}{ 87.810 $\pm$ 0.025} &\multicolumn{1}{c}{ 87.77 $\pm$ 0.03 } \\
\multicolumn{1}{c}{$t_0$ (BJD TDB - 2400000.5)} &\multicolumn{1}{c}{ 59770.83563 $\pm$ 0.00001 } &\multicolumn{1}{c}{ 59770.83566 $\pm$ 0.00001 } &\multicolumn{1}{c}{ 59791.11202 $\pm$ 0.00002 } &\multicolumn{1}{c}{ 59791.11213 $\pm$ 0.00003 } \\
\multicolumn{1}{c}{$a$ (DN/s)} &\multicolumn{1}{c}{ 6.84573 $\pm$ 0.00009 $ \times 10^8$ } &\multicolumn{1}{c}{ 1.70584 $\pm$ 0.00003 $ \times 10^8$ } &\multicolumn{1}{c}{ 1.60933 $\pm$ 0.00002 $ \times 10^6$ } &\multicolumn{1}{c}{ 8.7623 $\pm$ 0.0002 $ \times 10^5$ } \\
\multicolumn{1}{c}{$b$ (s$^{-1}$)} &\multicolumn{1}{c}{-0.0057 $\pm$ 0.0003} &\multicolumn{1}{c}{-0.0081 $\pm$ 0.0003 } &\multicolumn{1}{c}{ $0.000008^{ + 0.000127}_{- 0.000007} $} &\multicolumn{1}{c}{ 0.0005 $\pm$ 0.0003 } \\
\multicolumn{1}{c}{$c$ (s$^{-2}$)} &\multicolumn{1}{c}{ 0.0066 $\pm$ 0.0008} &\multicolumn{1}{c}{ 0.009 $\pm$ 0.001} &\multicolumn{1}{c}{ $-0.0002^{+ 0.0002}_{ - 0.0008} $} &\multicolumn{1}{c}{$ -0.00002^{ + 0.00002}_{ - 0.00026} $} \\

\multicolumn{1}{c}{shift} &\multicolumn{1}{c}{ N/A } &\multicolumn{1}{c}{ N/A } &\multicolumn{1}{c}{$ 0.00103^{+ 0.00002}_{ - 0.00003} $ } &\multicolumn{1}{c}{$ 0.00062^{+ 0.00005}_{ - 0.00006} $} \\
\hline
\hline
\end{tabular}
\end{center}
\end{table*}
 %%%%%%%%%%%%%%%%%%%%%%%%%%%%%%%%%%%%%%%%%%%%%%%%%
%%%%%%%%%%%%%%%%%%%%%%%%%%claret figures

%%%%%%%%%%%%%%%%%%%%%%%%%%

\begin{table}
\caption{White light curve goodness-of-fit for 4-factor model LDCs vs a quadratic LDC fit.}
\label{table: wlc goodness of fit}
    \centering
    \begin{tabular}{lccc}
    \hline    \hline
        ~ & Noise  & Baseline   & Model      \\  
                ~ & (DN/s) &  (quad. LDC fit)   &   4-factor LDC    \\   
                ~ &   &  $\chi^2_\nu$ &  $\chi^2_\nu$  \\ 
        \hline
        Prism (all wavelengths) & 82194.90 & 1.011 & 1.234 \\ 
        Prism ($>$ 2 \textmu m) & 23817.34 & 1.011 & 1.033 \\  
        G395H NRS1 & 226.93 & 1.028 & 1.075 \\ 
        G395H NRS2 & 170.87 & 1.023 & 1.028 \\      \hline    \hline
    \end{tabular}
\end{table}

\begin{figure}
	\includegraphics[trim={0cm 0cm 0cm 0cm}, clip,width=1\columnwidth]{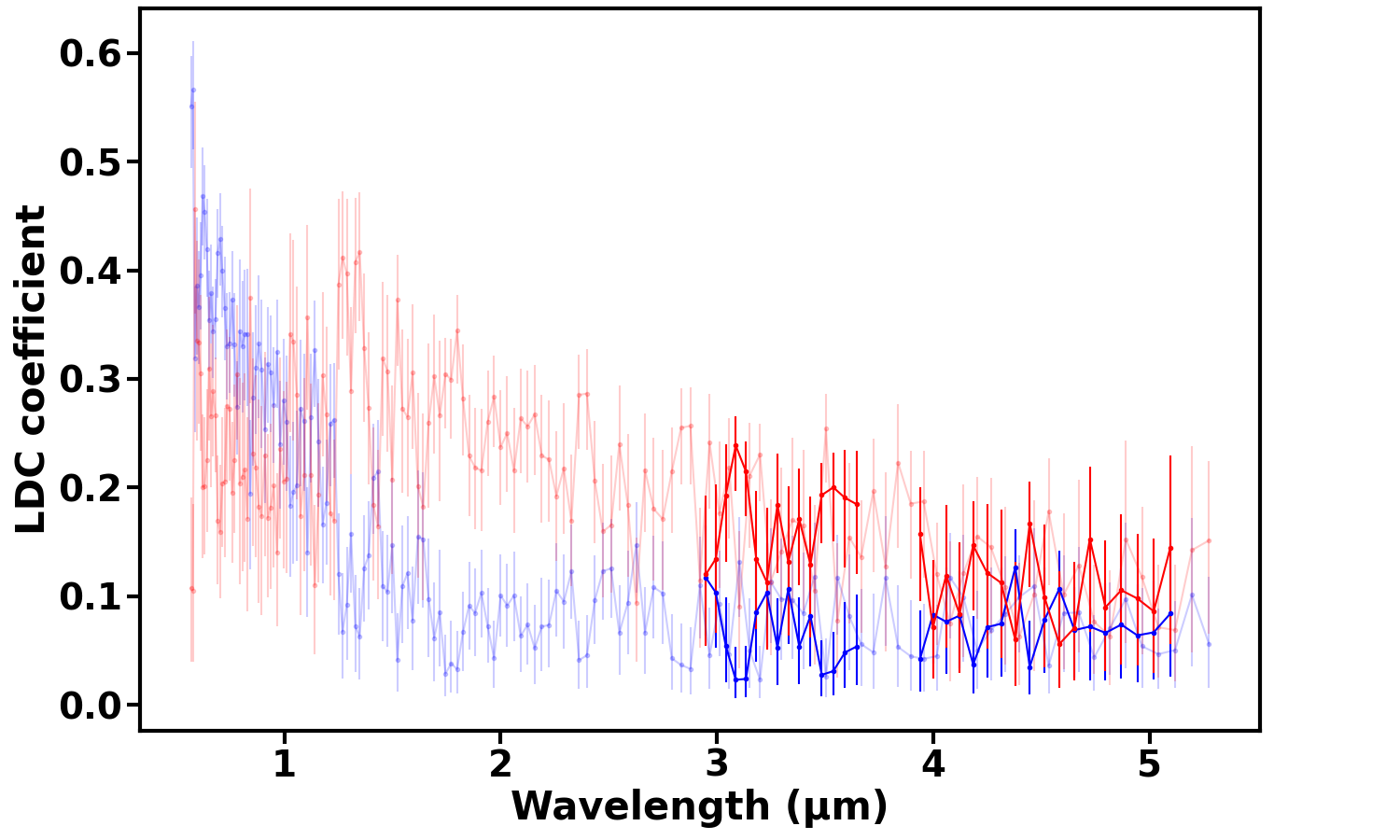}
    \caption{Empirically-derived quadratic LDCs from R=66 Prism data (faded red and blue points), and those from R=66 G395H data (bolder red and blue points).  $c_1$ is in blue and $c_2$ in red.  The values between Prism and G395H match within the 1$\sigma$ error bars.
    }
    \label{fig:LDC comparison}
\end{figure}

\begin{figure}
	\includegraphics[trim={0cm 0cm 0cm 0cm}, clip,width=1\columnwidth]{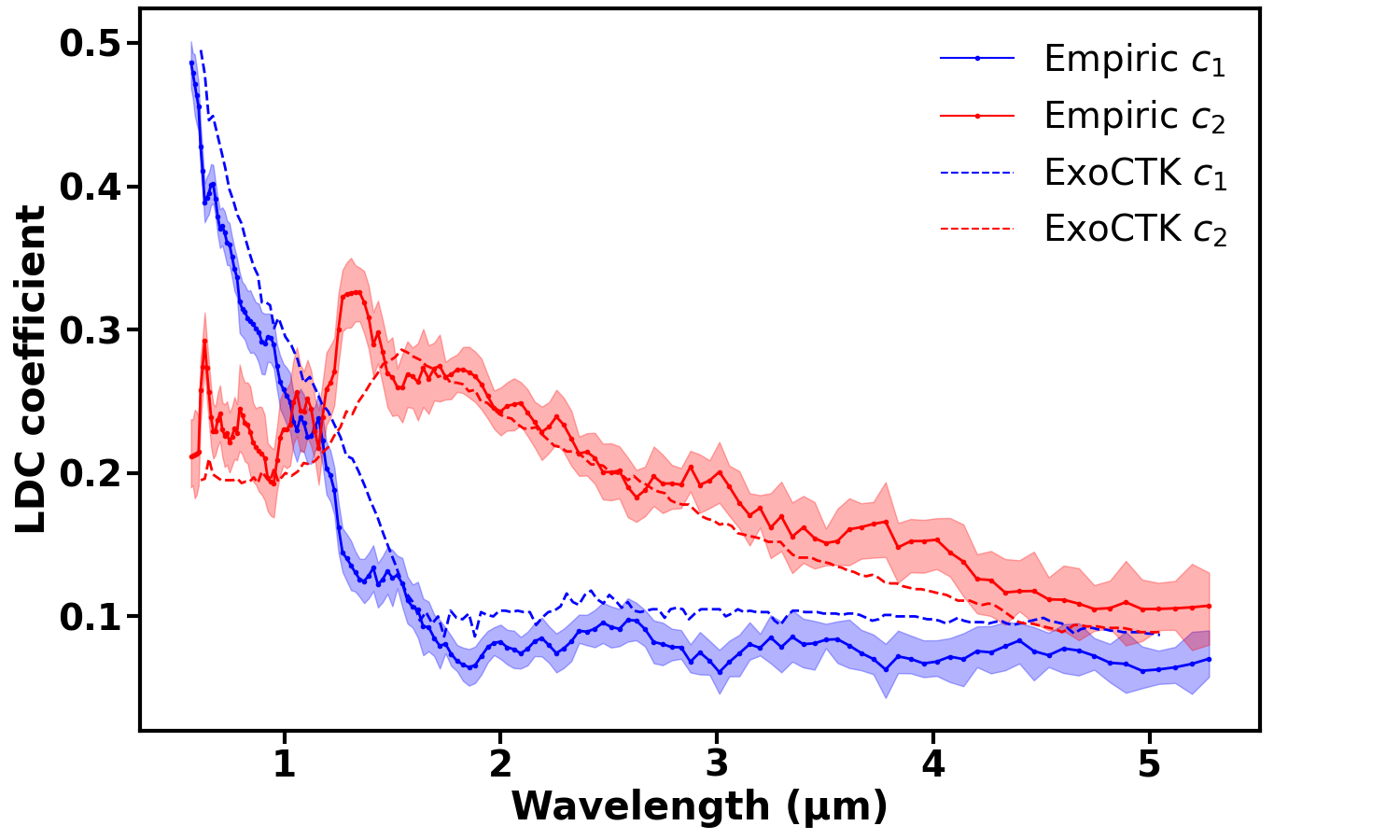}
    \caption{Empirically-derived quadratic LDCs from Prism data and those from the ExoCTK website model.  
    The latter uses the Kurucz ATLAS9 model grid ($T_\mathrm{eff}$=5000K, log$g$=4.45, Fe/H =0.01). Shaded regions give the 1$\sigma$ error on the empiric values.
    }
    \label{fig:LDC}
\end{figure}

\begin{figure}
	\includegraphics[trim={0cm 0cm 0cm 0cm}, clip,width=1\columnwidth]{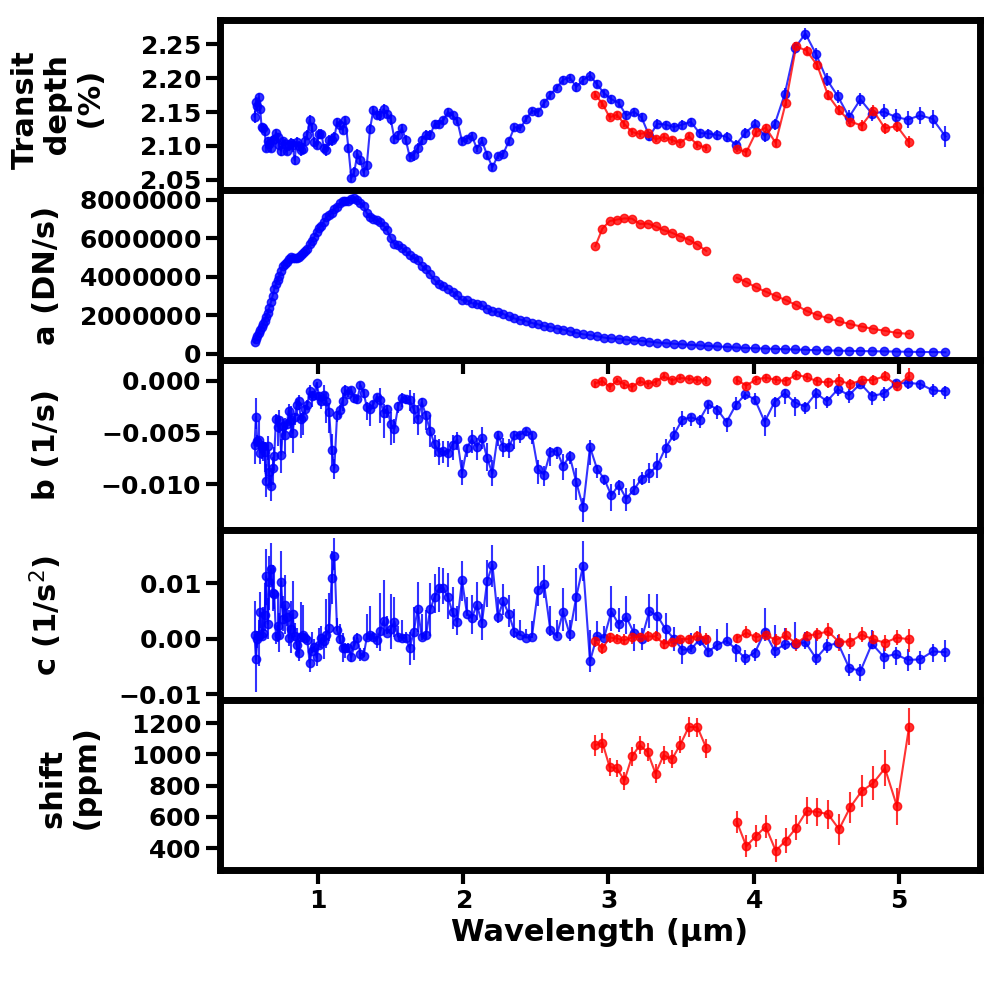}
    \caption{Systematic coefficients. Data have been binned to R=60.  Blue = Prism.  Red = G395H.  Transmission spectra are shown for comparison in top panel. For G395H, values for $a$ have been increased by a factor of 5000 for clarity. The light curve shift applied during light curve fitting for G395H to correct for the tilt event is shown in the bottom panel.
    }
    \label{fig:coeffs}
\end{figure}

\subsubsection{Spectral light curve fitting}
\label{Spectral light curve fitting}

We obtained model quadratic LDCs from the ExoCTK website, which were used for preliminary studies and to give initial values for fits, however for the final analysis we chose to obtain empiric LDCs for the full wavelength range covered by both Prism and G395H from light curve fitting.  This is partially motivated by the fact that for white light curves the empiric quadratic fit gave a better fit to data than the 4-factor model LDC coefficients.  Empiric 4-factor fits would be challenging due to the additional two free-parameters required, and quadratic fits have been used in previous fits to WASP-39 data \citep{Rustamkulov2023, Alderson2023, Ahrer2023}.  In addition, \cite{Rustamkulov2023} found that WASP-39 is 6\% brighter at the limb than predicted by models, indicating an empiric approach is preferred.

The spectra from Prism and G395H were binned to R=66 (slightly lower than the lowest native spectral power across the Prism subarray) to optimise the SNR.    We then fitted each binned spectral light curve as described below and obtained the two limb darkening coefficients, $c_1$ and $c_2$ for the binned wavelengths.  Figure \ref{fig:LDC comparison} shows the extracted LDCs with error bars.  Despite differences in instrument transmission, we find the empiric LDCs for G395H match those from Prism within the 1$\sigma$ errors.  As a result we proceed using just the Prism-derived LDCs for both modes.

A smoothing function was then applied to the Prism LDC vs wavelength plots to obtain the final LDCs\footnote{This involved a convolution with a 10-point wide box function, with five points at each end filled in with values from a polynomial fit.}  Figure  \ref{fig:LDC} shows the final empirically derived LDCs and those from the ExoCTK model.  On running the final spectral light curve fits for Prism and G395H, the LDCs were obtained for each spectral channel  by interpolating the empiric LDC vs wavelength plots to the central wavelength of each light curve.  

We fit the spectral light curves for both Prism and G395H at their native (pixel column level) resolutions. To allow comparison with the Prism, G395H spectral light curves were also binned to the Prism resolution. We exclude NRS2 wavelengths $>$ 5.1 \textmu m, due the rapid fall off in the transmission above that wavelength resulting in low SNR (Figure \ref{fig:stellar spectra}).  Example posterior plots and light curve fits for full resolution cases, are shown in Figure \ref{fig:corner slc example} ). 
In fitting the spectral light curves for each configuration (and NRS detector) we fix $t_0$ to the values in Tables \ref{table: wlc table prism} and \ref{table: wlc table grating} (i.e. to the Prism $>2$ \textmu m case, and the G395H fixed-to-Prism cases).  We fix the values for $a'/R_s$ and $i$ to those obtained from the Prism white light curve fit ($>2$ \textmu m case) and fix the period to 4.0552941 days \citep{Mancini2018}. Eccentricity is set to 0 and argument of periastron to 90$^\circ$. For Prism and both G395H detectors we fit for $R_p/R_s$ and three polynomial coefficients for the systematic fit, $a$, $b$ and $c$. Additionally for G395H we fit for the light curve `shift' correcting for the mirror tilt event.

Figure \ref{fig:coeffs} shows how the systematic coefficients $a$, $b$ and $c$ and the shift parameter in G395H vary with wavelength. The results have been binned to R=60 for clarity.
$a$ follows the shape of the stellar spectrum, giving an out-of-transit baseline flux.  The variation of $b$ and $c$ is complex in Prism, and appear negatively correlated to each other in wavelength, but not obviously to spectral features.  In G395H, $b$ and $c$ are within 1$\sigma$ of zero, consistent with no time-dependent systematic trend.  The shift parameter appears fairly constant with wavelength in NRS1, but displays an increasing trend with wavelength in NRS2.

\begin{figure*}
\includegraphics[trim={0cm 1cm 2cm 0cm}, clip,width=\textwidth]{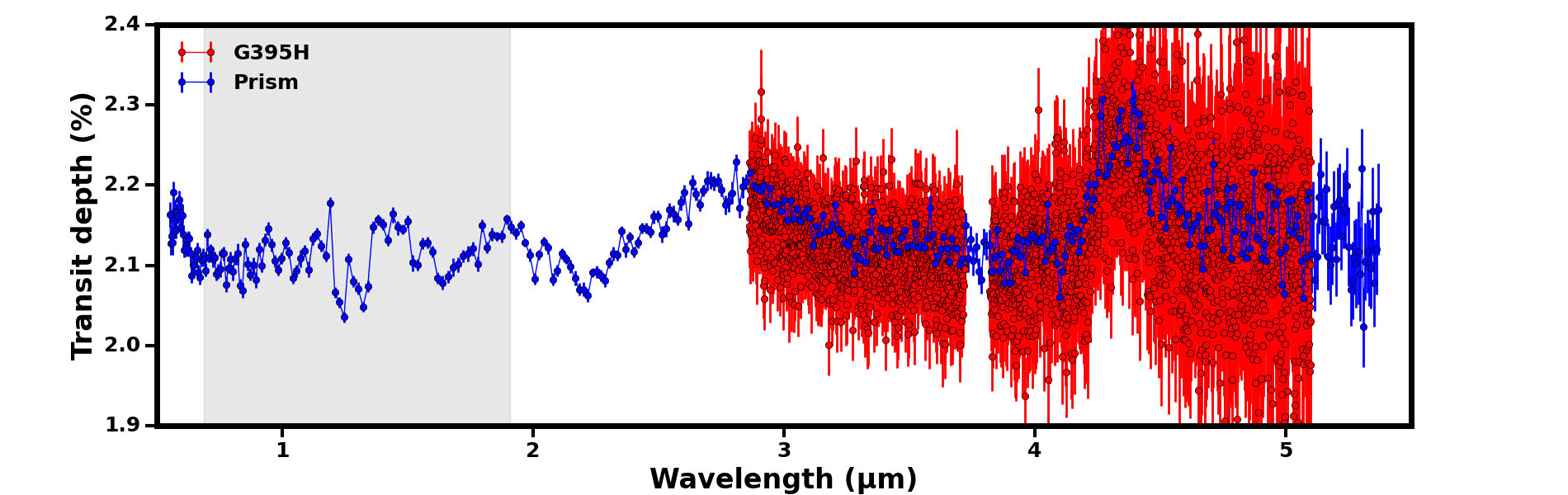}
\includegraphics[trim={0cm 1cm 2cm 0cm}, clip,width=\textwidth]{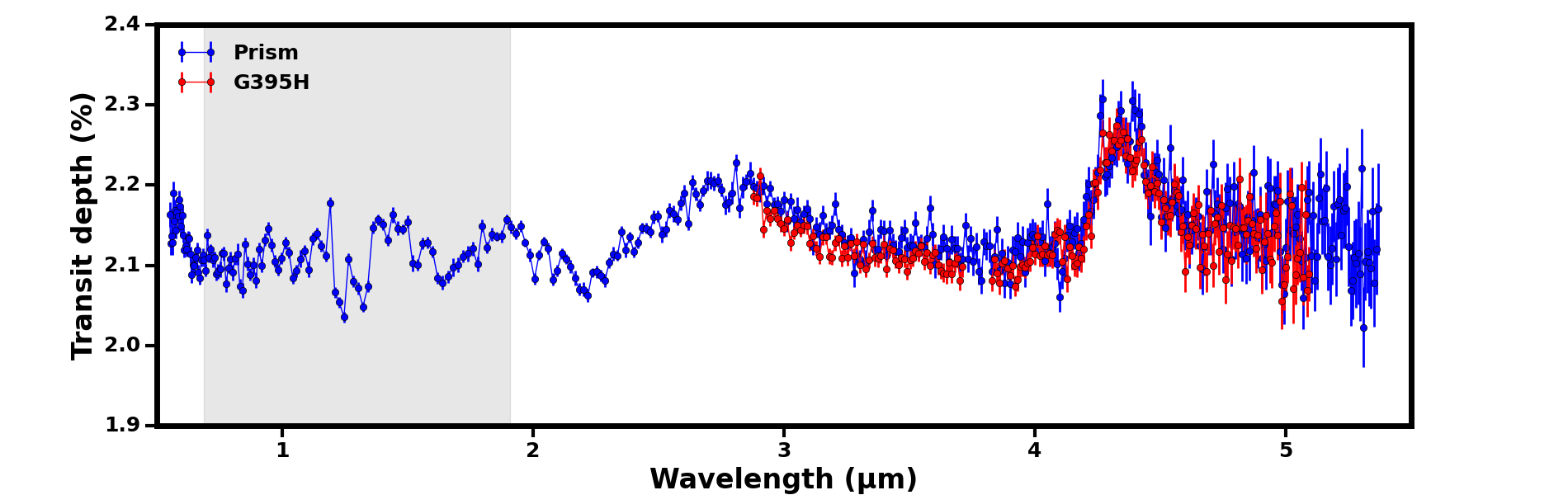}
\includegraphics[trim={0cm 1cm 2cm 0cm}, clip,width=\textwidth]{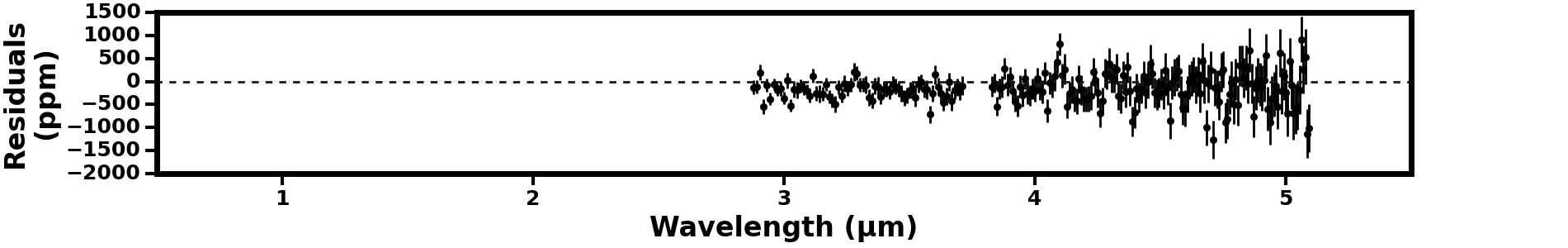}
\includegraphics[trim={0cm 0cm 2cm 0cm}, clip,width=\textwidth]{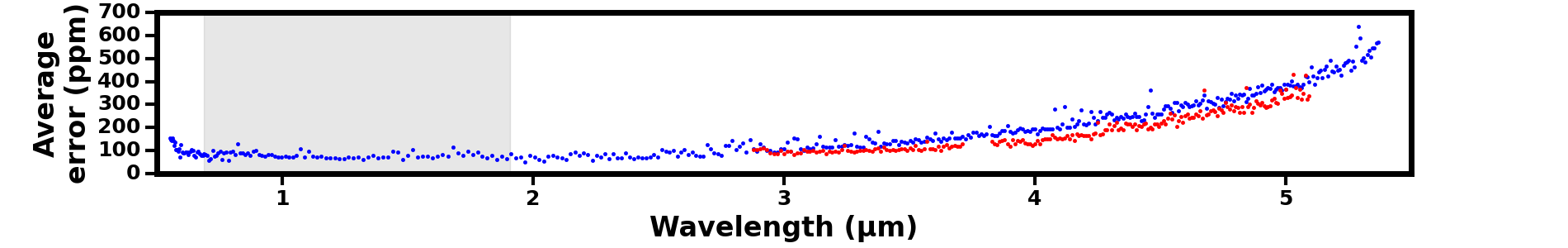}
    \caption{Transmission spectra of WASP-39 b obtained using JexoPipe.  Uppermost plot shows the spectra obtained at the native resolution without any wavelength binning.  The shaded grey area indicates the region of persistent saturation for the Prism. Central points are the median of the posterior distributions and the error bars span the 16th-84th percentile range.  In the second plot down, the Prism spectrum is shown at native resolution, while the G395H spectra are binned to the Prism wavelength grid to allow comparison of the transit depths with wavelength.  To permit a difference comparison of central points and quadrature sum of errors, the error bars show the average error (half the 16th-84th percentile range) and the central points are the average of the 16th and 84th percentiles in the posterior distribution.  The third plot gives the difference between the binned G395H spectrum and the Prism spectrum central points (i.e. residuals = G395H minus Prism).  The error bars on the residuals are the quadrature sum of the error bars as shown in the second plot.  The fourth plot gives the average error on the transit depth at the native resolution of the Prism.}
    \label{fig:final spectrum}
\end{figure*}

\section{Transmission spectra}

The final transmission spectra obtained are shown in Figure \ref{fig:final spectrum}.  The uppermost plot shows the unbinned `native resolution' spectra.  In the second plot the G395H spectra are rebinned to the resolution of the Prism spectrum, with the residuals between the two configurations and the average errors shown in the lower two plots.  

\subsection{Prism}

\begin{figure*}
	\includegraphics[trim={0cm 0cm 0cm 0cm}, clip,width=\columnwidth]{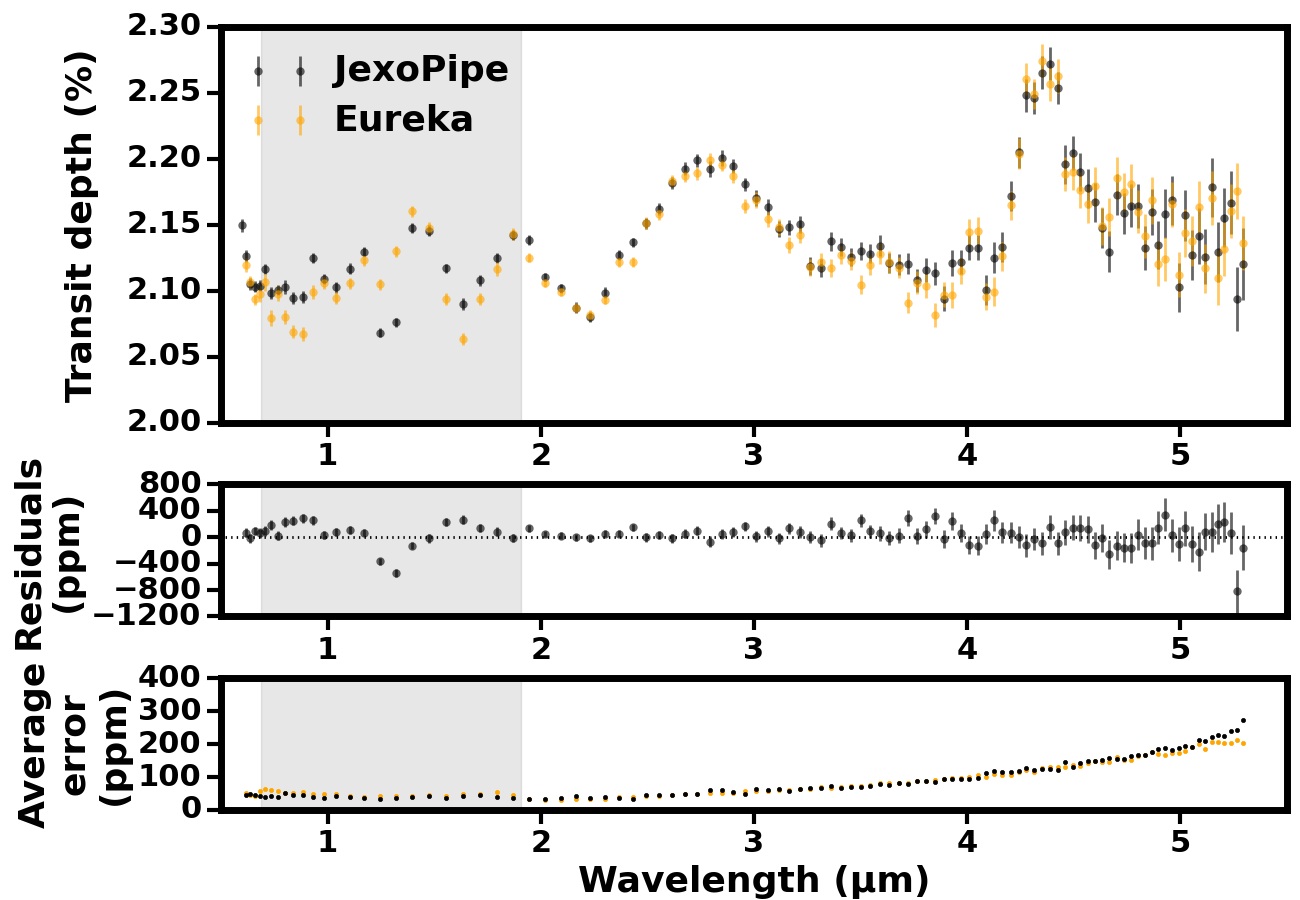}
 	\includegraphics[trim={0cm 0cm 0cm 0cm}, clip,width=\columnwidth]{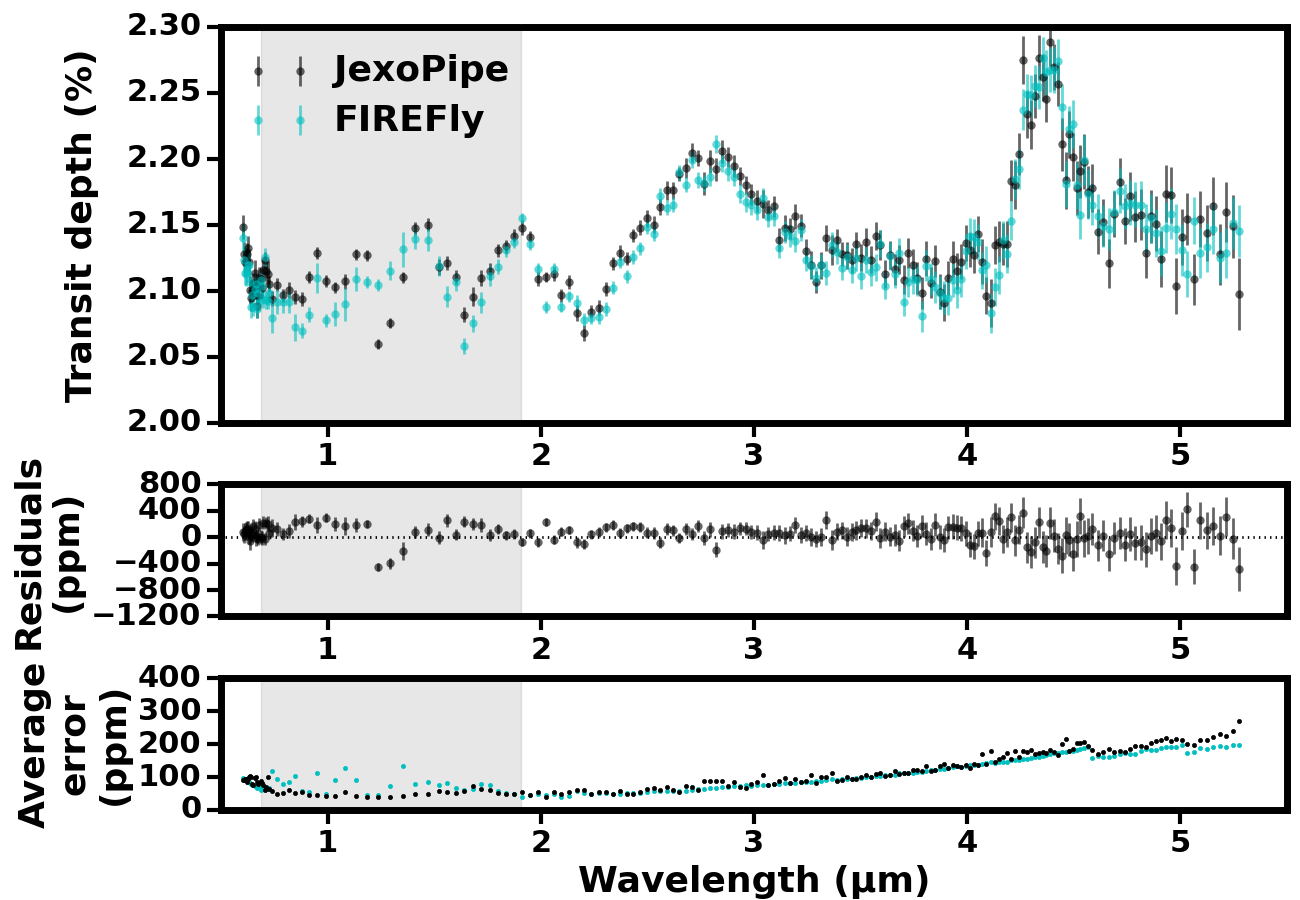}
	\includegraphics[trim={0cm 0cm 0cm 0cm}, clip,width=\columnwidth]{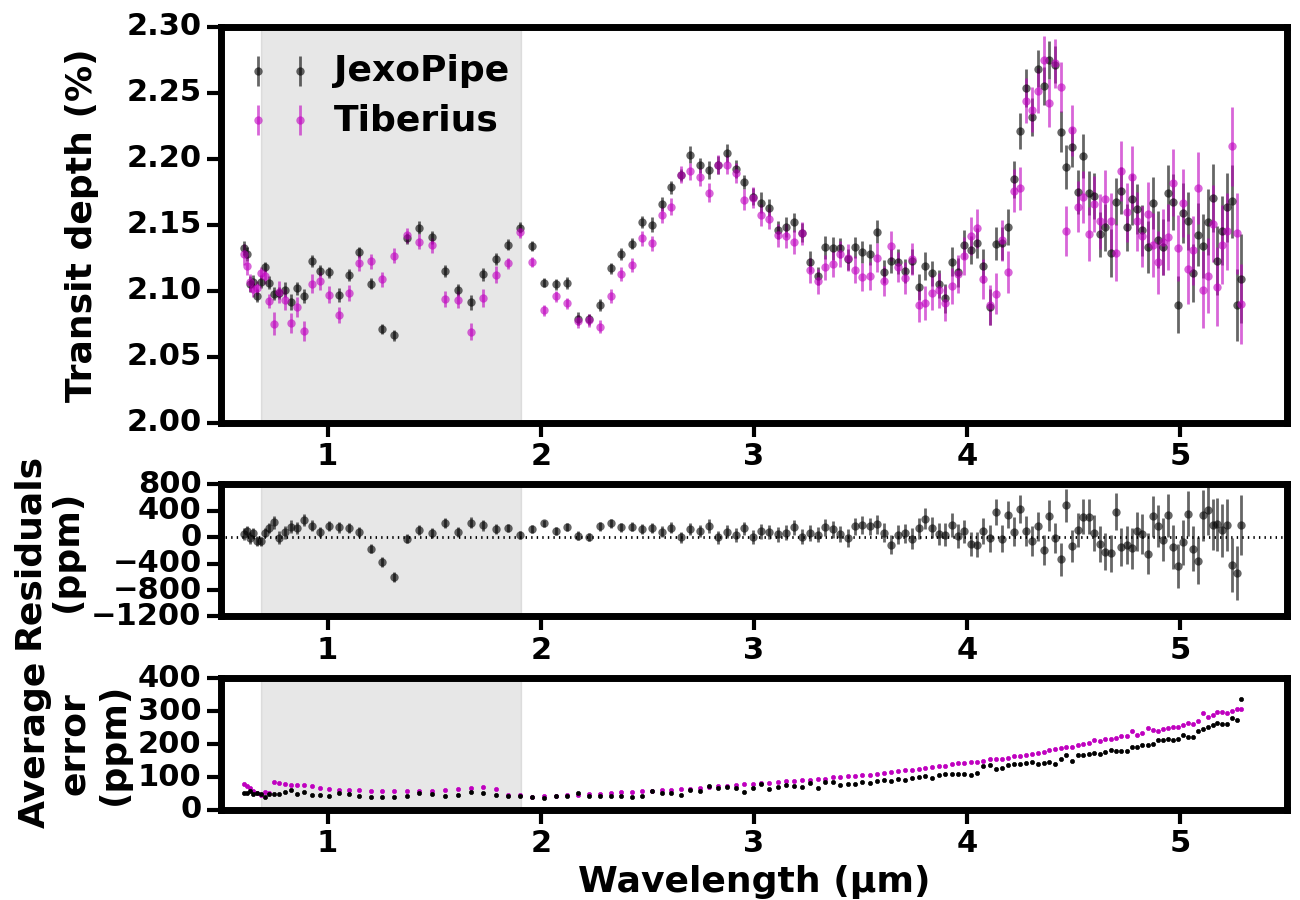}
 	\includegraphics[trim={0cm 0cm 0cm 0cm}, clip,width=\columnwidth]{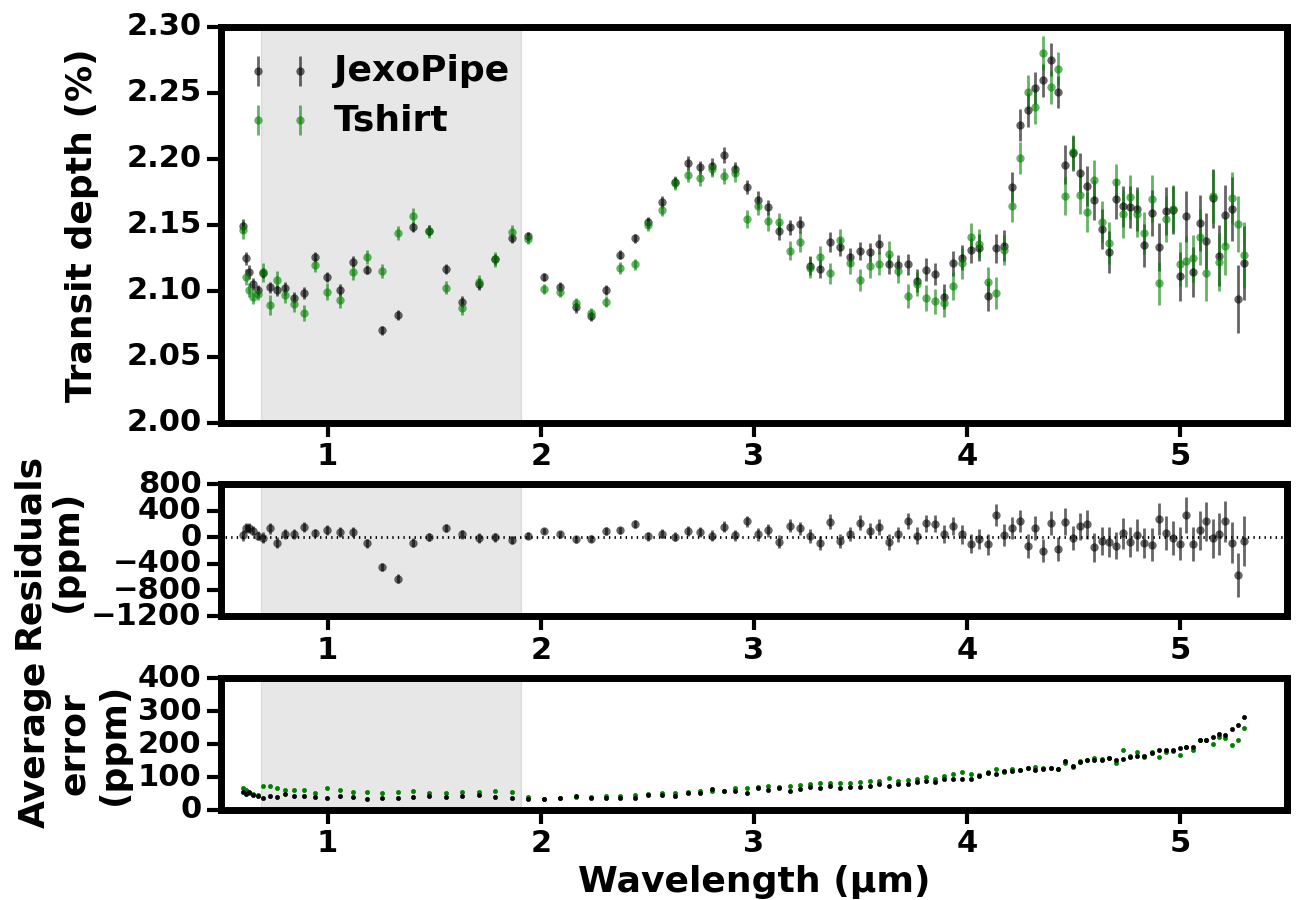}

    \cprotect\caption{Comparison of our Prism result with other Prism pipelines presented in \cite{Rustamkulov2023}. The JexoPipe result is rebinned to the wavelength grid of the comparison pipeline's publicly available spectrum. The grey shaded area indicates the region of persistent saturation.  Residuals are JexoPipe minus the comparison pipeline spectrum. The average errors on the transit depths for the comparison spectrum and the JexoPipe spectrum when binned to the resolution of the comparison spectrum are shown in the lowest plots.}
    \label{fig:compare prism}
\end{figure*}

Using the strategy of group control and increasing \verb'n_pix_grow_sat' to 3, we were able to recover data from the persistently saturated region including the 1.4 \microns H$_2$O feature. However, given there is no consensus on the optimal strategy to manage the saturated region, the results in this region should be considered tentative.  We recover the previously observed water features at  1.8 and 2.8 \textmu m, the large CO$_2$ feature at 4.3 \textmu m and the likely SO$_2$ feature at 4 \textmu m.  Figure \ref{fig:compare prism} compares our Prism spectrum with those publicly-available spectra from the four pipelines used in the study by \cite{Rustamkulov2023}.  %There appears to be a reasonably good degree of agreement between our result and these previous results, in terms of amplitude of the features and baseline level.   T
The errors on the residuals are the quadrature sum of the 1$\sigma$ error bars from the comparison pipeline spectrum and those from the re-binned JexoPipe spectrum.
There is some disagreement within the persistently saturated region, where JexoPipe gives a  shallower transit depth compared to the other pipelines on the blue-ward side of the 1.4 \microns water feature, but a deeper transit depth over the rest of the region.   This may reflect the difference in approach taken  in processing the saturated region between JexoPipe and the four other pipelines.
Another consideration for the difference in the saturated region could be possible correlations between our empirically-derived LDCs and transit depth which could have resulted in a transit depth bias in the saturated region.  Further investigation is needed to fully understand the differences.

Excluding the saturated region by comparing wavelengths above 2 \textmu m only, if we consider the residuals (i.e. JexoPipe minus the alternate pipeline), the closest match is to Eureka where the average residual above 2 \textmu m is 28$\pm$20 ppm.  When comparing the average errors on the transit depths (at the binned resolution of the comparison pipeline spectrum), the smallest difference is with Eureka (JexoPipe average error above 2 \textmu m is 5 ppm higher), and the largest difference is with Tiberius (JexoPipe average error is 24 ppm lower).   

Limb Darkening: Given the structure of the residuals presented in Figures 8 and 11 (no evident symmetry around the mid-time), I believe that the differences between the baseline analysis and the quadratic or 4-factor model LDC cases are due to the correlated noise. In practice, by fitting for empiric LDCs, the model is probably absorbing part of the correlated noise caused by the saturated pixels, resulting in better statistics. At a first glance this can be overlooked but I am concerned that it can cause biases in the final spectrum. Such a bias can be seen in the saturated region of the Prism data. In Figure A1, we can see that c1 is correlated with depth, while c2 and depth are anti-correlated. Around 1.4um the empiric c1 is significantly lower than the theoretical c1 and the empiric c2 is significantly higher than the theoretical c2. By combining the two observations we can expect that if the LDC calculations are biased, then the spectrum will be negatively biased. There are two data points in the current analysis that are consistently lower compared to any other result close to 1.4 um (Figure 16). Therefore, I believe that at least these two points are biased due to the choice of empirical LDCs instead of model LDCs. I do not expect the authors to change their analyses based on  this observation, but they could note it in the text as something to further investigate in the future.

\subsection{G395H}

The much higher native resolution of G395H compared to Prism is evident in Figure \ref{fig:final spectrum} (upper plot).  To make it easier to visualise the spectral features and to allow comparison to the Prism, the spectral light curves were rebinned to the Prism native resolution (Figure \ref{fig:final spectrum}, second plot).  The SO$_2$ and CO$_2$ features are evident. The gap in the middle of the spectrum is due to the physical division between the NRS1 and NRS2 detectors.  

\begin{figure*}
	\includegraphics[trim={0cm 0cm 0cm 0cm}, clip,width=\columnwidth]{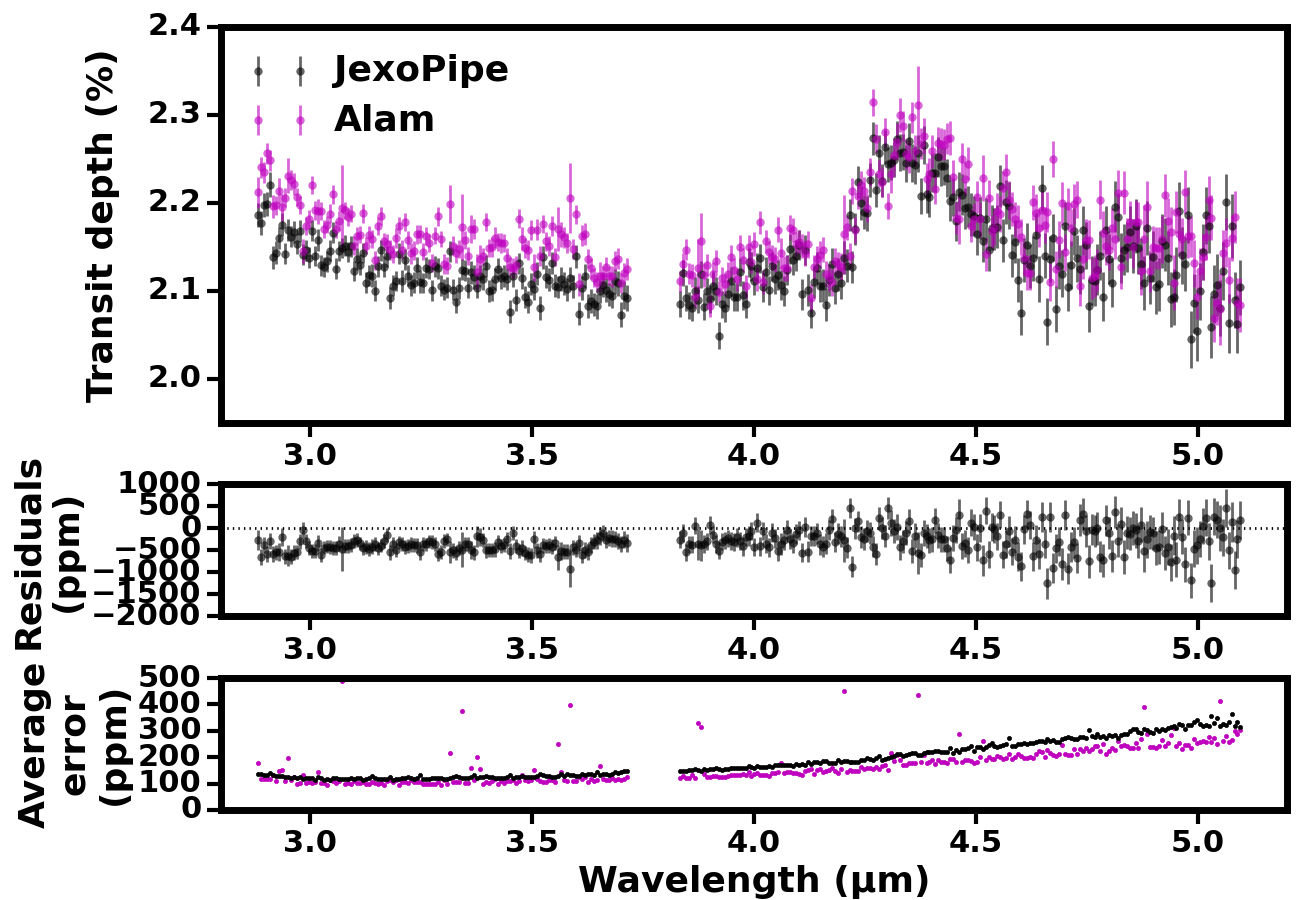}
 	\includegraphics[trim={0cm 0cm 0cm 0cm}, clip,width=\columnwidth]{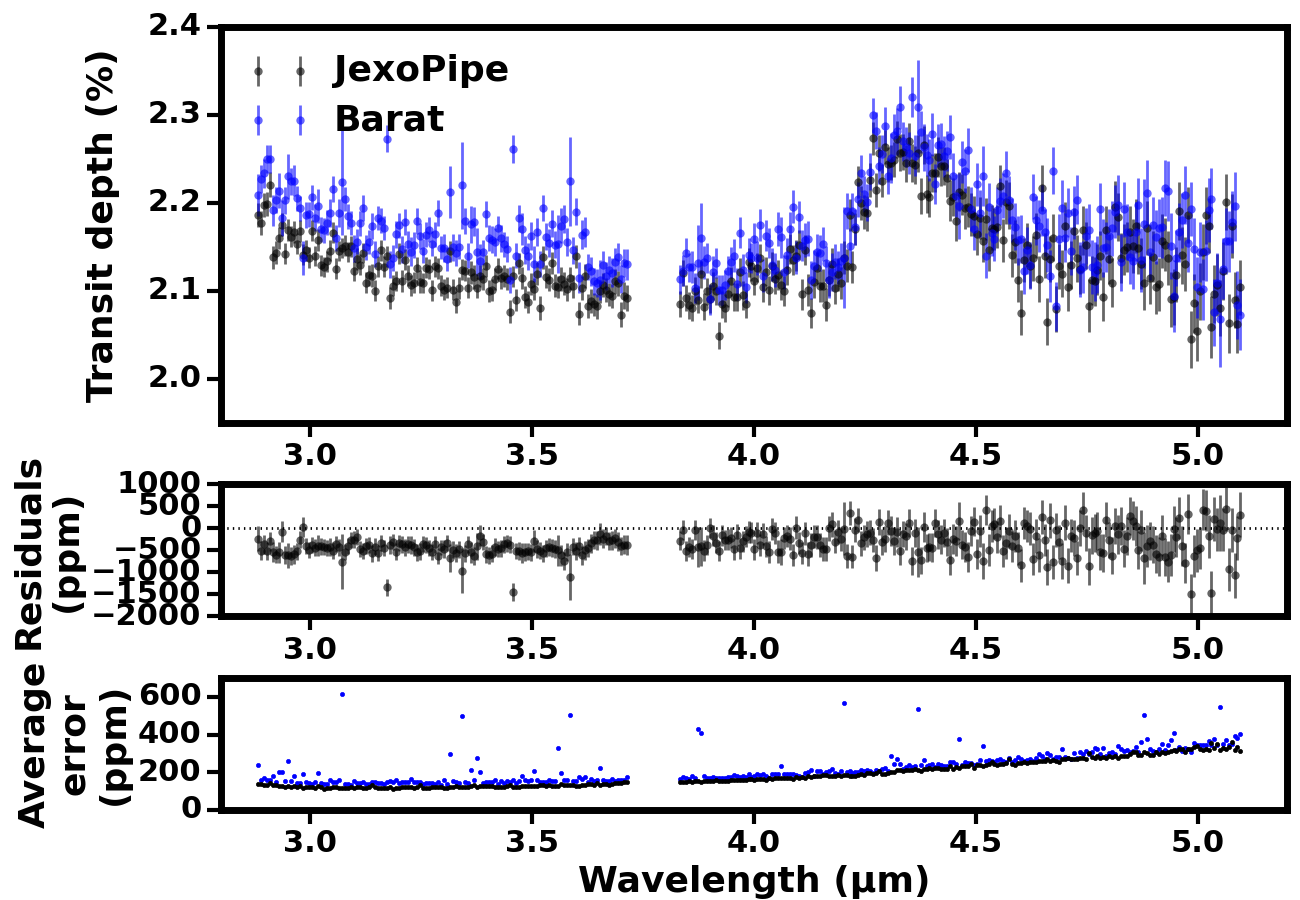}
 	\includegraphics[trim={0cm 0cm 0cm 0cm}, clip,width=\columnwidth]{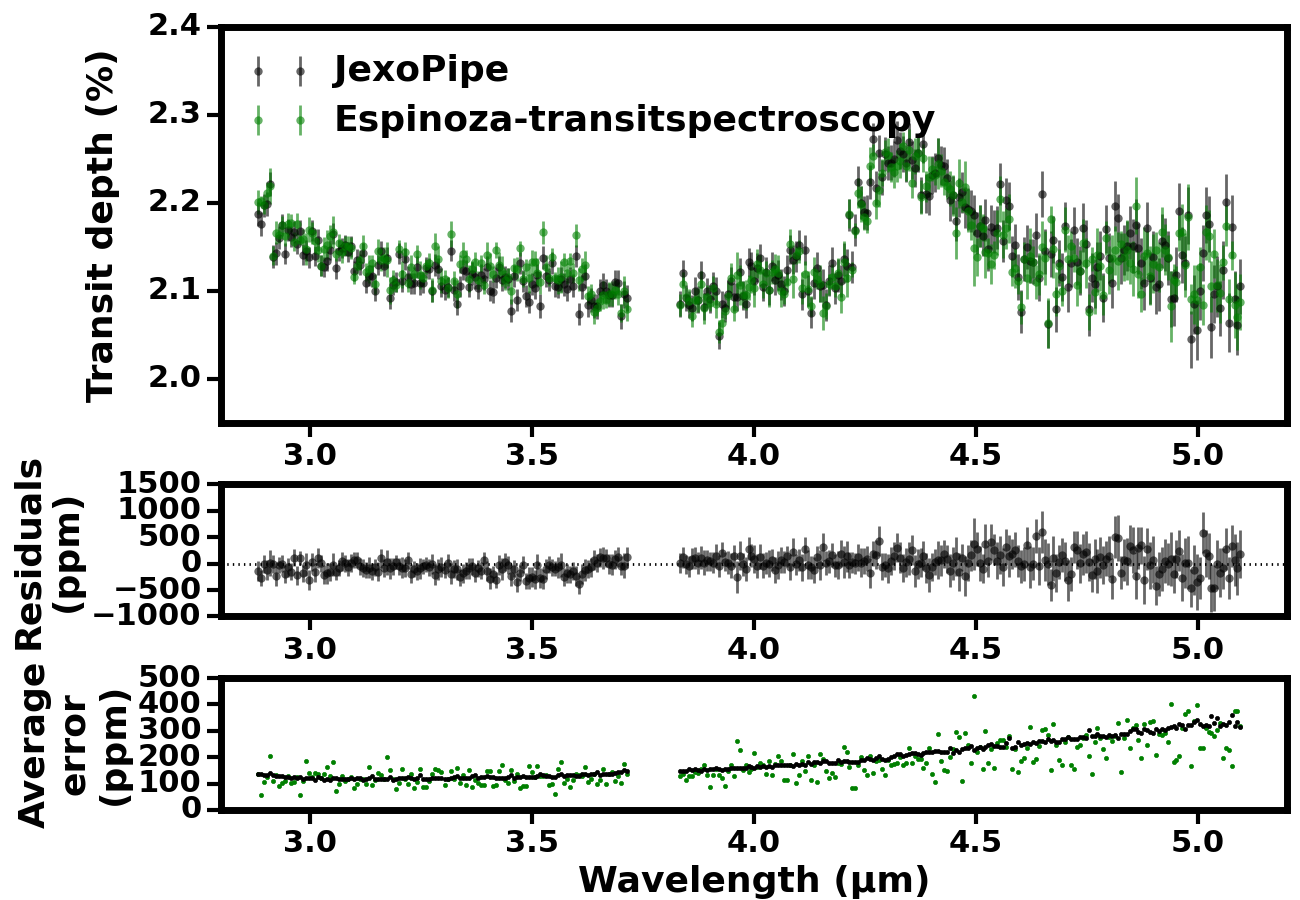}
   	\includegraphics[trim={0cm 0cm 0cm 0cm}, clip,width=\columnwidth]{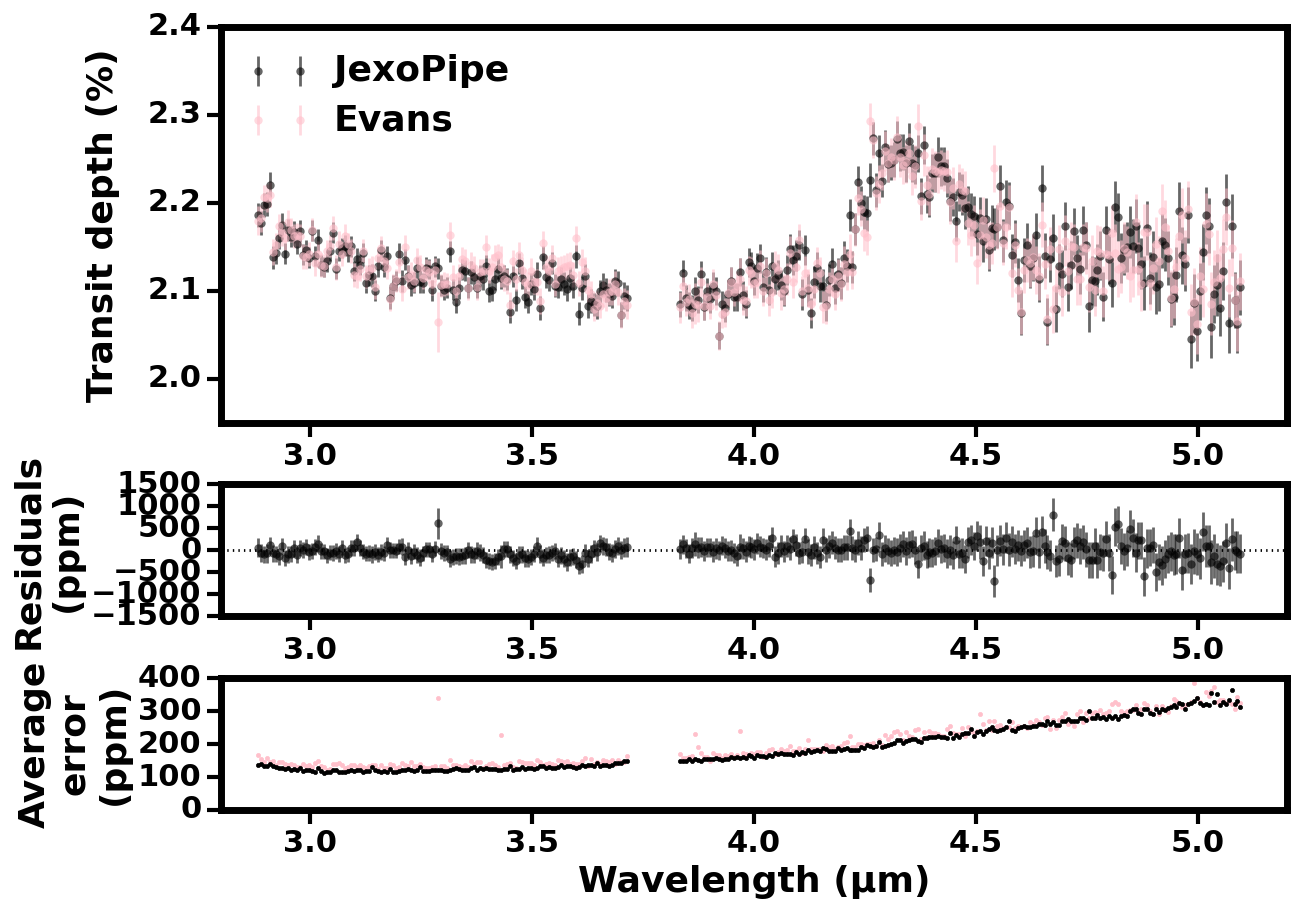}
        \includegraphics[trim={0cm 0cm 0cm 0cm}, clip,width=\columnwidth]{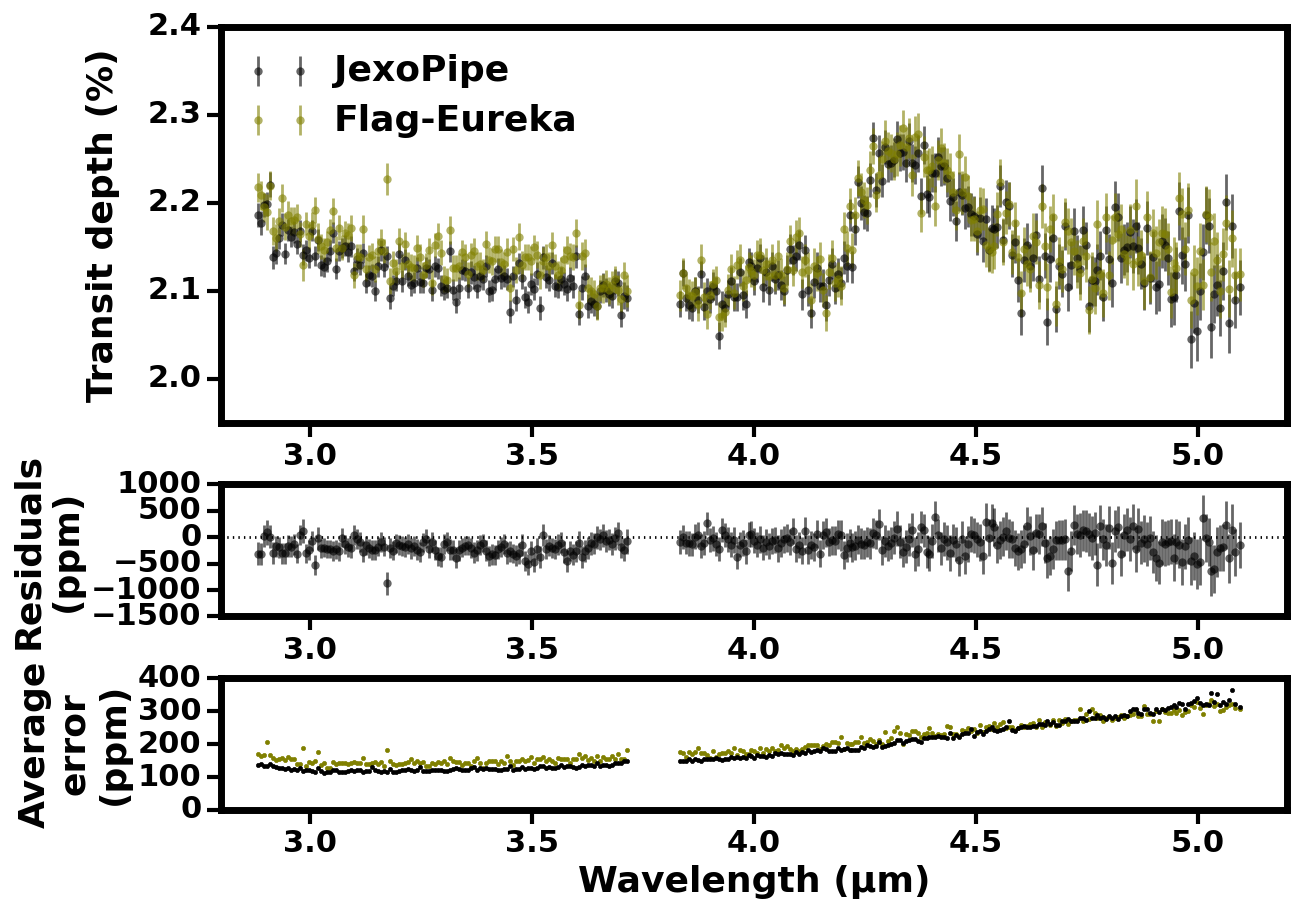}
       \includegraphics[trim={0cm 0cm 0cm 0cm}, clip,width=\columnwidth]{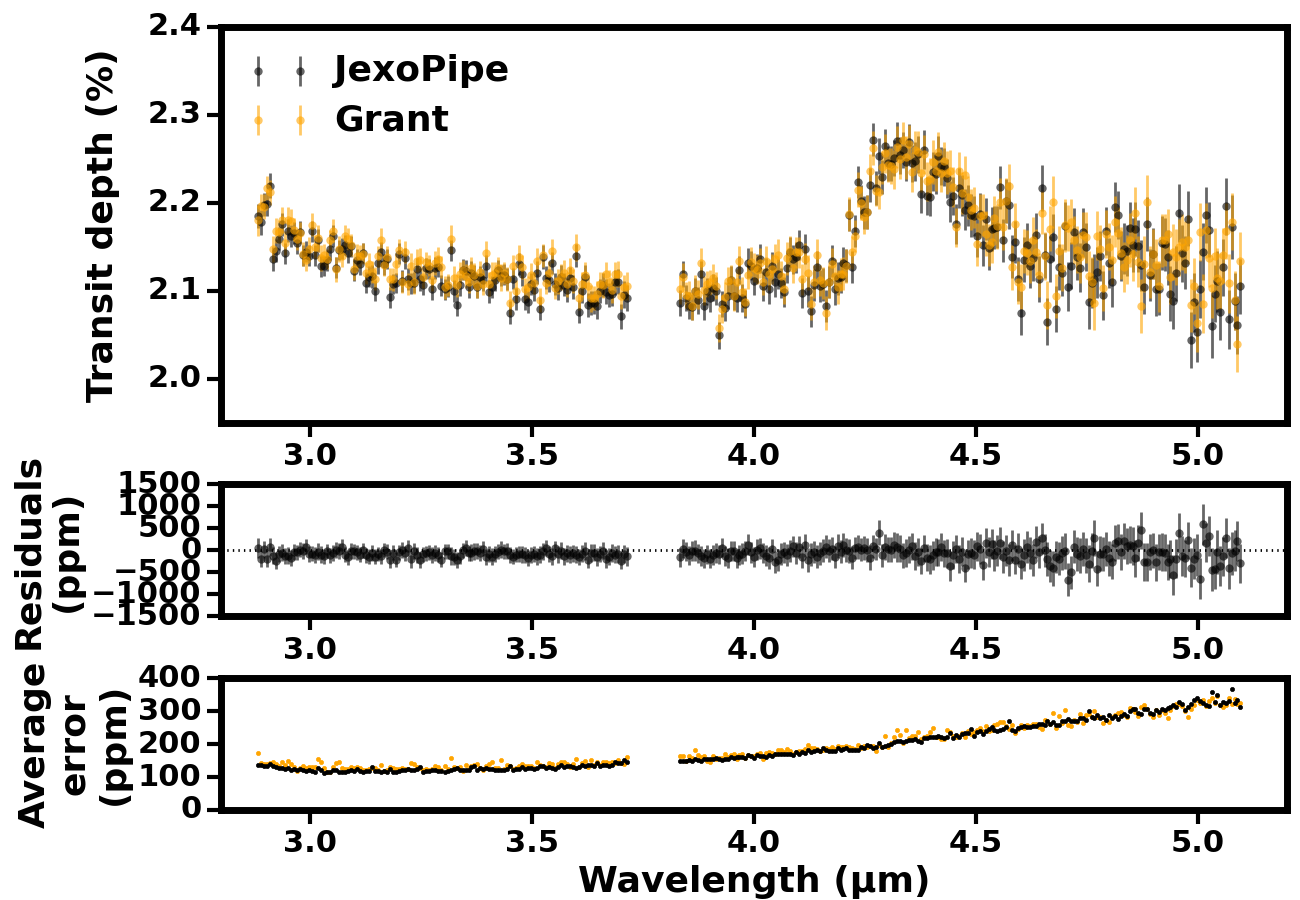}
    \cprotect\caption{{Comparison of our G395H result with six other G395H pipelines (listed by author and/or pipeline name) included in the study by \cite{Alderson2023}.  The JexoPipe result is rebinned to the wavelength grid of the comparison pipeline's publicly available spectrum.  Residuals are JexoPipe minus the comparison pipeline spectrum.  The average errors on the transit depths for the comparison spectrum and the JexoPipe spectrum when binned to the resolution of the comparison spectrum are shown in the lowest plots.
    }
}
    \label{fig:compare grating 1}
\end{figure*}

\begin{figure*}
 \includegraphics[trim={0cm 0cm 0cm 0cm}, clip,width=\columnwidth]{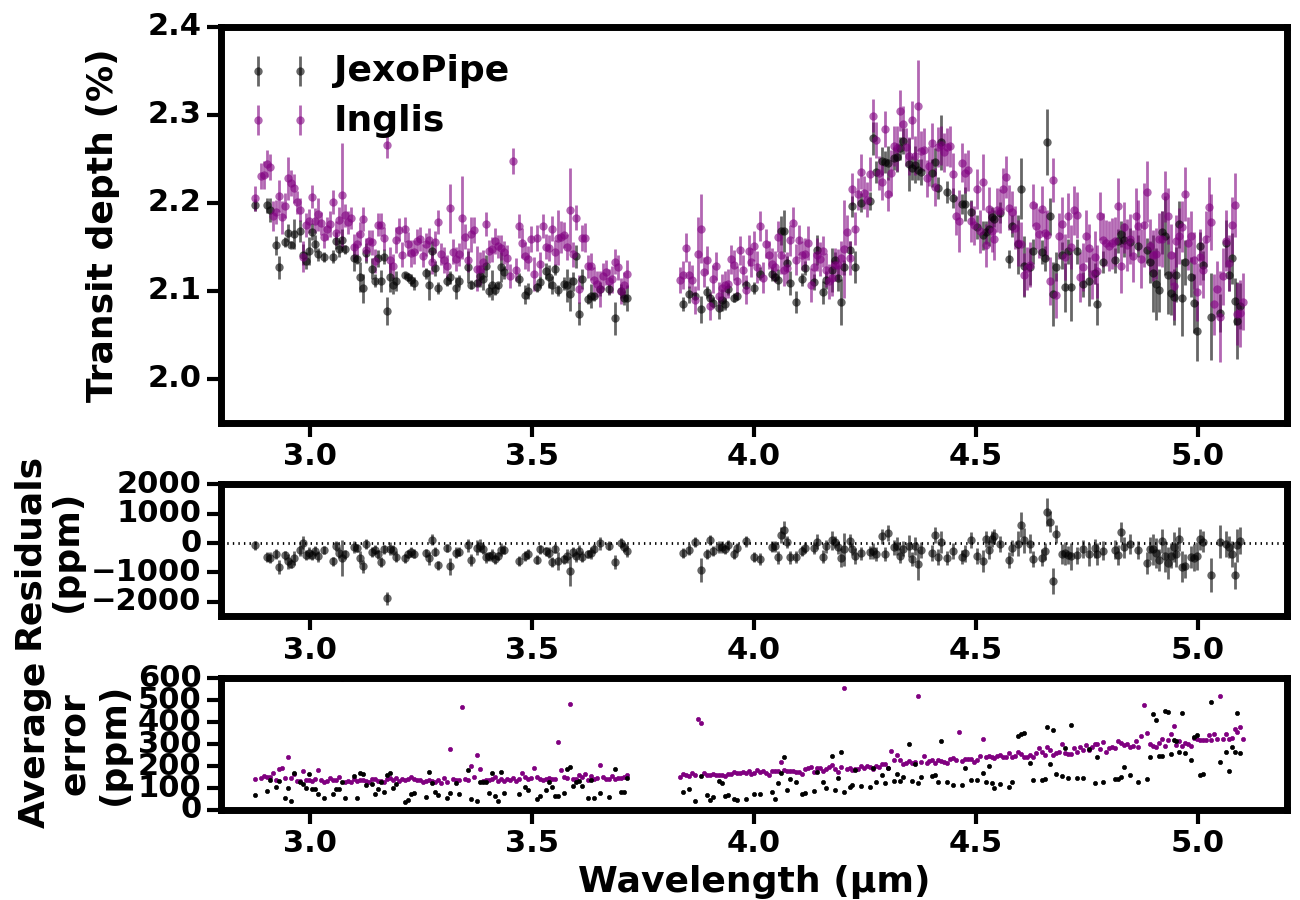}
        \includegraphics[trim={0cm 0cm 0cm 0cm}, clip,width=\columnwidth]{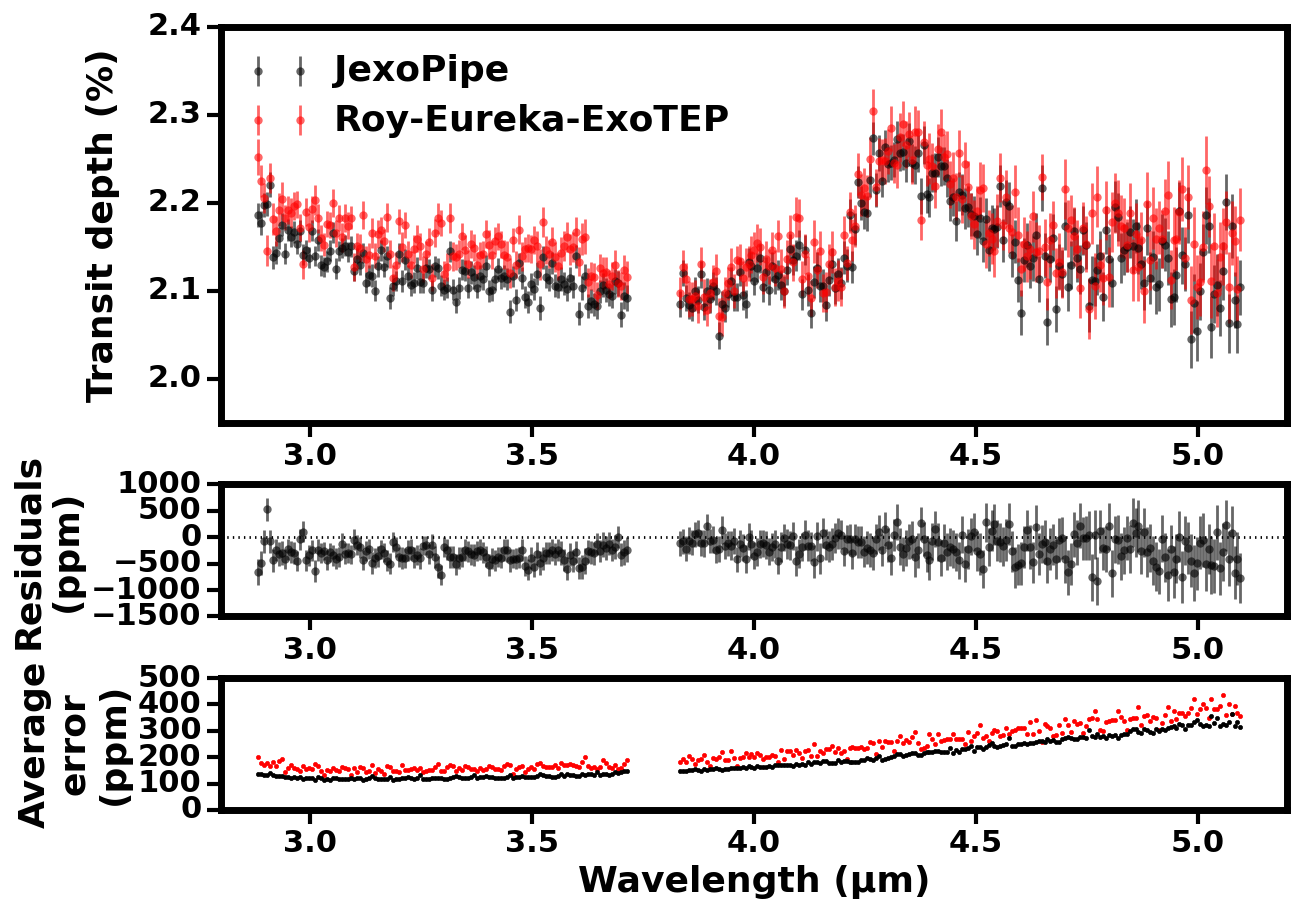}
        \includegraphics[trim={0cm 0cm 0cm 0cm}, clip,width=\columnwidth]{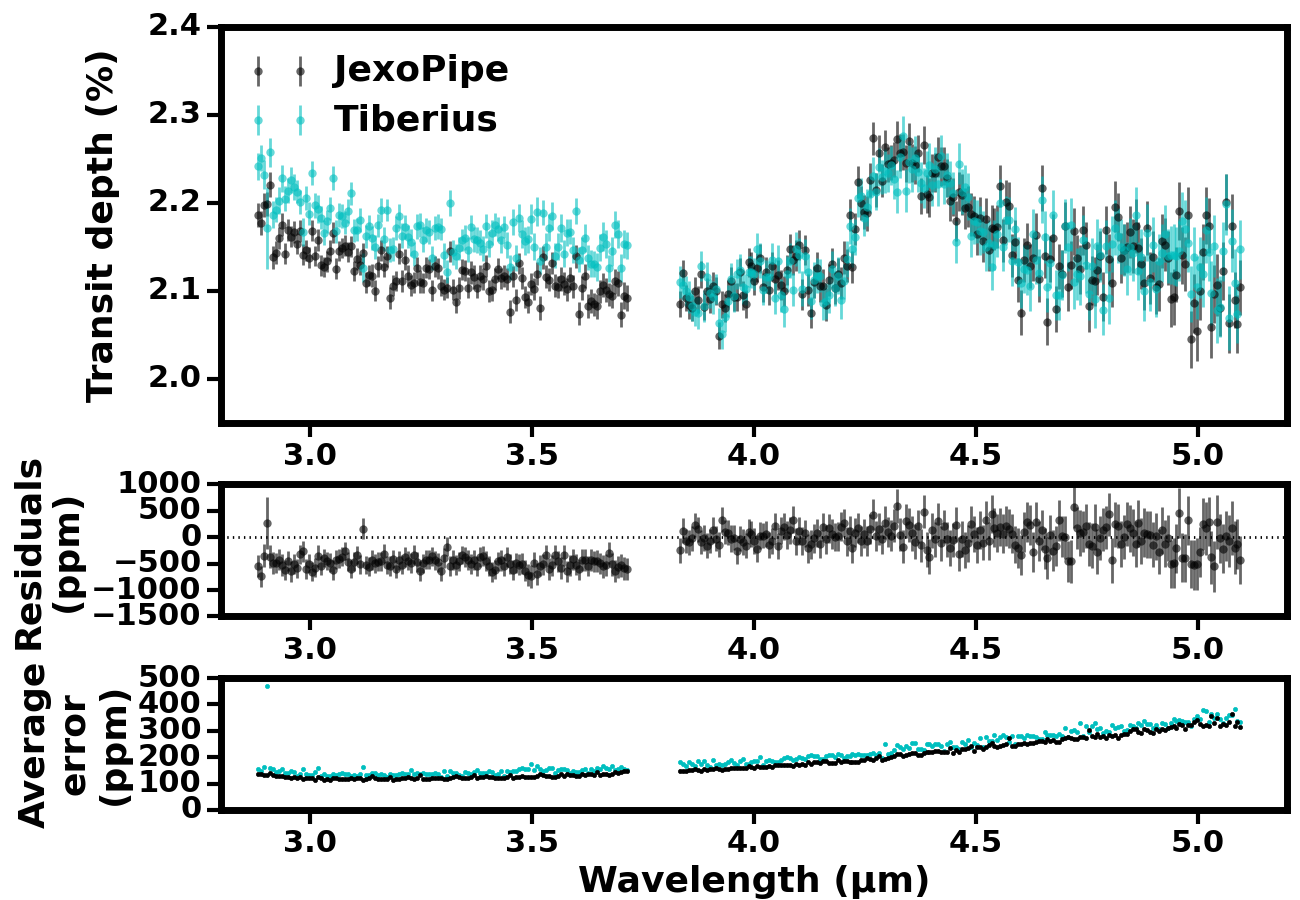}
       \includegraphics[trim={0cm 0cm 0cm 0cm}, clip,width=\columnwidth]{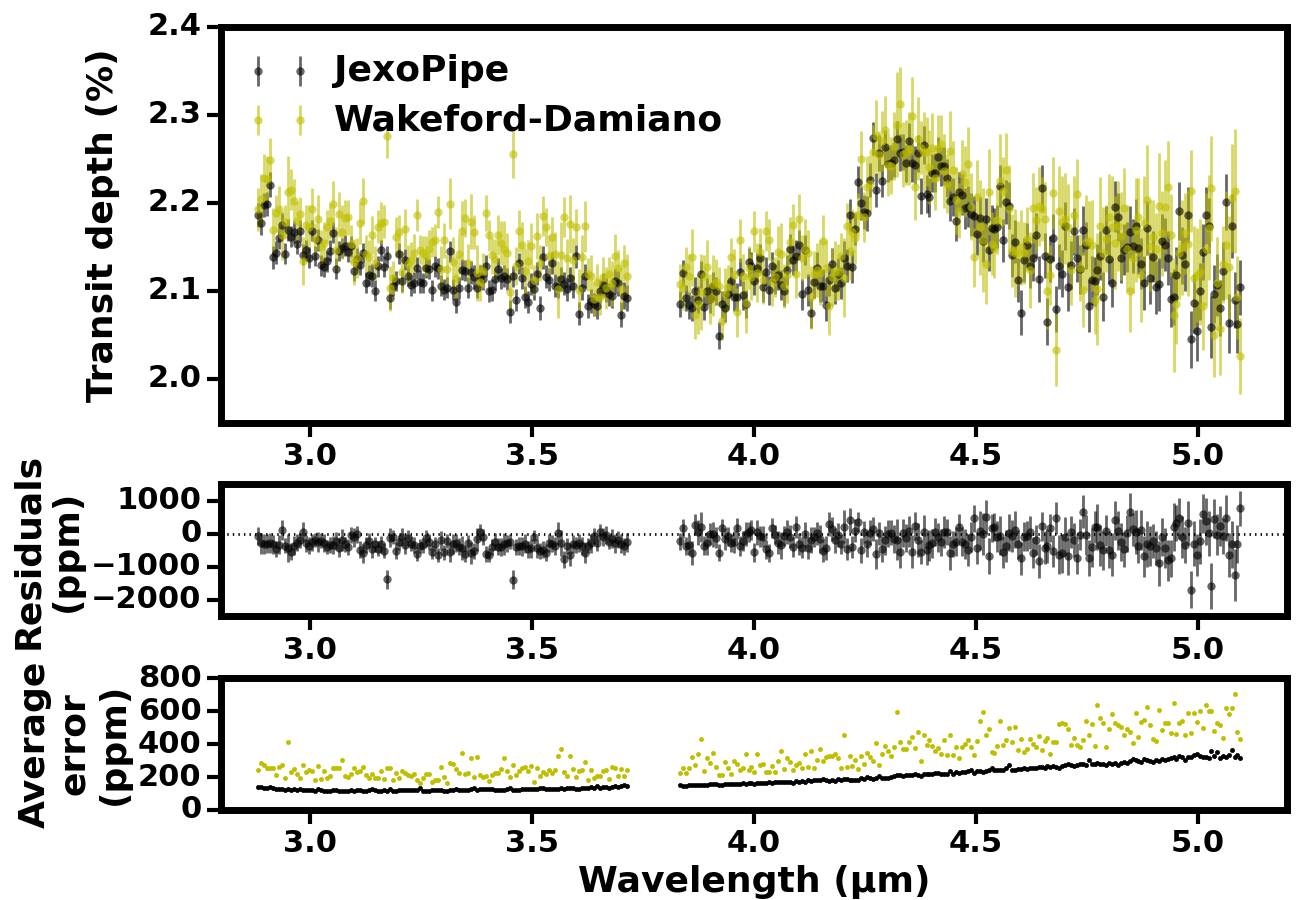}
       \includegraphics[trim={0cm 0cm 0cm 0cm}, clip,width=\columnwidth]{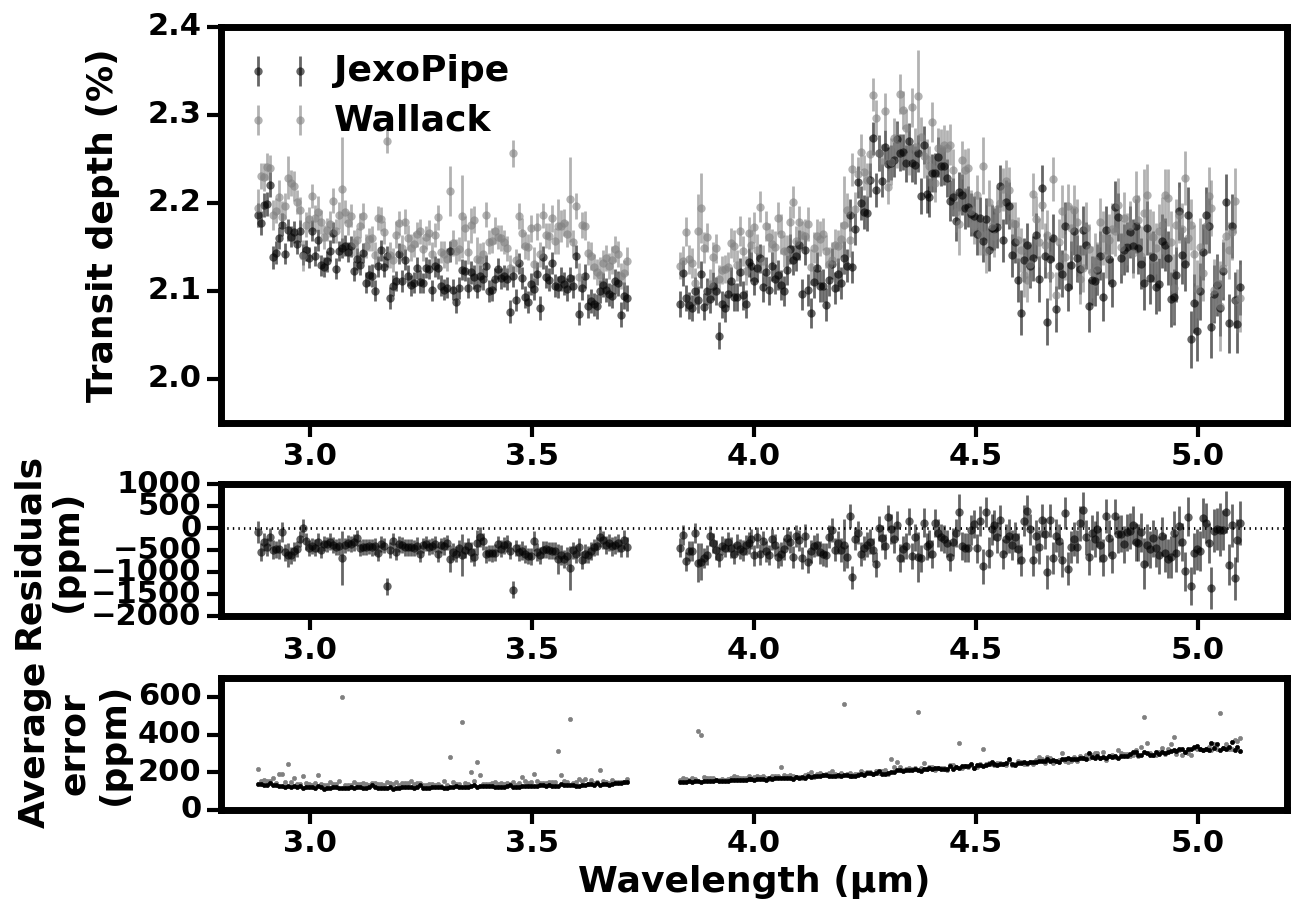}
              \includegraphics[trim={0cm 0cm 0cm 0cm}, clip,width=\columnwidth]{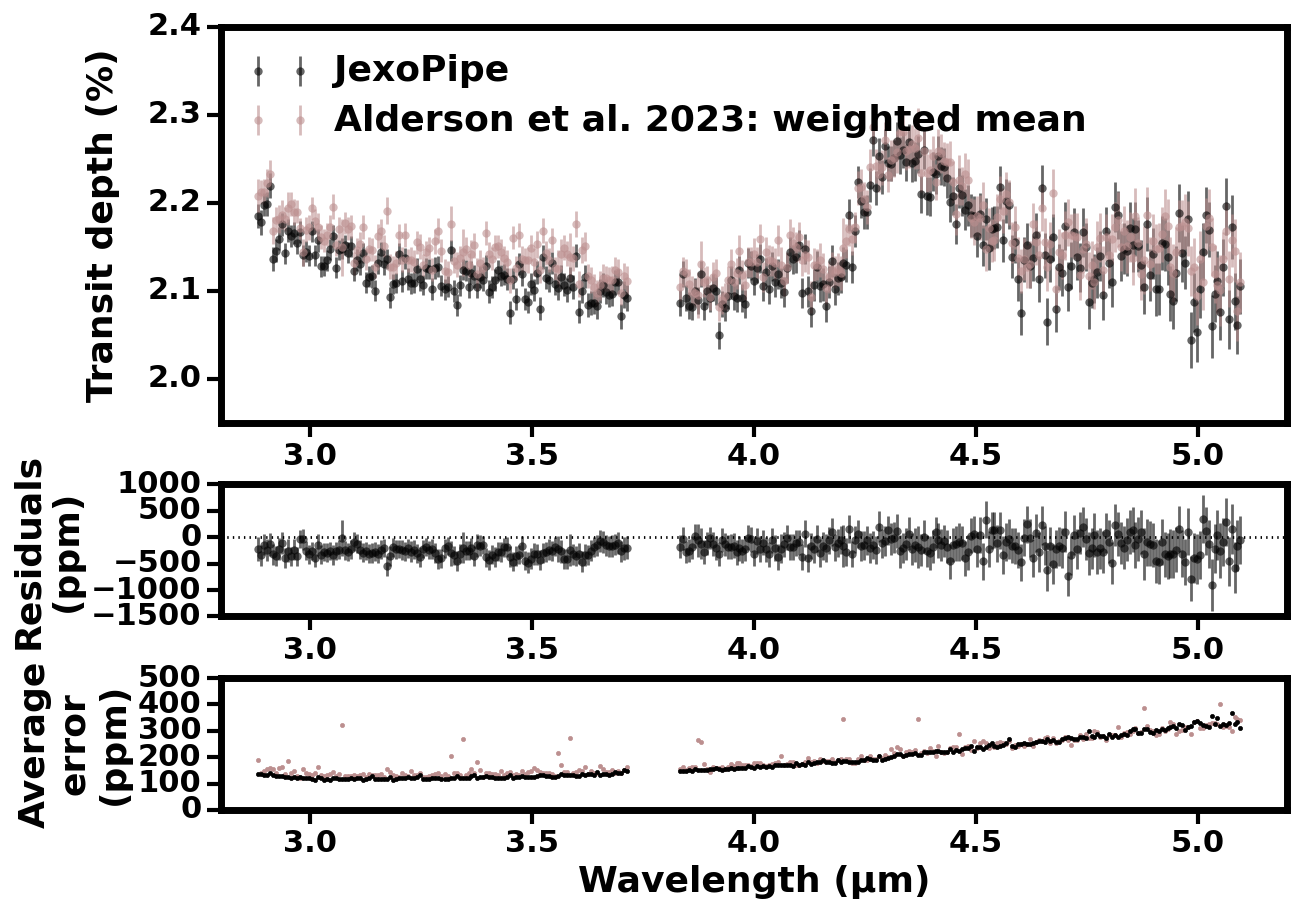}
    \cprotect\caption{Comparison of our G395H result with five further G395H pipelines (listed by author and/or pipeline name) included in the study by \cite{Alderson2023} and the weighted mean result in that study.  The JexoPipe result is rebinned to the wavelength grid of the comparison pipeline's publicly available spectrum.  Residuals are JexoPipe minus the comparison pipeline spectrum. 
     The average errors on the transit depths for the comparison spectrum and the JexoPipe spectrum when binned to the resolution of the comparison spectrum are shown in the lowest plots.}

    \label{fig:compare grating 2}
\end{figure*}

In Figures \ref{fig:compare grating 1} and \ref{fig:compare grating 2} we compare the JexoPipe spectrum to 12 publicly-available spectra from the 11 pipelines and weighted mean used in the study by \cite{Alderson2023}. The errors on the residuals are the quadrature sum of the 1$\sigma$ error bars from the comparison pipeline spectrum and those from the re-binned JexoPipe spectrum. We find good agreement with the amplitudes of spectral features obtained, however there are differences in the spectrum baseline that vary depending on the pipeline being compared. There is a broad range of spectrum baselines across the 11 comparison pipelines.  The closest matches in terms of average offset are with Espinoza-transitspectroscopy (-8$\pm$16 ppm) and Evans (-4$\pm$17 ppm).
The JexoPipe result has a lower spectrum baseline than the weighted mean result with an average offset of -194$\pm$17 ppm. Comparing average errors on the transit depths, JexoPipe has  similar errors to the weighted mean result, the average error being 12 ppm lower in JexoPipe.
 
\section{Comparison of Prism and G395H spectra}

We now examine the differences in the baseline Prism and G395H spectra (Figure \ref{fig:final spectrum}).  An offset is visible between the Prism and G395H result. The average offset (G395H-Prism) is -188$\pm$20 ppm for NRS1 and -139$\pm$31 for NRS2.  We can see that when binned to the same resolution, the transit depth precision is higher for G395H than for Prism (Figure \ref{fig:final spectrum}, lowest plot).  The average difference in noise compared to Prism is -29 ppm in G395H NRS1 and -50 ppm in G395H NRS2. The noise in Prism is on average $\sim$ 1.3 $\times$ that in G395H at the Prism resolution.
 
%There was overall a good agreement between the Prism spectrum and the G395H spectrum, with 65\% of the residuals points being consistent with zero within the error bars. If we were to draw two randomised samples from two Gaussian distributions with the same mean and examine their residuals, we would expect 68\% of the residuals to have zero within their error bars, so the difference between the Prism and G395H spectra is close to that expected statistically. The agreement is about the same over the two NRS detectors: 66\% of residuals comparing G395H NRS1 to Prism are consistent with zero within the error bars, and 65\% for NRS2.  Of the points that do not have zero within the error bars, 83\% and 65\% have a negative residual value (i.e. G395H transit depth is lower than Prism) in NRS1 and NRS2 respectively.  

The cause of the offset between the two modes is not immediately apparent.  Since we homogenised this study by using the same LDCs and system parameters, $a'/R_s$ and $i$, for both datasets, we can attribute any differences to instrumental or astrophysical variations between the two observations or differences in data processing.

We kept the data processing differences to a minimum, however there were differences in Stage 1, where the \textit{Reference Pixel Correction} stage was applied in the G395H pathway but not in the Prism pathway, and where the Prism pathway applied the \textit{Group Control} step and a different value of \verb'n_pix_grow_sat'.  Group control however is only applied to the persistently saturated region of Prism, and so would not be a cause of the offset seen with G395H which are at wavelengths beyond this region.

We investigated the effects of various changes to pipeline processing and light curve fitting from the baseline case.  The results are summarised in Table \ref{table: spectrum comparisons}.  The light curve fits for these comparisons were performed as previously described but with modified MCMC parameters (400 burn-in/ 2000 production steps, 32 walkers), and Prism spectra were obtained only at $>$ 2\textmu m, with G395H binned to Prism resolution to allow comparison. 
We re-ran the baseline case with the changed MCMC run parameters and obtained an offset of -185$\pm$ 20 ppm between G395H NRS1 and Prism as compared to -188$\pm$ 20 ppm mentioned above, while the offset between G395H NRS2 and Prism is unchanged.

The noise in all the cases is comparable to the baseline cases.  In all cases the G395H NRS2-Prism offset is greater than the G395H NRS1-Prism offset, indicating an offset between the two G395H NRS detectors which is $\sim$ 40-50 ppm in most cases, including the baseline case.  We note this is consistent with inter-detector offsets reported in previous studies \citep{Moran2023, Madhusudhan2023}.

An interesting finding is that if background subtraction is performed just before the linearity correction step\footnote{with dark current not subtraction to prevent the dark current being subtracted twice}, then the offset between Prism and G395H is greatly reduced or eliminated. Compared to the baseline, the G395H NRS1-Prism offset falls from   -185$\pm$20 to -96$\pm$20 ppm and the G395H NRS2-Prism offset falls from -139$\pm$31 to -1$\pm$31 ppm (Table \ref{table: spectrum comparisons}).  This reduction in offset is due to a change in the Prism spectrum from the baseline case (changing on average by -102$\pm$21 ppm).  There is no significant change in the G395H spectra from baseline. The reason for this change in Prism (and not G395H) is not immediately clear.

Omitting the reference pixel stage impacts G395H NRS1 by causing a significant negative offset compared to the baseline case of -74$\pm$18 ppm, and increasing the G395H NRS1-Prism offset to -259$\pm$21 ppm. 
However omitting the reference pixel stage does not seem to impact NRS2, where the change from the baseline case is not significant (-2$\pm$29 ppm).  The reason for this discrepant response between the two NRS detectors is not clear.  Applying \verb'n_pix_grow_sat' =3 to G395H also does not significantly change the offset (see Table \ref{table: spectrum comparisons}).
 The other variations in Table \ref{table: spectrum comparisons} do not show significant changes from the baseline case in terms of average offsets and noise.  These include using a `null' systematic for G395H where no time-dependent trends are included (i.e., we fit only for systematic coefficient $a$ and not $b$ or $c$) and the use of model LDCs.

We thus find offsets between Prism and G395H and also between the two G395H detectors.
The causes of the inter-modal and inter-detector offsets are not immediately clear.   The G395H-Prism offsets could be due to astrophysical or instrumental variations between the two observations.  However they may also be due to differences in data processing, e.g. use of different reference files.  
Alternate systematic models to the second-order polynomial used here may be investigated to see if these might affect the offsets seen, though as noted above use of a null systematic model (with no time-dependent trend) in G395H gives similar results to the baseline case.  While the offset between Prism and G395H has several possible causes, we can rule out astrophysical variation for the G395H inter-detector offset between NRS1 and NRS2, which we detect here through the comparison with the Prism spectrum.  While the cause of offsets like these need continued investigation, they represent a systematic uncertainty that needs to be accounted for when analysing such spectra, and when comparing spectra taken at different times and/or with different instrument modes.
 
%While the baseline offset is reduced using the linear systematic fit in G395H, a small difference in baseline remains.  The same system parameters: $a/R_s$, $i$, $\omega$ and $e$ were used for the Prism and G395H spectral light curve fits, and the the same set of LDCs were adopted.  Thus we can rule out these factors as the cause of the difference in baseline between the Prism and G395H results.  We used a quadratic polynomial fit for the Prism systematic trend, but when we use a quadratic fit for the G395H data, the difference in baseline is greater than if we used a linear fit.  One possible reason for the baseline offset could be that the out-of-transit trend was not of the same form during the two observations, resulting in differences manifesting in the spectrum baseline.
%Other possible reasons are the differences in pipeline processing prior to light curve fitting (Stages 1 and/or 2), and/or changes in astrophysical conditions (e.g. changes in star spot coverage or planet cloud coverage) between the two observations as the causes.  While astrophysical variation possible (the observations were separated by 20 days), it is probably less likely to be the cause of the baseline offset than differences in systematic trend or processing pathways, especially given the short period of time elapsed between the two observations. The high sensitivity to processing appears to be a by-product of the exceptional precision and stability of JWST raw data.

\begin{table*}
\caption{Effects of changing data reduction and light curve fitting steps on relative spectral transit depths.  Baseline case is described in the text and corresponds to spectra shown in Figure \ref{fig:final spectrum}*.  First three columns give the mean difference in ppm between spectral data points (binned to the Prism wavelength grid) between that case and the baseline case.  The next two columns give the average offset between Prism and G395H spectra binned to the Prism resolution, for the given case, to enable a statistical comparison between the two spectra (Prism vs G395H).    The average noise (when binned to Prism resolution) on the spectrum transit depth is given in the final three columns.
Changes in stage 1 reduction steps are shown above the line, and changes in stages 3-4 extraction and light curve fitting are shown below the line.  The Prism results are for wavelengths $>$ 2 \textmu m only.  `Background pre-linearity' applies the background subtraction stage just prior to the linearity correction and omits dark current subtraction. }
\label{table: spectrum comparisons}
\setlength\tabcolsep{0pt} % make LaTeX figure out width of inter-column spaces
`\begin{tabular*}{\textwidth}{l @{\extracolsep{\fill}}
                            *{8}{S[table-format=1.4]}} 
   % \begin{tabular}{|l|c|c|c|c|c|c|c|c|}
\toprule
& \multicolumn{3}{c}{Change from baseline (ppm)} & \multicolumn{2}{c}{Comparison with Prism (ppm)} & \multicolumn{3}{c}{Average noise (ppm)} \\
\cmidrule{2-4} \cmidrule{5-6} \cmidrule{7-9}
& {Prism} & {G395H NRS1} & {G395H NRS2} & {G395H NRS1} & {G395H NRS2} & {Prism} & {G395H NRS1} & {G395H NRS2} \\ 
\midrule
Baseline case & N/A   & N/A  & N/A   & {-185$\pm$20} & {-139$\pm$31} & {234} & {104}   & {232}   \\  
Background pre-linearity & {-102 $\pm$ 21} & {1 $\pm$  17} & {1 $\pm$  28} & {-96 $\pm$  20} & {-1 $\pm$  31} & {237}  & {100}  & {217}   \\ 
No dark subtraction & {-7 $\pm$  22} & {-4 $\pm$  18} & {2 $\pm$  29} & {-191 $\pm$  20} & {-103 $\pm$  32} & {239}   & {102}   & {232}  \\  
No reference pixel stage & N/A  & {-74 $\pm$ 18} & {-2 $\pm$  29} & {-259 $\pm$  21} & {-142 $\pm$  32} & {234}  & {107}   & {233}  \\
n\_pix\_grow\_sat=3 (G395H) &N/A  & {-1 $\pm$  18} & {-11 $\pm$  29} & {-186 $\pm$  20} & {-151 $\pm$  32} & {234}   & {104}   & {233}   \\ 
CR rej. threshold=4$\sigma$ (G395H) & N/A & {-1 $\pm$  18} & {1 $\pm$  29} & {-186 $\pm$  20} & {-138 $\pm$  31} & {234}   & {104}   & {230}  \\ 
Custom superbias &  {28$\pm$ 	21} &	{6	$\pm$ 18} & {2$\pm$ 	29}	& {-223	$\pm$ 20} &	{-138$\pm$ 	32}	& {236}	  &	{105}	 	& {231} \\
\hline
Optimal extraction & {6$\pm$  21} & {2 $\pm$  18} & {5 $\pm$  28} & {-189 $\pm$ 21} & {-140 $\pm$  30} & {227}   & {104}   & {228}  \\  
Quadratic model LDC & {-12 $\pm$  21} & {-7 $\pm$  18} & {-4 $\pm$ 29} & {-183 $\pm$  20} & {-140 $\pm$  32} & {235} & {104}   & {232}   \\  
4-factor model LDC & {11 $\pm$ 21} & {19 $\pm$  18} & {12 $\pm$  29} & {-183 $\pm$  20} & {-138 $\pm$  32} & {234}   & {105}   & {232}   \\  
Null systematic (G395H) & N/A  & {-6 $\pm$  18} & {-5 $\pm$  28} & {-191 $\pm$  20} & {-144 $\pm$  31} & {234}   & {102}  & {226} \\  
\bottomrule
\end{tabular*} 
*To permit comparison with the other cases the baseline case was repeated here with the shorter MCMC runs used for the comparisons; hence the G395H NRS1 offset is slighly different from that quoted in the text for the longer MCMC run used for Figure \ref{fig:final spectrum}.
\end{table*}

\section{Atmospheric Modelling}
\label{sec:modelling}

\begin{figure*}
 \includegraphics[width=\textwidth]{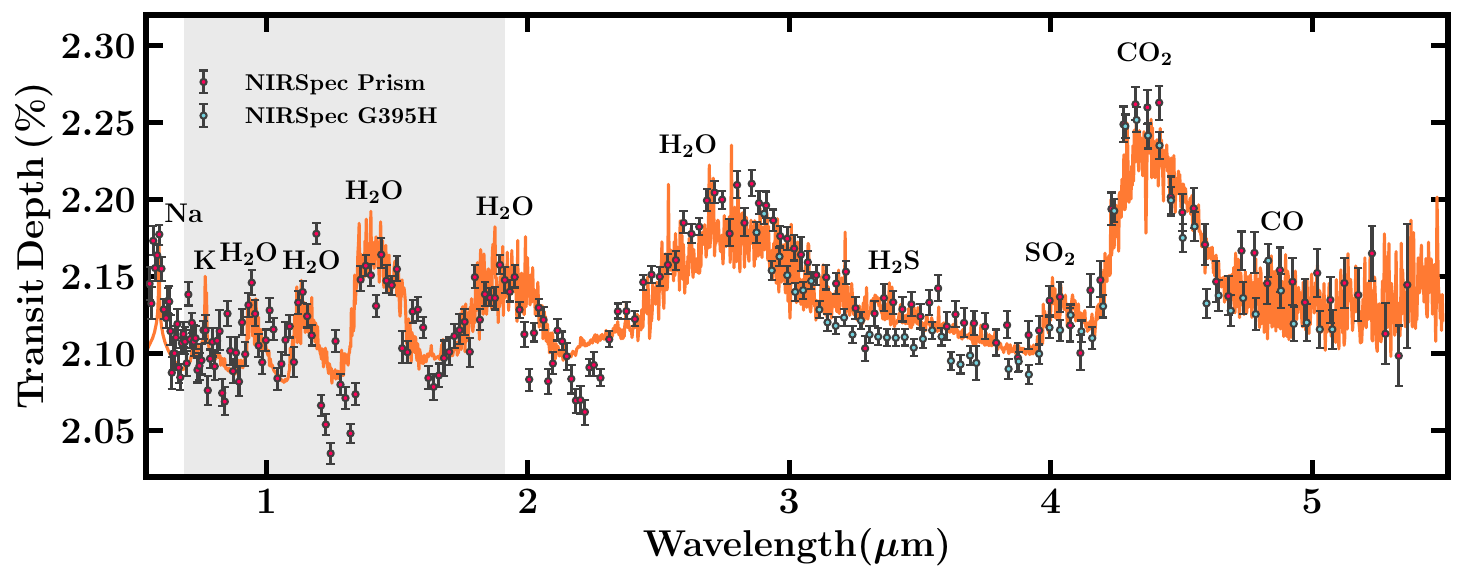}
    \caption{The obtained NIRSpec Prism and G395H transmission spectra of WASP-39 b, shown as errorbars with green and blue centres, respectively. Also shown is a nominal model assuming 10$\times$~solar elemental abundances and Mie scattering aerosols, as discussed in section \ref{sec:modelling}.  The observations are binned to R$\sim$100 for visual clarity, while the grey area denotes the persistent saturation region for NIRSpec Prism as shown in Figure \ref{fig:PSR}}
    \label{fig:fwd_model}
\end{figure*}

We use the AURA framework \citep{Pinhas2018} to generate a number of simulated transmission spectra to compare to the obtained JWST NIRSpec observations. AURA treats the terminator as a 1D plane-parallel 
atmosphere in hydrostatic equilibrium with a uniform composition. For the models considered in this work, we treat the atmospheric regions giving rise to the transmission spectrum as isothermal.

We consider atmospheric opacity contributions from a number of gaseous species previously reported in the atmosphere of WASP-39~b \citep[e.g.][]{Nikolov2016, Fischer2016, Wakeford2018, Kirk2019, ERS2023, Alderson2023, Ahrer2023, Constantinou2023, Feinstein2023, Rustamkulov2023}. We specifically consider H$_2$O \citep{Barber2006, Rothman2010}, CO \citep{Li2015}, CO$_2$ \citep{Tashkun2015}, SO$_2$ \citep{Underwood2016}, H$_2$S \citep{Azzam2016, Chubb2018}, Na \citep{Allard2019} and K \citep{Allard2016}. We also include atmospheric extinction arising from Mie scattering through  ZnS \citep{Querry1987} aerosols, again motivated by prior findings \citep{Constantinou2023, Feinstein2023}. We lastly include opacity contributions from H$_2$-H$_2$ and H$_2$-He collision-induced absorption \citep{Borysow1988, Orton2007, Abel2011, Richard2012}, which set the spectral baseline in the absence of aerosols.
%MgSiO$_3$ \citep{Dorschner1995} and

For the nominal model shown in figure \ref{fig:fwd_model}, the abundances of all gaseous species except SO$_2$ correspond to a 10$\times$ enhancement over solar values under thermochemical equilibrium \citep{Burrows1999, Lodders2002, Madhusudhan2011a, Moses2013}. For SO$_2$, which is the product of disequilibrium processes \citep{Zahnle2009, Wang2017, Polman2022, Tsai2023}, we use a volume mixing ratio of 10$^{-5}$, which is largely consistent with literature findings \citep{Constantinou2023, Rustamkulov2023, Tsai2023}. We set the atmosphere's isothermal temperature to 900~K. The ZnS aerosols have a 40\% coverage fraction of the terminator and full vertical extent, with a mixing ratio of 5$\times10^{-7}$ and a modal particle radius of 0.01 \microns.

The forward model shown in Figure \ref{fig:fwd_model} provides a good fit to the observations and displays prominent spectral features consistent with previous studies \citep{Alderson2023, Rustamkulov2023, Tsai2023, Constantinou2023} . The data and model show absorption features from H$_2$O near 0.9, 1.1, 1.4, 1.9 and 2.9 \microns. The latter feature also contains secondary spectral contributions from CO$_2$ and H$_2$S. Additionally, the blue end of the observations prominently shows a significant absorption feature from Na. Both NIRSpec Prism and G395H datasets show a highly prominent CO$_2$ absorption feature at $\sim$~4.3 \microns, which is reproduced by the forward model. There are also spectral contributions from SO$_2$, CO and H$_2$O  on either side of the large CO$_2$ feature. Specifically, SO$_2$ is responsible for a small peak at 4 \microns, while CO and H$_2$O provide atmospheric opacity towards the red end of the spectrum. Lastly, ZnS aerosols are responsible for significant truncation of spectral features, particularly those of H$_2$O at wavelengths smaller than 2 \microns.

While the above forward modelling does not provide a definitive retrieval of the present dataset, the findings are broadly consistent with prior analyses of JWST observations of WASP-39 b with NIRSpec \citep{ERS2023, Rustamkulov2023, Alderson2023, Constantinou2023, Niraula2023} and other instruments \citep{Ahrer2023, Alderson2023, Feinstein2023}. Specifically, prior works also find highly prominent spectral features from H$_2$O and CO$_2$, with additional contributions from SO$_2$, CO and H$_2$S. Moreover, many of the above works find that the observations are best explained by a super-solar atmospheric metallicity, with additional spectral contributions from non-grey clouds. A more comprehensive retrieval analysis of the present data can confirm this agreement with prior works.

As noted earlier, we find the greatest time-correlated noise in the Prism white light curve composed of wavelengths between 0.65-2 \microns, which would suggest that the transit depth error bars in this range may be underestimated, precisely where we find largest discrepancy between the forward model and the data.

%acknowledge worse fit in saturated range? mention non-isothermal temperature profile effects?

%FROM COBRA0 FOR REFERENCE
%We additionally include opacity contributions arising from H$_2$-H$_2$ and H$_2$-He collision-induced absorption \citep{Richard2012}, as well as ZnS  \citep{Querry1987} and MgSiO$_3$ \citep{Dorschner1995} aerosols. Our choice for these two aerosol species is driven by both thermochemical expectations for the condensates based on the terminator temperature \citep{Morley2013} and indicative constraints obtained with our maximal model retrievals. The absorption cross-sections for the gaseous species are derived following \citet{Gandhi2017}, using line lists of H$_2$O, CO and CO$_2$ from \citet{Rothman2010} and \citet{Li2015}, CH$_4$ from \citet{Yurchenko2014}, NH$_3$ from \citet{Yurchenko2011}, HCN from \citet{Harris2006} and \citet{Barber2014}, C$_2$H$_2$ from \citet{Chubb2020} SO$_2$ from \citet{Underwood2016} and H$_2$S from \citet{Azzam2016} and \citet{Chubb2018}.

\vspace{-0.3cm}
\section{Conclusions}
 In this work, we present JexoPipe, a newly developed JWST pipeline for exoplanet tranist spectroscopy. We applied it to observations of the warm Saturn WASP-39 b obtained in the ERS program 1366 using the NIRSpec instrument in two contrasting configurations, Prism and G395H. We use  JexoPipe to apply consistent pipeline procedures, LDCs and system parameters to both datasets, enabling a comparative analysis of the spectra and instrument configurations.
%allowing us to probe potential differences in the final spectra. 

We find a significant offset between Prism and G395H spectra which is more pronounced for G395H NRS1.  The Prism data baseline also reveals an offset between G395H NRS1 and NRS of the order of 40-50 ppm.  This is consistent with an intra-detector offset reported by \cite{Moran2023} and supports the incorporation of detector offsets when interpreting such spectra \citep{Madhusudhan2023}.  We cannot rule out astrophysical causes for the G395H-Prism offset, however instrumental changes or differences in processing may be potential causes.

We note the effect of omitting the reference pixel stage on the spectral baseline was more pronounced for NRS1 than NRS2.  We also found that performing the group-level background subtraction before the linearity correction step resulted in a significant fall in the Prism spectrum baseline, such that the offsets with G395H were reduced (NRS1) or eliminated (NRS2), but no such change is seen for G395H itself.  Further investigation of these differences is warranted.

We address the Prism saturation through a combination of increasing the \verb|n_pix_grow_sat| argument to counteract detector `blooming' and a custom group control stage.  In the Prism, we find `flipping' noise, which appears to result from the variation in number of groups with integration.  Group control is used to mitigate this.

On choosing between Prism and G395H each offers advantages and disadvantages.  The Prism avoids any inter-detector offsets, but the final spectra have somewhat more noise, and correlated noise was detected in this study.  The average error on the spectrum transit depths are about 1.3 $\times$ higher in Prism compared to those on G395H when binned to the Prism resolution.
It also saturates easily.  G395H is superior in terms of noise and allows for higher resolution spectra, however there is the potential for inter-detector spectrum baseline offsets of the order of 10s of ppm.  

Using a nominal forward atmospheric model with 10$\times$ solar elemental abundances we show that we recover water peaks at 1.1, 1.4 and 1.9 \textmu m within the saturated region, although the scatter and deviation from the model is somewhat higher in this region than outside.  We note this is also the region where time-correlated noise is highest likely leading to an underestimation of the error bars on the spectrum and which might explain at least some of this added deviation.  It remains to be seen if the method used in this paper provides a more or less accurate recovery of the saturated region than previous methods.

The examination of JexoPipe involved a comparison with spectra  obtained from previously developed pipelines. The greatest disagreement with pipelines used in \cite{Rustamkulov2023} occurs in the persistently saturated region. JexoPipe has reasonably good agreement with two pipelines used in \cite{Alderson2023} but has appreciable baseline differences with others.   
   
In this early stage of JWST observations, development of independent pipelines such as JexoPipe allows us to compare results, ultimately leading to more robust scientific conclusions and the elucidation of optimal and best practice approaches for the processing of data from JWST exoplanet transit observations.

\section*{Acknowledgements}
The authors acknowledge
the JWST Transiting Exoplanet Community ERS team (PI: Batalha)
and the ERO team (PI: Pontoppidan) for developing their observing
programs with a zero-exclusive-access period.
We thank NASA, ESA, CSA, STScI and everyone whose efforts
have contributed to the JWST, and the exoplanet science community
for the thriving current state of the field. N.M. and M.H. acknowledge support from the MERAC Foundation, Switzerland, and the UK Science and Technology Facilities Council (STFC)
Center for Doctoral Training (CDT) in Data intensive science at
the University of Cambridge (STFC grant number ST/P006787/1), towards the doctoral studies of M.H.  
We thank the anonymous reviewer for their helpful comments and suggestions.
%%%%%%%%%%%%%%%%%%%%%%%%%%%%%%%%%%%%%%%%%%%%%%%%%%
\section*{Data Availability}
This work is based on observations made with
the NASA/ESA/CSA JWST. The publicly available data were obtained from the Mikulski Archive for Space Telescopes at the Space
Telescope Science Institute, which is operated by the Association of
Universities for Research in Astronomy, Inc., under NASA contract
NAS 5-03127 for JWST. These observations are associated with programs 1091, 1366, 1541, and 2734. 
The data underlying this article will be shared on reasonable request to the corresponding author.

%%%%%%%%%%%%%%%%%%%%%%%%%%%%%%%%%%%%%%%%%%%%%%%%%%
%light curve system, trial ,PRISM : wide wavelength coverage in one go disadvtange: no ref pixels, jump detection problems saturation easy ; dealing with sat region not straght forward.  This highlights the improved duty cycle with the grating compared to the PRISM, due to the large number of groups possible per integration cycle.  The risk of saturation inevitably reduces the total number of groups per integration in the PRISM, so that the single reset group accounts for a larger amount of the total exposure time. GRATING:step effect ? possible mirror segement?- not same in all pixels, seems to occur at same time; corresponds to flux change in guidestar? mirror flopping?  corrected by... - examine guidestar flux to detect possible steps; need to examine engineering data for HGA movements and changes in flux in the guide star


%\input{paper.bbl}
%\bibliographystyle{mnras} 

\bibliographystyle{mnras}
%\nocite{*}
\bibliography{main.bib}

%\bibliographystyle{mnras}
%\nocite{*}

%%%%%%%%%%%%%%%%%%%%%%%%%%%%%%%%%%%%%%%%%%%%%%%%%%

%%%%%%%%%%%%%%%%% APPENDICES %%%%%%%%%%%%%%%%%%%%%
\vspace{-0.5cm} 
\appendix
%\section{Extra figures}
\section{White Light Curve Posteriors}
\begin{figure*}
	\includegraphics[trim={0cm 0cm 0cm 0cm}, clip,width=0.99\columnwidth]{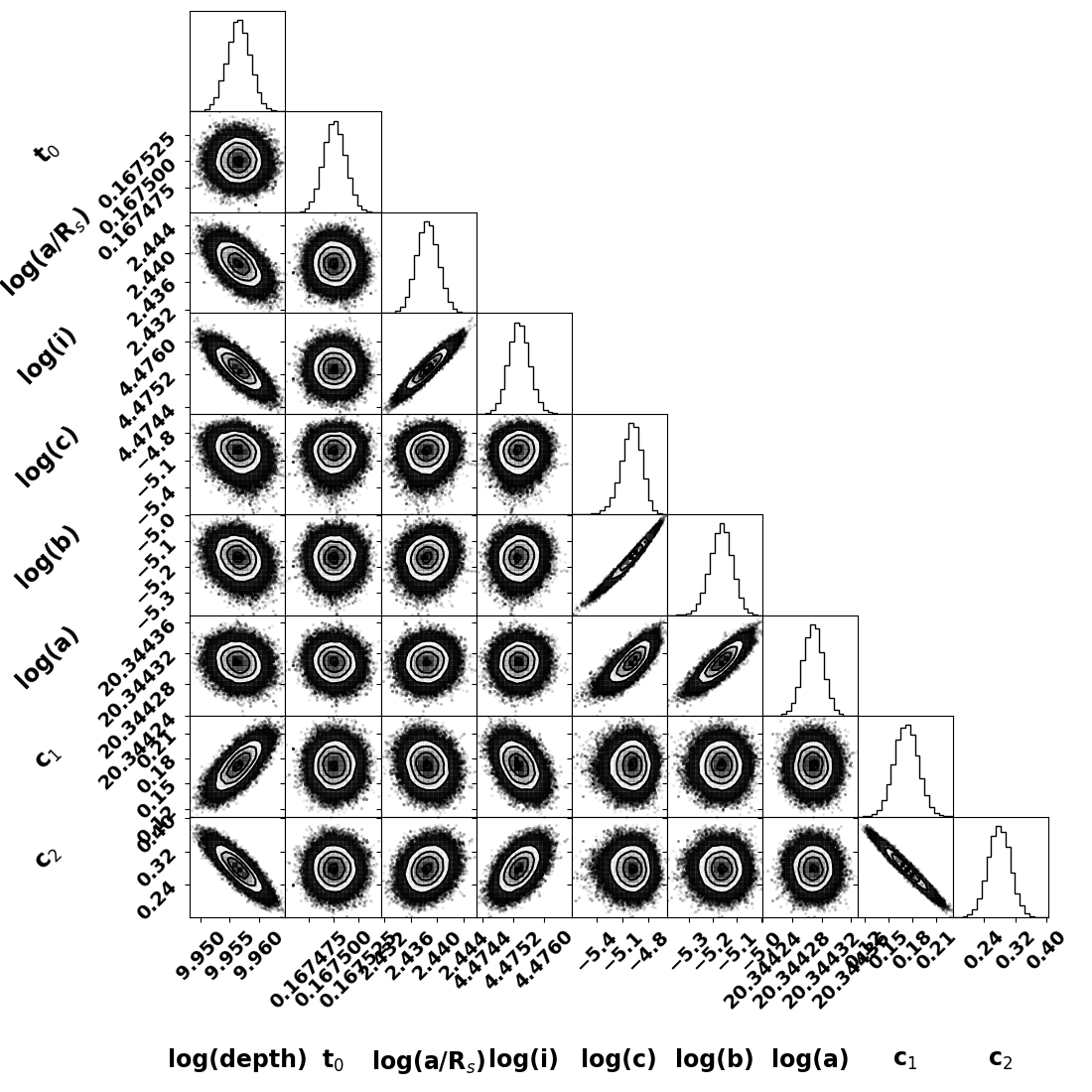}
	\includegraphics[trim={0cm 0cm 0cm 0cm}, clip,width=0.99\columnwidth]{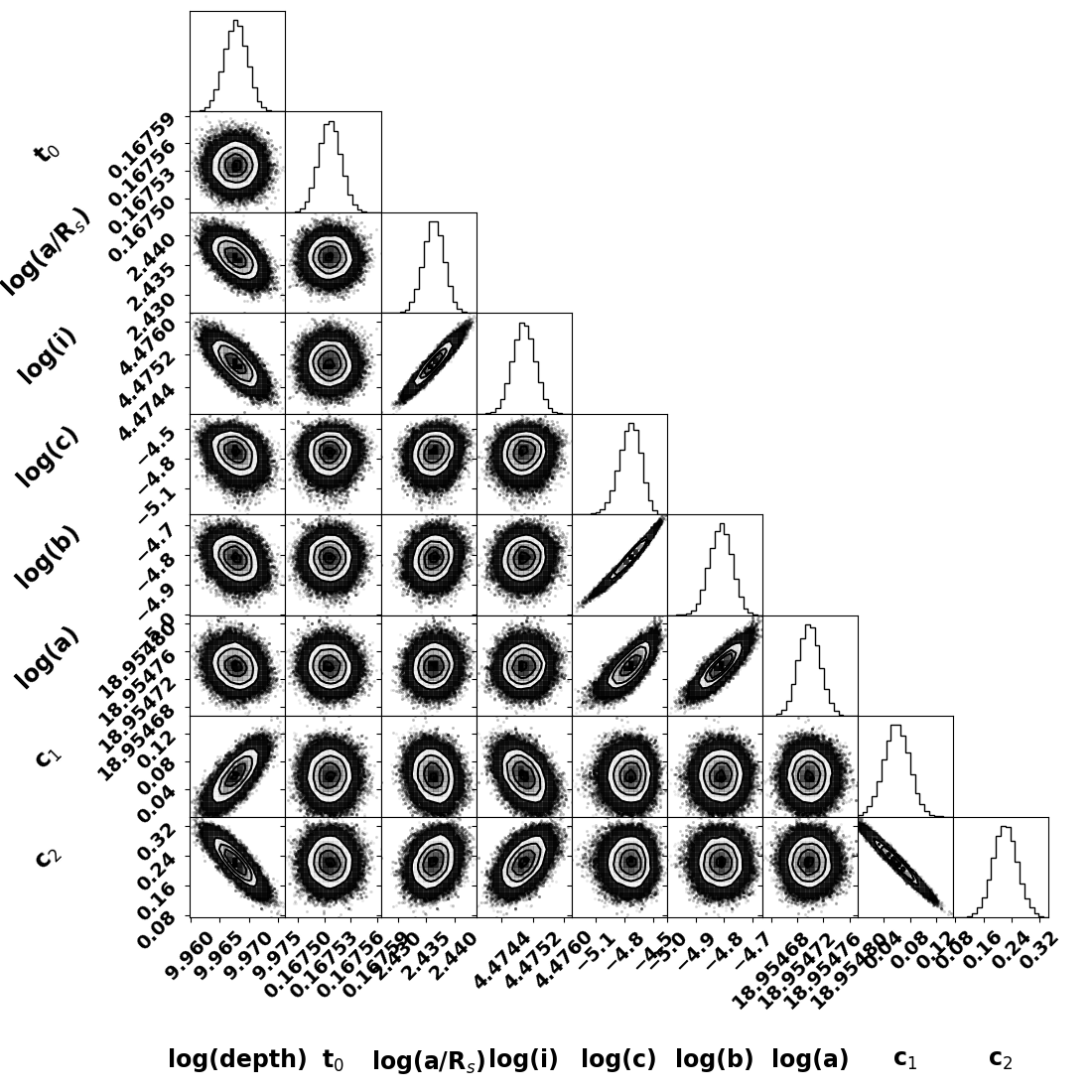}
			{\\ { }  Prism (all wavelengths) { } { } { } { } { } { } { } { } { } { } { } { } { } { } { } { } { } { } { } { } { } { } { } { } { } { } { } { } { } { } { } { } { } { } { } { } { } { } { } { } { }   Prism (>2 \textmu m)}
    \caption{Joint posterior probability distributions from white light curve MCMC processing for Prism.}
    \label{fig:corner prism wlc}
\end{figure*}

\begin{figure*}
	\includegraphics[trim={0cm 0cm 0cm 0cm}, clip,width=0.99\columnwidth]{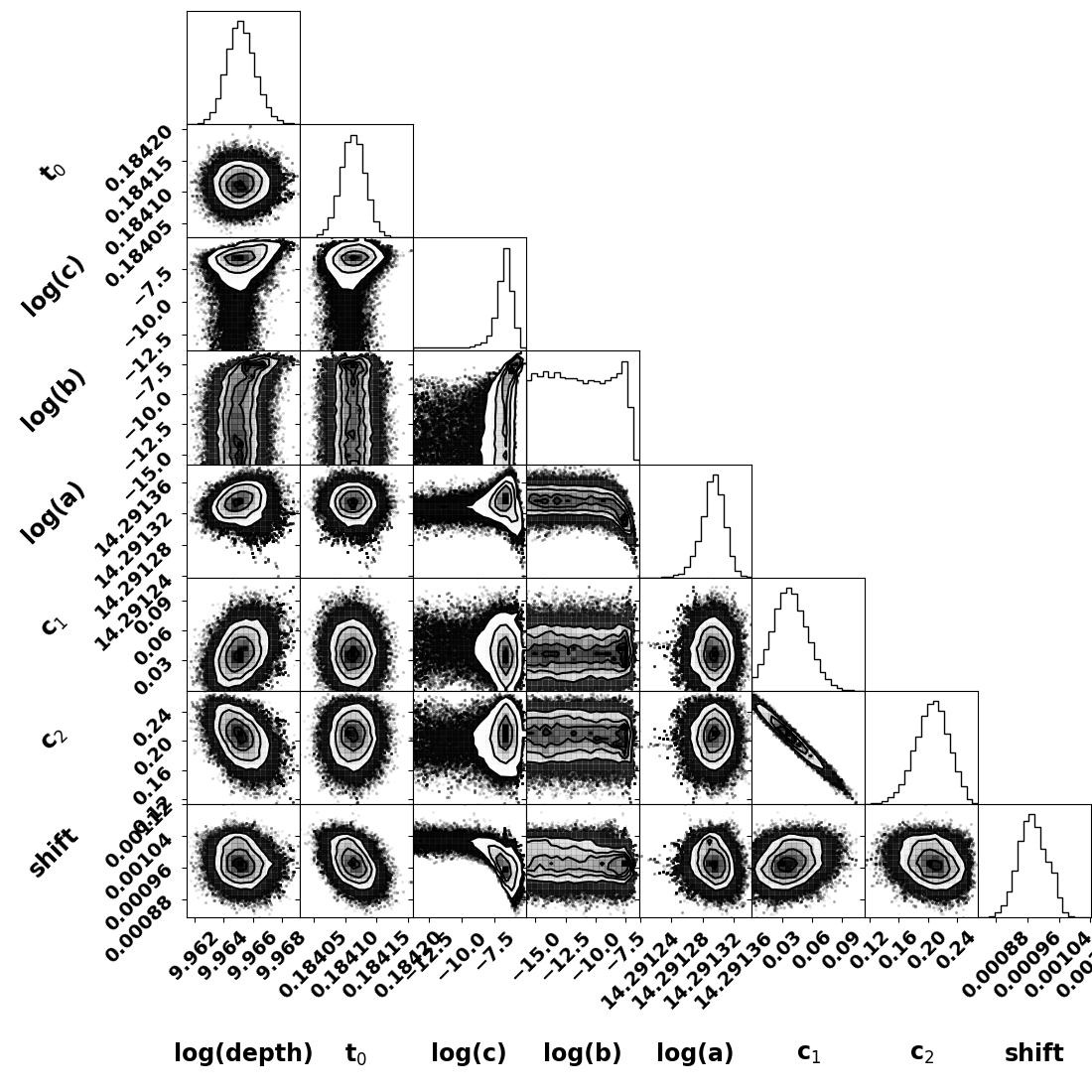}
	\includegraphics[trim={0cm 0cm 0cm 0cm}, clip,width=0.99\columnwidth]{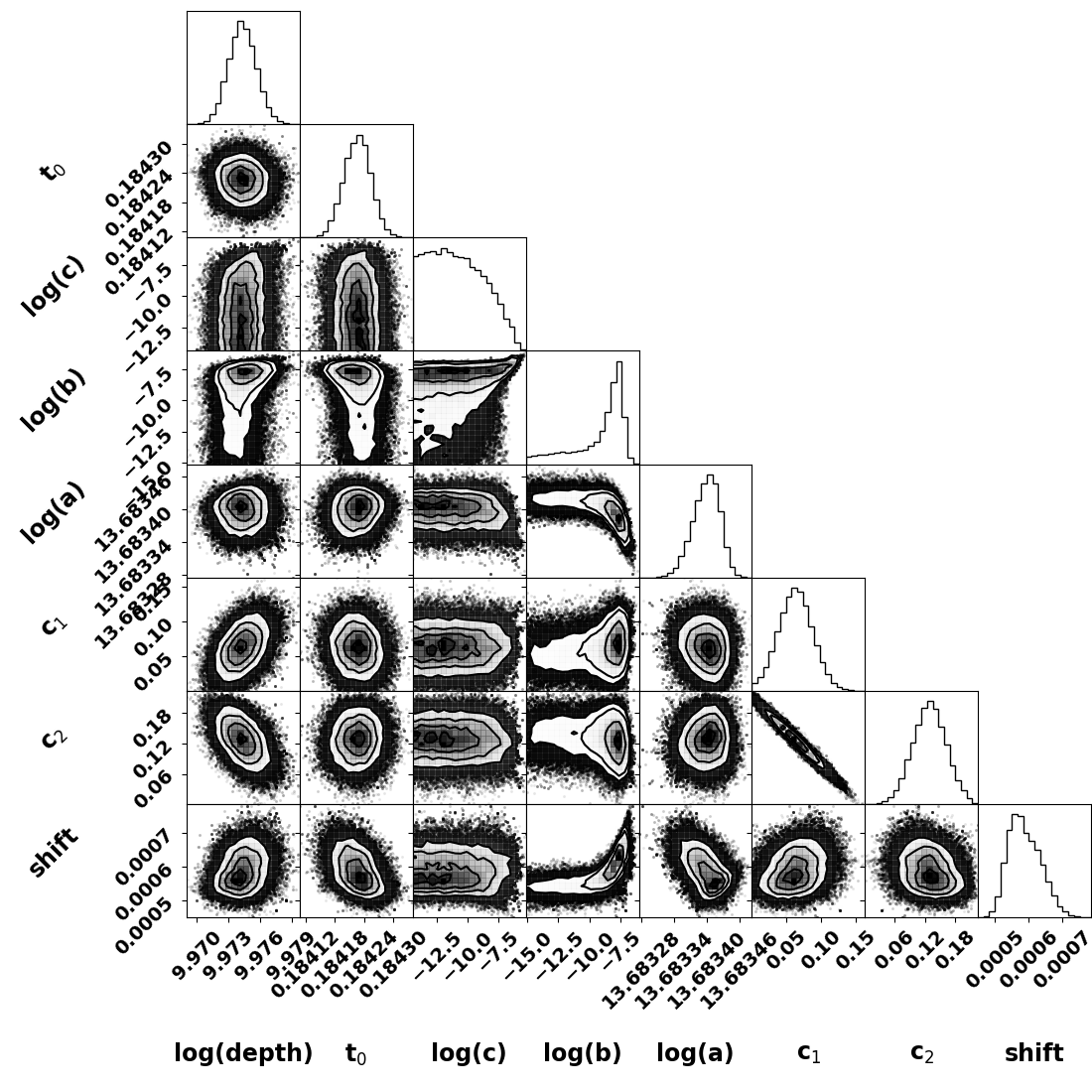}
			{\\ { } { } { } { } G395H NRS1 (baseline) { } { } { } { } { } { } { } { } { } { } { } { } { } { } { } { } { } { } { } { } { } { }   { } { } { } { } { } { } { } { } { } { } { } { } { } { } { } G395H NRS2 (baseline)}
	\includegraphics[trim={0cm 0cm 0cm 0cm}, clip,width=0.99\columnwidth]{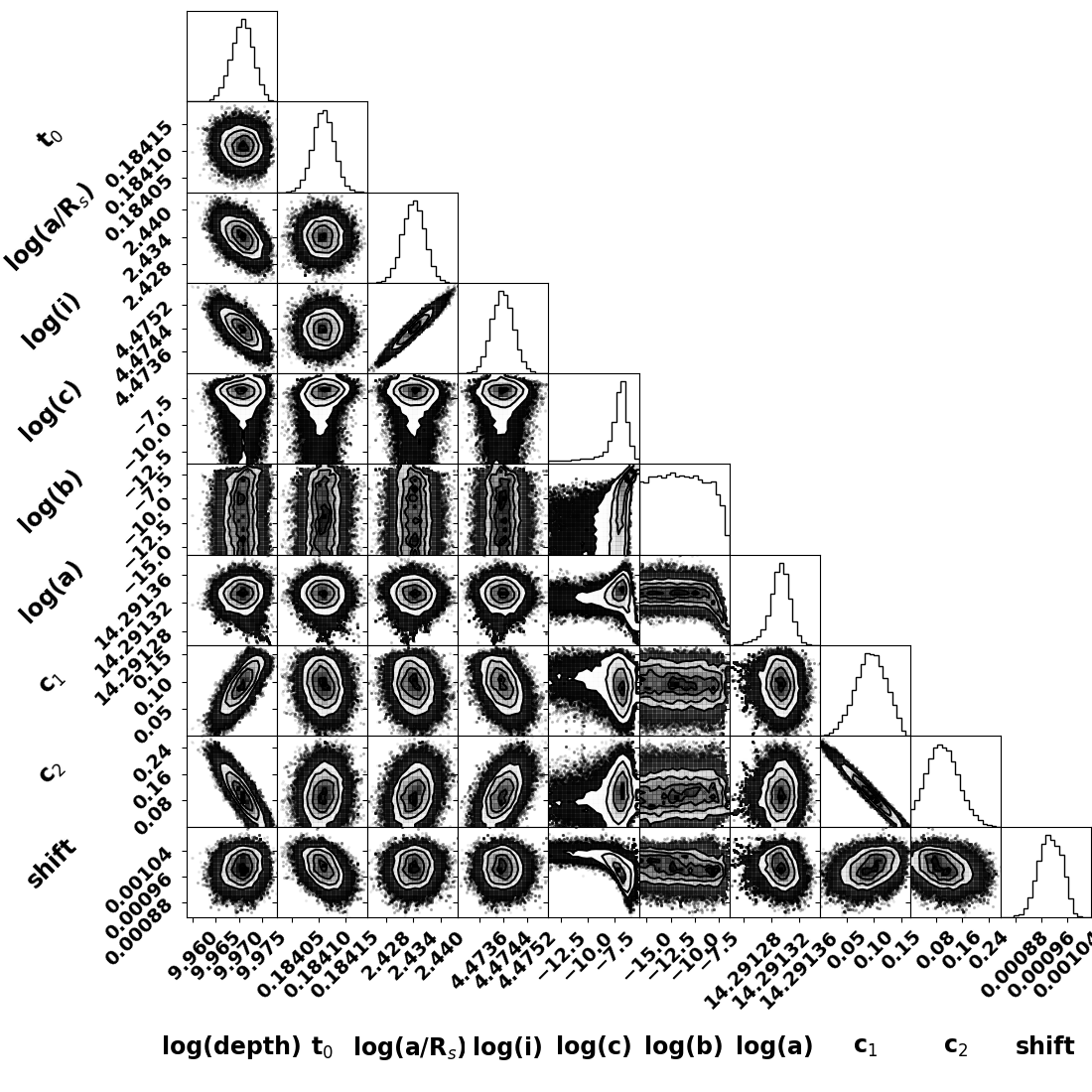}
	\includegraphics[trim={0cm 0cm 0cm 0cm}, clip,width=0.99\columnwidth]{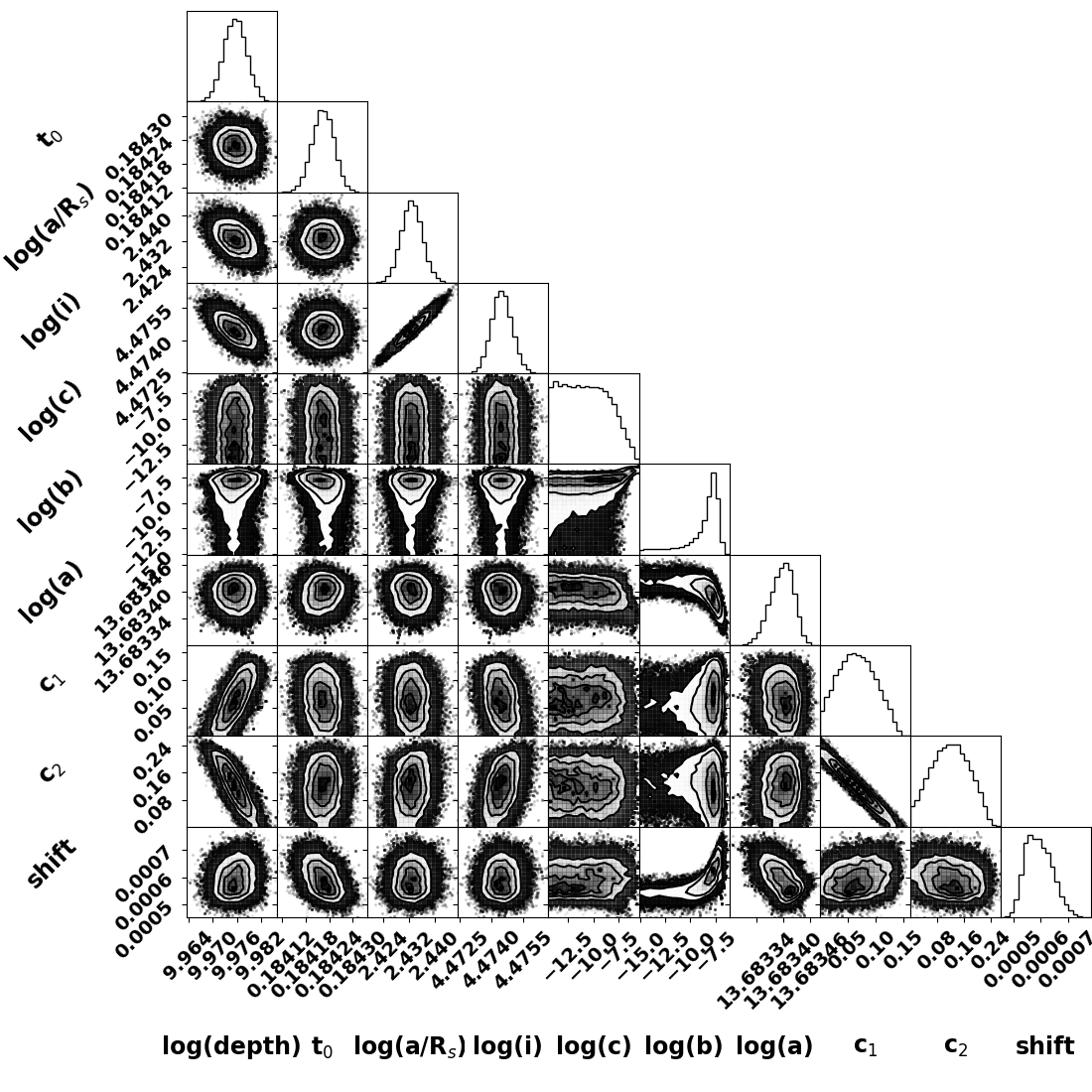}
			{\\ { } { } { } { } G395H NRS1 (independent) { } { } { } { } { } { } { } { } { } { } { } { } { } { } { } { } { } { } { } { } { } { }   { } { } { } { } { } { } { } { } { } { } { } G395H NRS2 (independent)}
    \caption{Joint posterior probability distributions from white light curve MCMC processing for G395H.}
    \label{fig:corner grating wlc}
\end{figure*}

\begin{figure*}
	\includegraphics[trim={0cm 0cm 0cm 0cm}, clip,width=0.99\columnwidth]{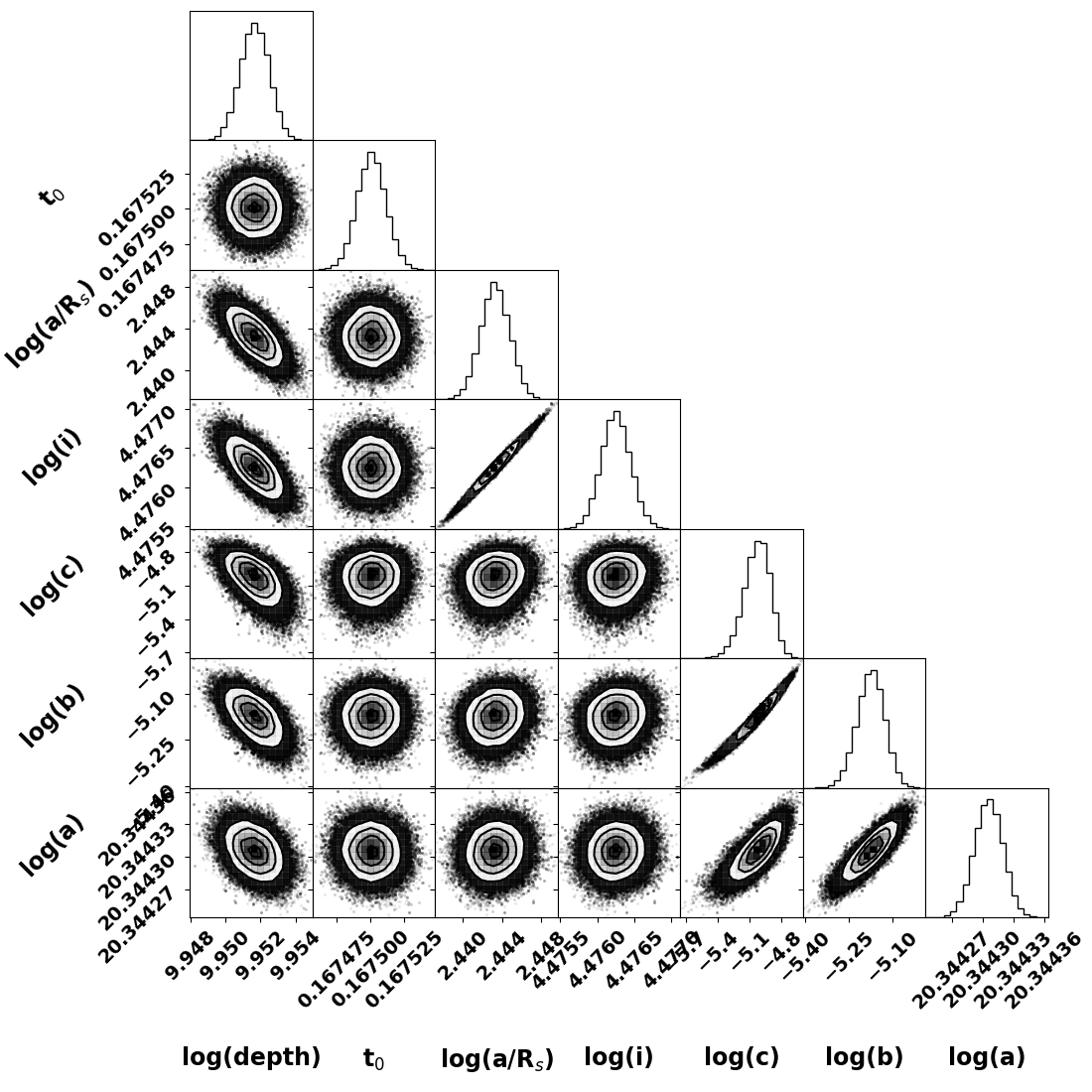}
	\includegraphics[trim={0cm 0cm 0cm 0cm}, clip,width=0.99\columnwidth]{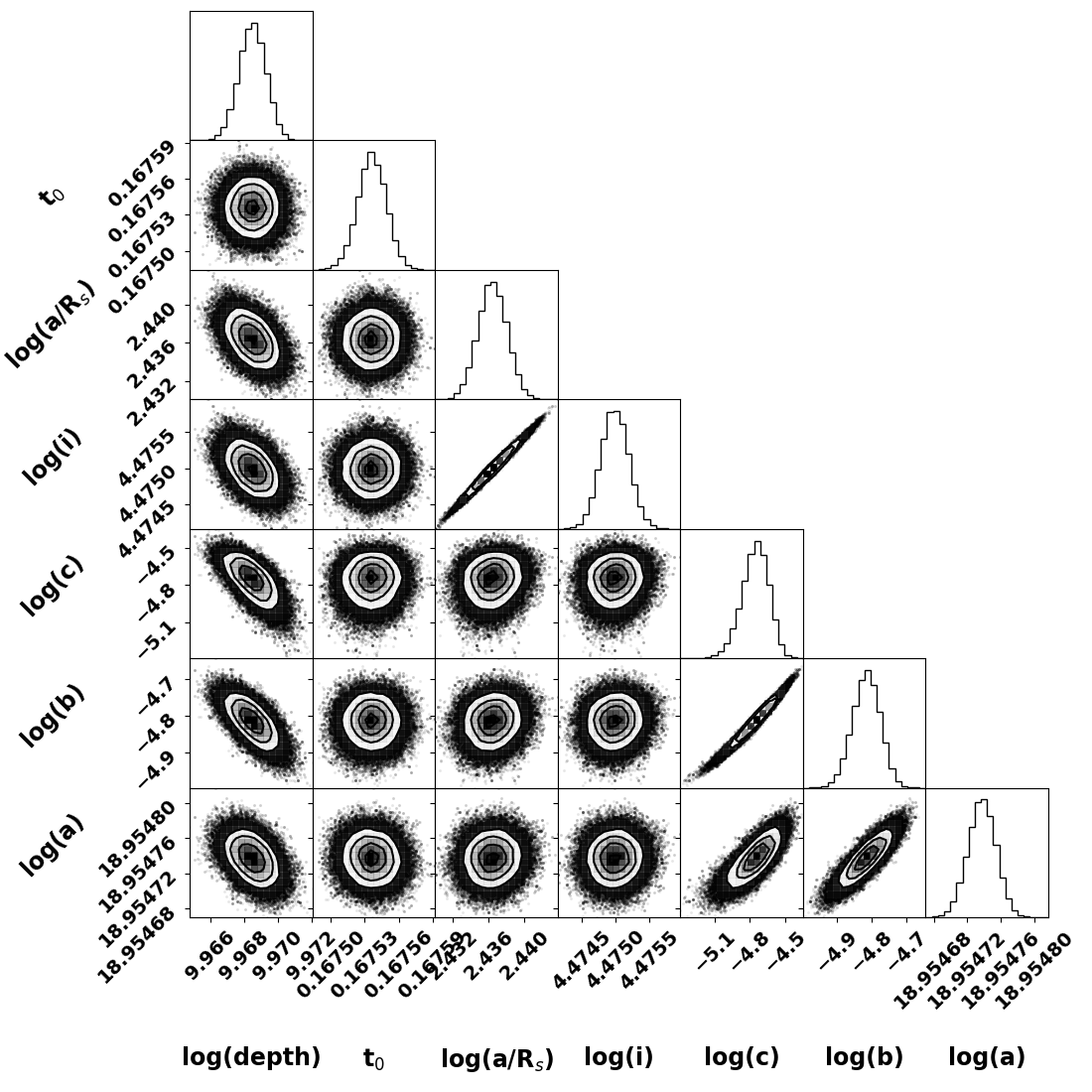}
			{\\ { }  Prism (all wavelengths) { } { } { } { } { } { } { } { } { } { } { } { } { } { } { } { } { } { } { } { } { } { } { } { } { } { } { } { } { } { } { } { } { } { } { } { } { } { } { } { } { }   Prism (>2 \textmu m)}
   
	\includegraphics[trim={0cm 0cm 0cm 0cm}, clip,width=0.99\columnwidth]{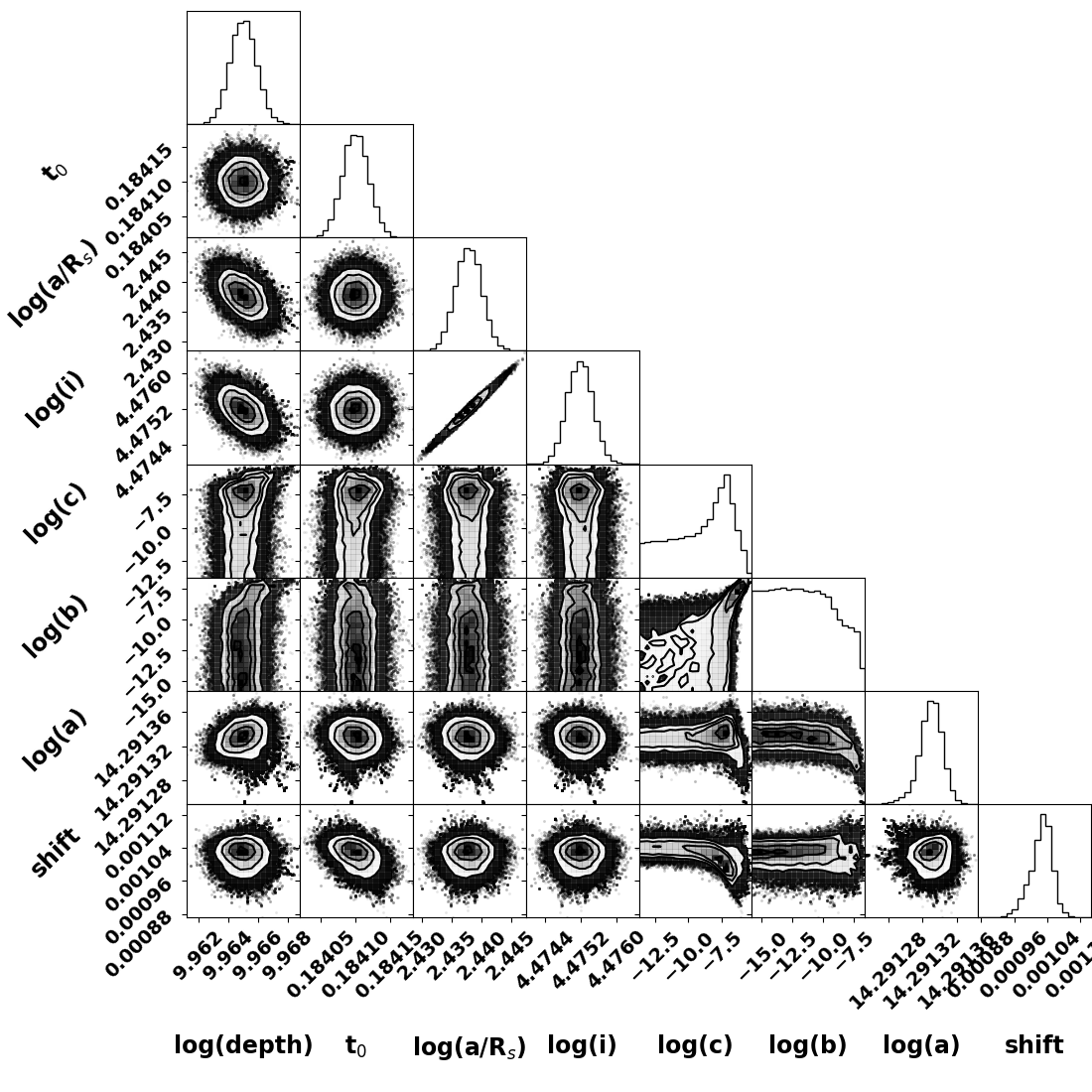}
	\includegraphics[trim={0cm 0cm 0cm 0cm}, clip,width=0.99\columnwidth]{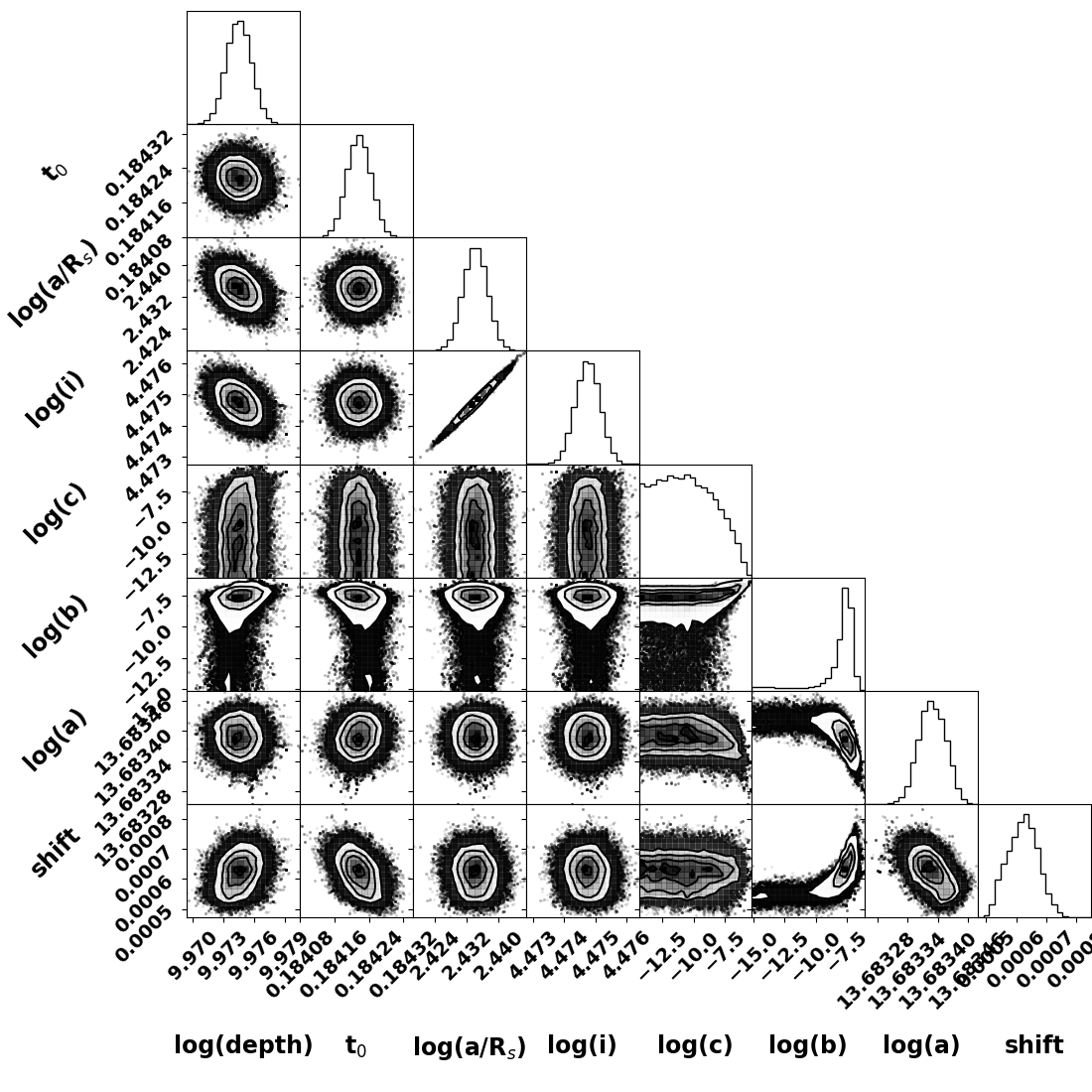}
			{\\ G395H NRS1 { } { } { } { } { } { } { } { } { } { } { } { } { } { } { } { } { } { } { } { } { } { } { } { } { } { } { } { } { } { } { } { } { } { } { } { } { } { } { } { } { } { } { } { }   G395H NRS2  }
    \caption{Joint posterior probability distributions from white light curve MCMC processing when using model 4-factor LDCs.}
    \label{fig:corner claret}
\end{figure*}

Here we show the joint posterior distributions from the MCMC parameter estimations using while light curves from the different instrument modes.   Figure \ref{fig:corner prism wlc} shows the results for the Prism baseline case with all wavelengths included and also when only wavelengths $>$ 2 \textmu m are included.   Figure \ref{fig:corner grating wlc} shows the results for the G395H baseline case (with $a'/R_s$ and $i$ fixed to Prism values), and for the `independent' cases (where $a'/R_s$ and $i$ and are fitted for).  Figure \ref{fig:corner claret} shows the results for cases where 4-factor model LDCs were used. 
 
%%%%%%%%%%%%%%%%%%%%%%%%%%%%%%%%%%%%%%%%%%%%%%%%%%

%%%%%%%%%%%%%%%%%%%%%%%%%%claret figures
%%%%%%%%%%%%%%%%%%%%%%%

\vspace{-0.5cm} 
\section{Spectral Light Curve Fits and Posteriors}
\begin{figure*}
	\includegraphics[trim={0cm 0.9cm 0cm 0.4cm}, clip,width=0.9\columnwidth]{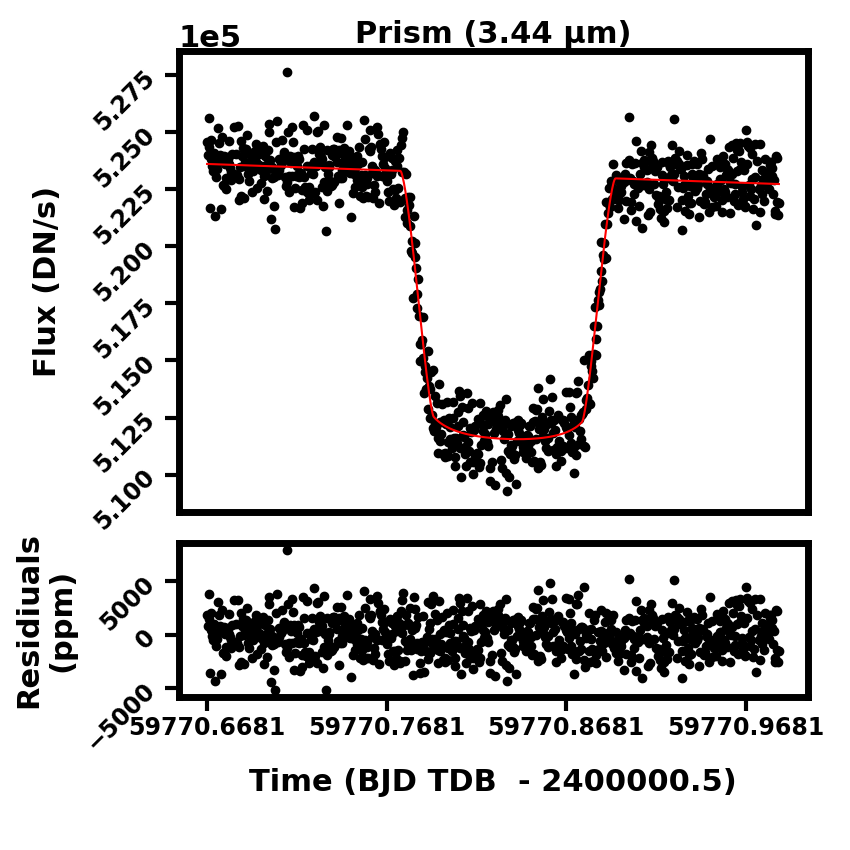}
	\includegraphics[trim={0cm 0cm 0cm 0cm}, clip,width=0.9\columnwidth]{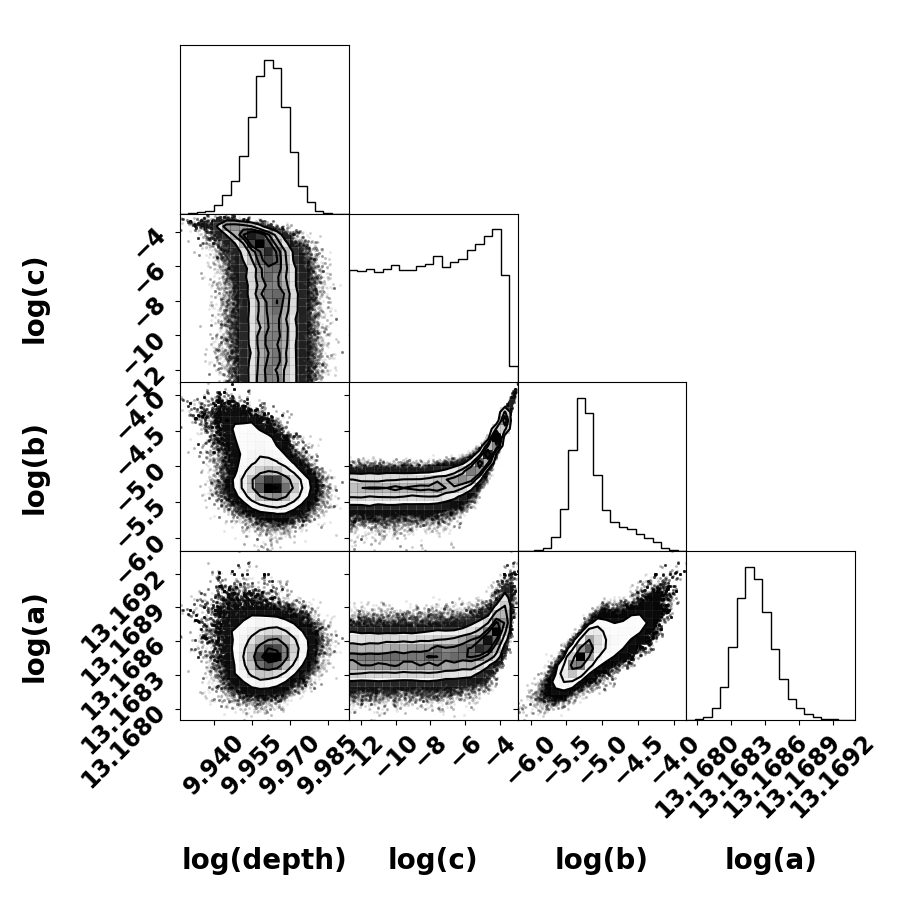}
	\includegraphics[trim={0cm 0.9cm 0cm 0.4cm}, clip,width=0.9\columnwidth]{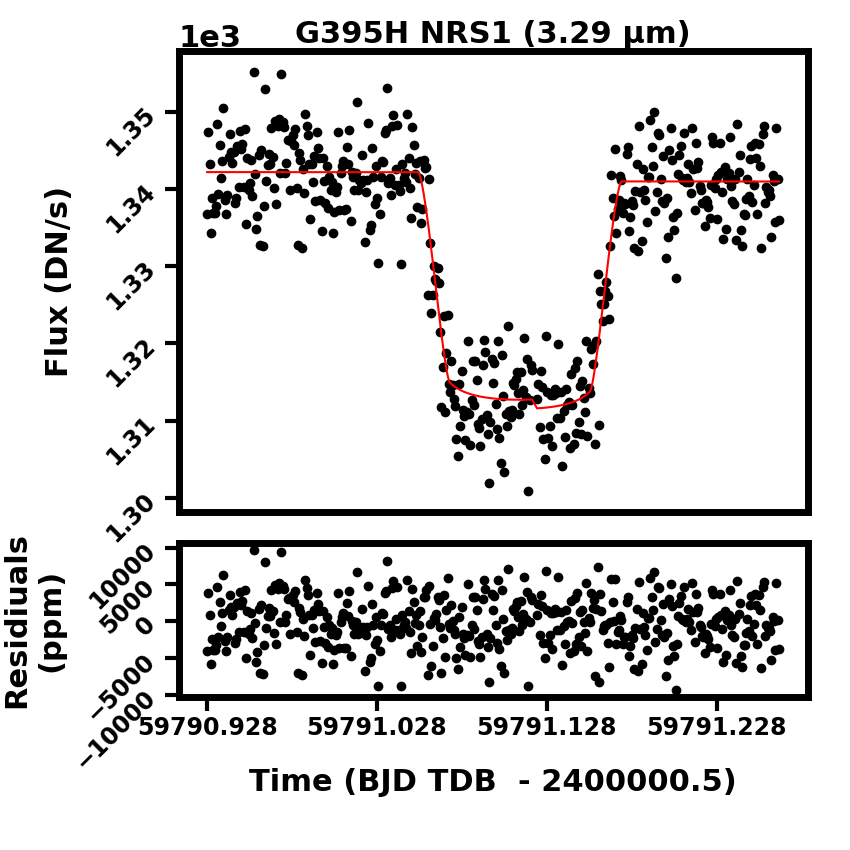}
 	\includegraphics[trim={0cm 0cm 0cm 0cm}, clip,width=0.9\columnwidth]{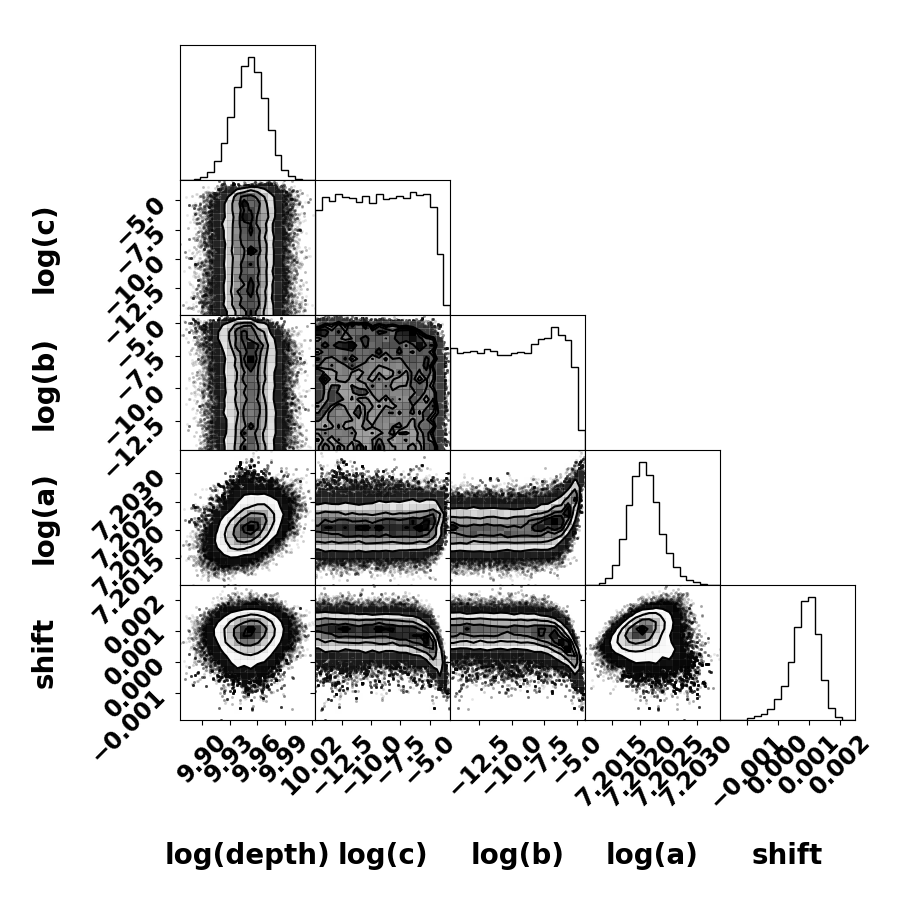}
  	\includegraphics[trim={0cm 0.9cm 0cm 0.4cm}, clip,width=0.9\columnwidth]{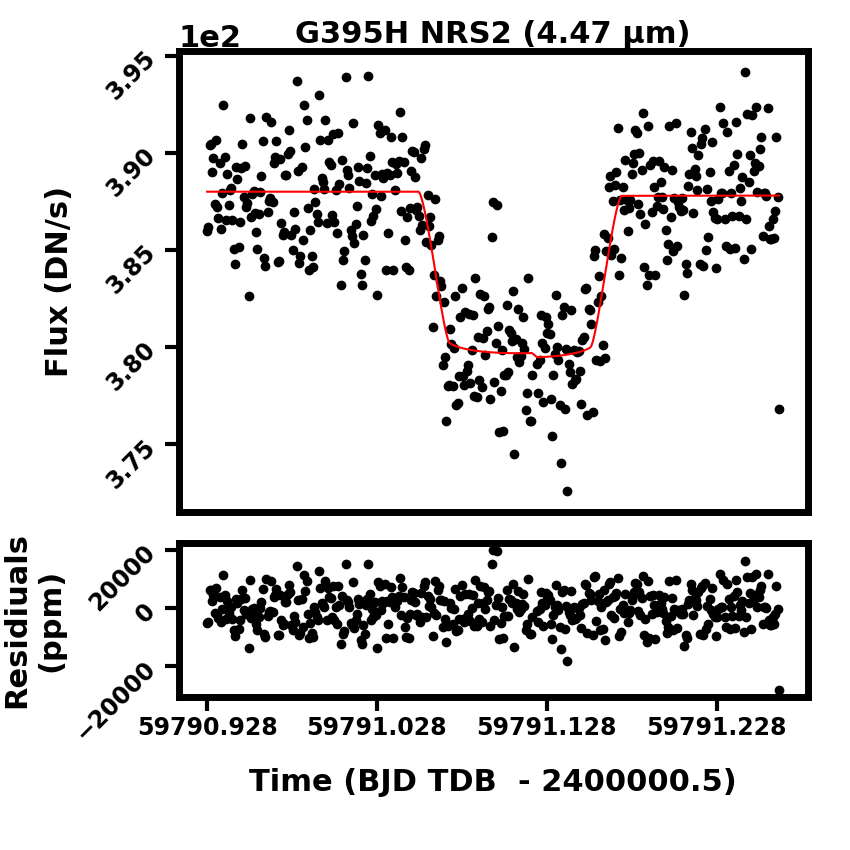}
	\includegraphics[trim={0cm 0cm 0cm 0cm}, clip,width=0.9\columnwidth]{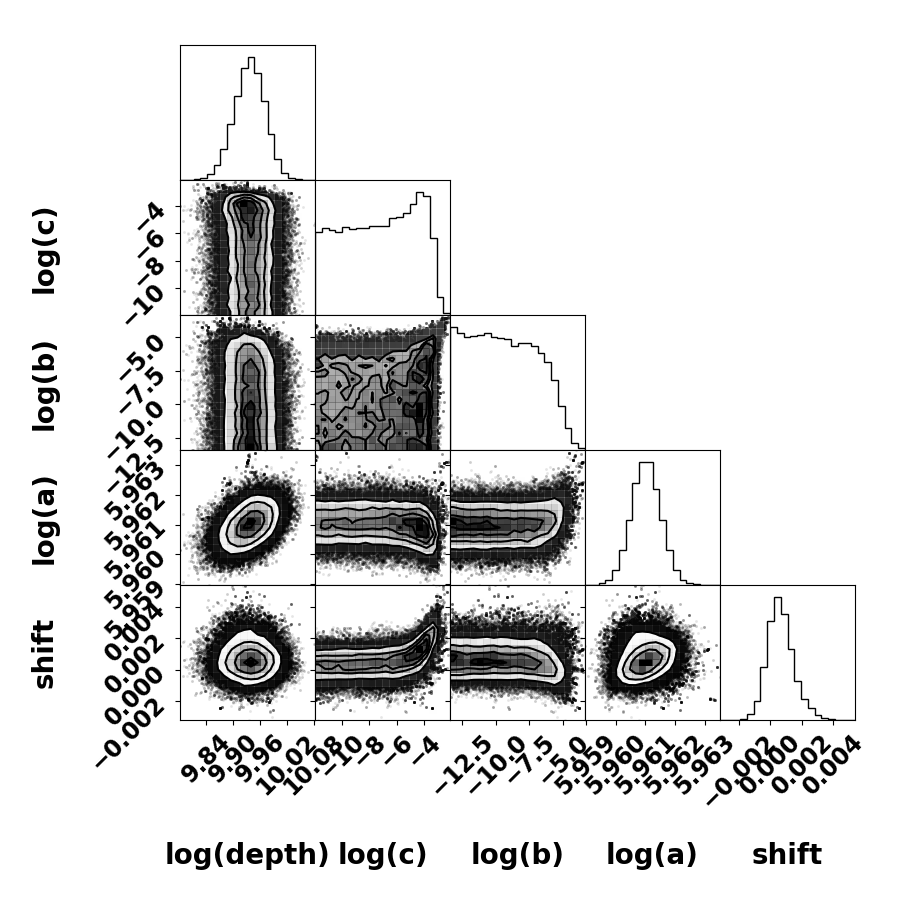}
    \caption{Joint posterior probability distributions and light curve fits for three example spectral light curve fits at full resolution.}
    \label{fig:corner slc example}
\end{figure*}

Here we show the light curve fits and corresponding joint posterior distributions from the MCMC parameter estimations for three example spectral light curves at full resolution.  Figure \ref{fig:corner slc example} shows an examples from the baseline cases for Prism (top), G395H NRS1 (middle) and G395H NRS2 (bottom) with central wavelengths shown above each light curve plot.

%%%%%%%%%%%%%%%%%%%%%%%%%%%

% Don't change these lines
\bsp	% typesetting comment
\label{lastpage}
\end{document}